\def\CleanVersion{1}
\documentclass[preprint,authoryear,12pt]{elsarticle}

\usepackage{amssymb}
\usepackage{amsmath}
\usepackage{amsfonts}
\usepackage{bm}
\usepackage{dsfont}
\usepackage{float}
\usepackage{xurl}
\usepackage[normalem]{ulem}
\usepackage{setspace}
\usepackage[expansion=false]{microtype}
\usepackage{hyperref}
\usepackage[title]{appendix}
\hypersetup{
    colorlinks = true,
    hypertexnames = false,
    allcolors = [rgb]{0,0.29,0.6}
    }
\usepackage{pdflscape}
  \usepackage{tikz}
  \usetikzlibrary{arrows.meta, decorations.pathreplacing}
\usepackage{enumitem} 
\usepackage{epstopdf}
\usepackage[official]{eurosym}
\usepackage{MnSymbol,wasysym}
\usepackage{caption}
\usepackage{subcaption}
\usepackage{graphicx}
\usepackage{rotating}
\usepackage{booktabs}
\usepackage{tabularx}
\usepackage{threeparttable}
\usepackage{adjustbox,lipsum}
\usepackage{algorithm,algpseudocode}
\usepackage{changepage}
\usepackage{xcolor}
\usepackage{multirow}
\usepackage{dcolumn}
\usepackage[a4paper, left=1.3in, right=1.3in, top=1in, bottom=1in]{geometry}
\DeclareMathOperator{\Var}{Var}
\DeclareMathOperator{\Cov}{Cov}

\newcolumntype{Y}{>{\centering\arraybackslash}X}

\usepackage{longtable}
\usepackage{array}

\usepackage{multibib}
\newcites{A}{Bibliography of the Appendices}

\newenvironment{vardictionary}{%
  \begin{itemize}[
    label={},
    leftmargin=0pt,
    itemindent=0pt,
    labelsep=0pt,
    itemsep=1.1em,
    topsep=0.4em,
    parsep=0.25em
  ]
}{%
  \end{itemize}
}

\newcommand{\varheading}[2]{%
  \textbf{#1}\par\nobreak
  {\small\itshape #2\par}\nopagebreak
  \smallskip
}

\usepackage{xcolor}

\ifdefined\CleanVersion
  \newcommand{\rev}[1]{#1}
  \newcommand{\old}[1]{}
  
\else
  \newcommand{\rev}[1]{\textcolor{red}{#1}}
  \newcommand{\old}[1]{\ifmmode\allowbreak\,\else\space\fi\textcolor{black}{(#1)}}
\fi

\begin{document}

\begin{frontmatter}



\title{What Drives Energy Use? Prices, Policy, and Inefficiency}


\author[david1]{David Benatia\corref{cor1}} %% author
\ead{david.benatia@hec.ca} %% Email of David

\author[po1]{Rémy Molinié} %% author
\ead{remy.molinie@hec.ca} %% Email of 

\author[po1]{Pierre-Olivier Pineau} %% author
\ead{pierre-olivier.pineau@hec.ca} %% Email of Pierre-Olivier

%% Corresponding author marker
\cortext[cor1]{Corresponding author}

% %% Author affiliations
% \affiliation[alicia1]{organization={CNRS, CREST (UMR 9194), ENSAE, Institut Polytechnique de Paris}, 
% addressline={5 Avenue Henry Le Chatelier}, 
% city={Palaiseau},
% postcode={91120}, 
% country={France}}

\affiliation[david1]{organization={HEC Montréal, Département d'Économie Appliquée}, 
addressline={3000 Chemin de la Côte-Sainte-Catherine}, 
city={Montréal},
postcode={H3T 2A7}, 
state={QC},
country={Canada}} 
\affiliation[po1]{organization={HEC Montréal, Département de Sciences de la Décision}, 
addressline={3000 Chemin de la Côte-Sainte-Catherine}, 
city={Montréal},
postcode={H3T 2A7}, 
state={QC},
country={Canada}} 

% \affiliation[david2]{organization={CREST (UMR 9194), ENSAE, Institut Polytechnique de Paris}, 
% addressline={5 Avenue Henry Le Chatelier}, 
% city={Palaiseau},
% postcode={91120}, 
% country={France}}

\begin{abstract}
Why is U.S.\ energy use so high, even among rich economies? Using a 2006--2022
panel of the 50 states and DC, an LMDI decomposition documents a 12.8\%
national decline alongside substantial cross-state heterogeneity. To explain
differences in energy-use levels across states, we estimate a microfounded
stochastic-frontier model and adapt the LMG relative-importance method to the
frontier and inefficiency components. Our approach assigns
56.3\% of variation to the frontier, 42.2\% to
inefficiency, and 1.5\% to noise. Prices and efficiency-policy environments are
the two leading observed contributors, jointly accounting for 31\%: prices
operate mainly through the benchmark, whereas policy contributes through both
components. Their leading ranks are stable across 21 specifications. Persistent low prices, uneven efficiency-policy environments, structural demand, and a sizeable composite inefficiency component help explain high U.S.\ energy use. Policy should therefore combine durable price signals, coherent efficiency programs, structural investment, targeted measures addressing avoidable energy use, and targeted household support.
\end{abstract}

\begin{highlights}
  \item A microfounded SFA distinguishes best-practice energy use from inefficiency and noise
  \item Adapted LMG method quantifies factor shares separately for frontier and inefficiency
  \item Prices and policies remain leading factors in 21 models and explain 31\% of variation
  \item Prices operate largely through the frontier; efficiency policy affects both channels
  \item High U.S. use reflects persistent low prices, policy gaps, structure and inefficiency
\end{highlights}


\end{frontmatter}

\pagebreak
\section{Introduction}

Per capita energy use in the United States is an outlier even among rich
  countries. Primary energy consumption per capita is nearly twice the OECD                   
  average and more than three times the global  average \citep{WorldBank_WDI_2025}. These gaps are often justified by differences in economic                          
  activity, climate, or industrial structure, yet other high-income                           
  economies sustain comparable standards of living with far less energy per                   
  capita \citep{CsereklyeiStern2015,SteinbergerRoberts2010}. This fact raises a question central to the design of energy and                     
  climate policy: what drives variations in energy use, and how much of                    
  it is attributable to factors beyond economic or geographic fundamentals?   

In this paper, we use variation across the 50 U.S.\ states and the District of
Columbia over 2006--2022 to assess whether high energy use mainly reflects
income, climate, and production structure, or whether energy prices, policy
environments, and energy-use efficiency also play central roles. We combine
Logarithmic Mean Divisia Index (LMDI) decomposition, stochastic frontier
analysis (SFA), and variable-importance methods. We find that high energy use in the United States reflects persistent
low-price environments, uneven efficiency-policy environments, and a sizeable
composite inefficiency component alongside structural demand. To quantify the relative importance of these
different sources, we extend variable-importance analysis to SFA, allowing each
factor's contribution to be measured separately for the benchmark level of
energy use and the composite inefficiency component.

\begin{figure}[H]
  \centering
  \caption{Per capita energy consumption in 2022 and change in per capita energy consumption by state and for the world, 2006--2022.}
  \includegraphics[width=0.80\textwidth]{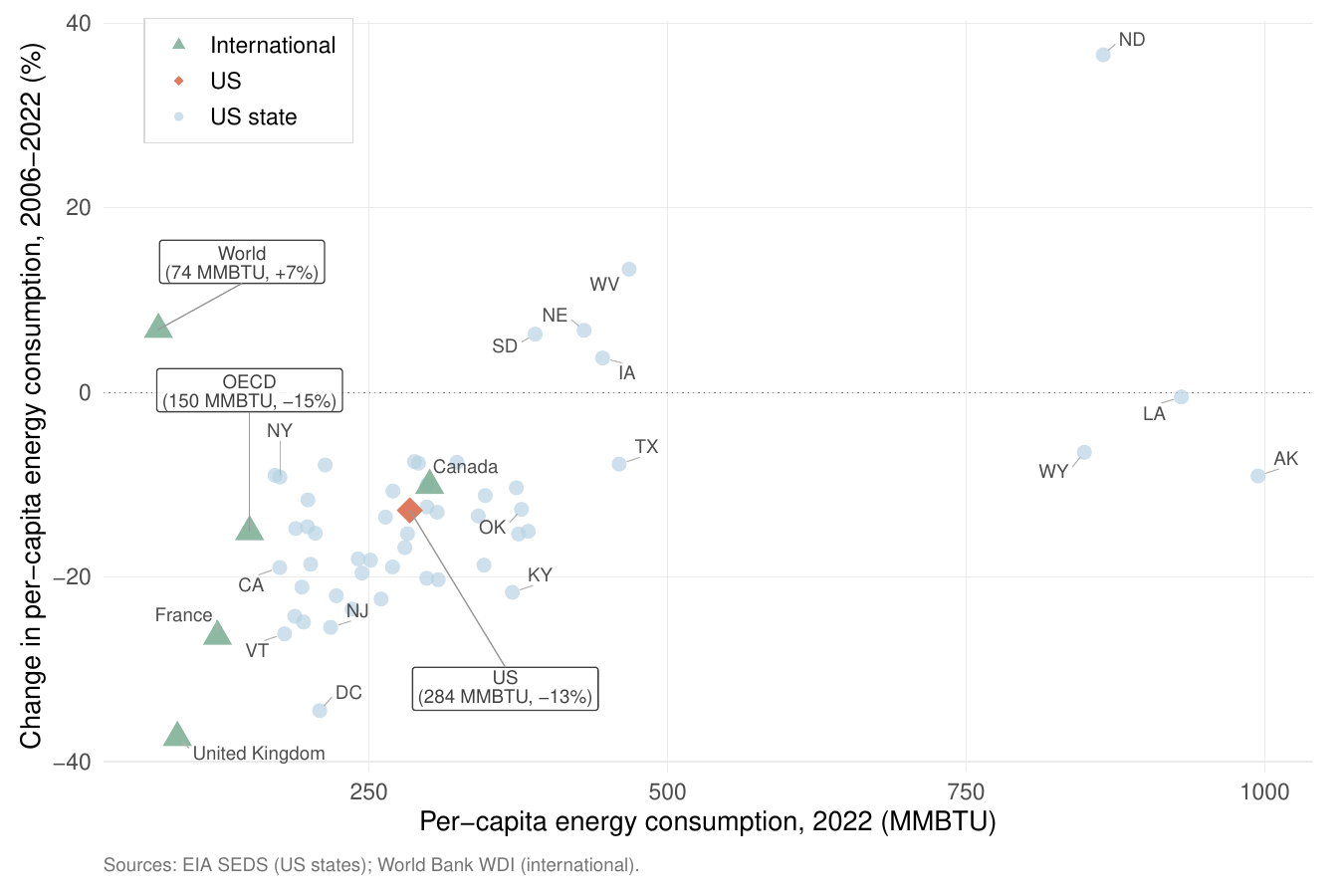}
  \label{fig:map_cons_change}
\end{figure}

Between 2006 and 2022, aggregate U.S.\ per capita total energy consumption by end-use sectors fell by approximately 12.8\%, a decline driven almost entirely by reductions in energy intensity rather than by variations in economic activity or climate. National aggregates, however, conceal substantial heterogeneity across states. Figure~\ref{fig:map_cons_change} shows the percentage change in per capita energy consumption across the 50 states and DC over this period.  Most states record a decline, typically
on the order of 10--20\%. In contrast, a small number of states experience
increases, where the rapid expansion of fossil-fuel
extraction coincides with a marked rise in per capita energy use.

The discrepancies shown in Figure~\ref{fig:map_cons_change} reflect several
factors. Climate, long passenger and freight distances, low population density,
resource extraction, and energy-intensive production all raise energy
requirements. These
pressures are especially strong in Canada and some U.S.\ states, where resource extraction,
raw-material processing, and long-distance freight contribute substantially to
territorial energy demand \citep{IEACanada2022}. Complementing these
factors, \citet{FilippiniHunt2011} place both countries toward
the lower-efficiency end of their OECD comparison after accounting for income,
climate, prices, and economic structure.

Buildings and mobility provide a second explanation. Although climate matters,
dwelling size, building characteristics, and comfort levels are also important
contributors to high U.S.\ energy use
\citep{IEAEnergyUse2007,OECDBrickByBrick2021}. Automobile dependence and high
vehicle use likewise make U.S.\ passenger-transport energy use unusually high
among rich economies \citep{IEAEnergyUse2007}. Within the United States,
differences in urban form and transit provision are associated with substantial
variation in household vehicle travel \citep{BentoEtAl2005}, while a comparison
with Germany highlights the roles of fuel taxation, parking and road policy,
public transport, and land-use planning \citep{Buehler2011}.

Energy prices offer a stylized comparative clue. Using a simple energy product index places
end-user prices in France and Germany at 2.03 times the United States--Canada
level, while the United States--Canada average for 2019 total energy supply per
person was 2.08 times the France--Germany average
\citep{IEAKeyWorldEnergy2021}.\footnote{The underlying series and construction
are described in Section~\ref{sec:policy}.} This inverse pattern makes
comparatively low prices a plausible determinant of high energy use in North
America. Consistent with this interpretation, \citet{Sterner2007} estimates
that European gasoline use would have been approximately twice as high under
low U.S.\ fuel-tax levels. More generally, energy demand responds more strongly
to prices over longer horizons, when households and firms have time to adjust
their consumption and investment decisions \citep{LabandeiraEtAl2017}.

Energy-efficiency policy can affect many of the same long-lived decisions, but
\rev{program} outcomes vary. Stricter Florida building codes reduced energy use in
new homes \citep{JacobsenKotchen2013}, whereas realized savings from California
codes and U.S.\ weatherization were substantially below their engineering
projections \citep{Levinson2016,FowlieEtAl2018}. Rebound offsets part, but
generally not all, of the savings from improved efficiency
\citep{GillinghamEtAl2016}. At a broader level, stronger energy-governance
environments are associated with greater efficiency across OECD
countries \citep{BarreraSantana2022EJ}. 

Taken together, this evidence identifies prices and efficiency policy as
plausible margins of adjustment. Aggregate trends likewise show that energy
use can become less tightly coupled to economic growth. U.S.\ final energy
consumption remained approximately flat from 2010 to 2022 while GDP expanded,
and Canada also experienced substantially slower growth in demand than in
output \citep{IEACanada2022,IEAUnitedStates2024}. Identifying what produced
this divergence requires separating changes in activity, structure, and
intensity. LMDI provides that accounting \citep{ANG2005}, but its intensity
component does not reveal the underlying mechanism. U.S.\ decompositions reach different conclusions: \citet{Metcalf2008}
attributes most of the 1970--2003 decline to falling energy intensity within
sectors, while \citet{SueWing2008} finds structural change dominant over
1958--2000 but within-industry improvements dominant after 1980. Cross-country results also vary
\citep{VoigtEtAl2014}, and \citet{Levinson2021} finds little evidence that
deindustrialization explains recent differences in state trends, alongside
mixed evidence on prices and regulation. Declining aggregate intensity
therefore documents improvement without settling why it occurred or why
U.S.\ energy use remains high.

The energy-demand-frontier literature addresses part of this problem by
comparing observed use with a benchmark conditional on income, prices,
climate, and economic structure. These studies show why raw energy intensity
is an incomplete measure of efficiency
\citep{FilippiniHunt2011,FilippiniHunt2012}, while subsequent work separates persistent
from transient efficiency, incorporates rebound, and introduces policy or
governance variables
\citep{FilippiniHunt2016,OreaLlorcaFilippini2015,
FilippiniHuntZoric2014,BarreraSantana2022EJ}. The literature nevertheless
concentrates mainly on coefficient estimates, efficiency scores, and rankings.
Predictive studies likewise rank energy-demand determinants without separating
benchmark energy use from inefficiency \citep{Korsavi2025}. What
remains missing is a determinant-level account of how observed factors
contribute separately to the conditional benchmark and the composite
inefficiency component.

Our paper bridges these strands of literature \rev{and makes three contributions: empirical, theoretical, and methodological, by focusing on the United States.} The state panel allows us to compare income, climate, production,
mobility, and the built environment with energy prices, energy-efficiency
policy environments, and composite inefficiency under a common national
setting. \rev{First, our empirical contribution is to show} that structural requirements
remain important but incomplete explanations of high U.S.\ energy use.
Persistent price environments and state energy-efficiency policy environments
emerge as the two leading observed contributors when both modeled components
are considered \rev{(jointly accounting for 31\% of total log-energy variation in the baseline and 24\%--39\% across 21 total-energy specifications)}, while mobility and fossil-fuel production remain key 
\rev{structural factors. The frontier and inefficiency receive 56.35\% and 42.20\% of total log-energy variance, respectively, with the remaining 1.45\% assigned to noise. Prices contribute mainly through the frontier, whereas policy contributes through both components. The sizeable composite inefficiency component}  provides a
further explanation beyond the energy use associated with observed state
fundamentals. Together with the international evidence, the results also offer
a coherent interpretation of Canada's high-use pattern: stronger geographic
and resource-production requirements are reinforced by a similarly low-price
North American energy environment.

\rev{Second,} building
on \citet{FilippiniHunt2015}, we develop a parsimonious microfoundation linking
the conditional benchmark and inefficiency component to demand for
energy services, substitution between energy and efficiency capital,
technology adoption, and usage. \rev{Third,} 
we adapt established variable-importance methods to SFA. Rather than reporting
only coefficients or state rankings, we measure how much each observed factor
contributes through the benchmark and composite-inefficiency components
separately, retain the unexplained part of the latter, and express both
allocations on a common descriptive scale. Exact LMG provides the main additive
attribution \citep{Groemping2006,Gromping2007}, while Random Forest importance
provides a separate frontier-only predictive check
\citep{Gromping2009,strobl2008}. This allows us to ask not only which factors
are associated with high energy use, but how much they matter and through which
component.

The remainder of this paper is organized as follows. Section~\ref{sec:lmdi}
presents the LMDI decomposition of U.S.\ energy-use trends.
Section~\ref{sec:eco_frame} develops the economic framework, and
Section~\ref{sec:data} describes the data. Section~\ref{sec:empirical}
presents the empirical framework and results, including the contribution
analysis. Section~\ref{sec:policy} discusses policy implications, and
Section~\ref{sec:conclusion} concludes.

\section{Energy use trends in the United States}
\label{sec:lmdi}
In the United States, total energy consumed by end-use sectors has been on a declining trend since the mid-2000s, despite continued growth in real output. This pattern is consistent with improvements in energy efficiency, gradual structural change, and policy-driven conservation efforts \citep{EIA_AEO_2025}. This aggregate measure of energy use is dominated by the industrial and transport sectors, with residential and commercial buildings accounting for the remainder. Over the last twenty years, these sectoral shares have changed only slowly, suggesting that the decline in per capita energy use reflects primarily changes within sectors rather than large shifts in the sectoral composition of demand.

In this section, we document recent trends in U.S. per capita energy consumption by end-use sector using an LMDI decomposition. The analysis has two objectives. First, it provides a descriptive accounting of how much of the observed historical change in per capita energy use can be attributed to activity, intensity, and climate effects across the main end-use sectors. Second, it motivates our subsequent stochastic frontier analysis by showing that changes in energy intensity are the dominant driver of recent aggregate trends, while concealing substantial cross-state heterogeneity of policy relevance. For readability, we keep the exposition in this section concise. Detailed information on the data is provided in Section~\ref{sec:data}, and the decomposition methodology is set out in \ref{app:lmdi}.

\subsection{LMDI decomposition framework}

We decompose the total energy consumed by end-use sector using annual data for the 50 U.S.\ states and the District of Columbia over the period 2006--2022, using both state-level and
national aggregates.\footnote{We use the EIA State Energy Data System (SEDS), per capita ``Total Energy Consumed by End-Use Sector'' expressed in million BTU and disaggregated into sectors. For each sector, this measure includes (i) the energy directly consumed within the sector (final energy) and (ii) the share of electricity system losses allocated to that sector’s electricity use.}  Total energy consumption is disaggregated into four
main sectors: residential, commercial, industrial, and transport. The sectoral activity indicators follow the methodologies outlined in recent decomposition studies for Europe
and the United States \citep{Reuter2019,SerranoPuente2021,Jiang2016,
Jiang2019}.

Following \citet{ANG2005,Ang2015}, we represent per capita energy consumption
in each sector $k$ as the product of an activity term $A_{k,\text{pc}}$, an
intensity term $\mathrm{INT}_{k}$, and, for the residential and commercial
sectors, a climate factor $\mathrm{CLI}_{k}$:
\begin{equation}
E_{k,\text{pc}} \;=\; A_{k,\text{pc}} \times \mathrm{INT}_{k} \times \mathrm{CLI}_{k}.
\end{equation}
We then decompose the observed
change in $E_{k,\text{pc}}$ between two dates into three contributions,
corresponding to changes in activity, intensity, and climate. Aggregating across
the four sectors yields an overall decomposition of the change in total
per capita energy use, $E_{\text{tot,pc}}$, into the same three effects.\footnote{The exact expressions for
the weights used in the decomposition are in  \ref{app:lmdi}. %In practice,
%once the database is organized by sector and factor, we implement the
%decomposition using the \texttt{LMDIR} package \citep{Heun_LMDIR_0_1_14_2024}.
}

To fix ideas, consider the residential sector. For each state and year, we
observe (i) per capita residential energy use, $E_{\text{res,pc}}$, (ii)
residential floor area per capita, $A_{\text{res,pc}}=\rev{\textit{ResSurf/pc}}$,
and (iii) a climate factor $\mathrm{CLI}_{\text{res}}$ constructed from heating
and cooling degree days (HDD and CDD, respectively). We then define residential intensity as energy use per
unit of floor area, net of climate:
\begin{align}
E_{\text{res,pc}} \;&=\;
A_{\text{res,pc}} \times \mathrm{INT}_{\text{res}} \times \mathrm{CLI}_{\text{res}},
\\
\mathrm{CLI}_{\text{res}} \;&=\;
\left(\frac{\text{HDD}+1}{\text{HDD}^{\text{ref}}+1}\right)^{\alpha_{\text{res}}}
\left(\frac{\text{CDD}+1}{\text{CDD}^{\text{ref}}+1}\right)^{\beta_{\text{res}}},
\\
\mathrm{INT}_{\text{res}} \;&=\;
\frac{E_{\text{res,pc}}/A_{\text{res,pc}}}{\mathrm{CLI}_{\text{res}}}.
\end{align}
Thus, changes in residential energy use per capita between an initial year $0$
and a final year $T$ can be viewed as arising from changes in floor area per
capita (activity), in climate conditions (HDD and CDD relative to their
reference values), and in the residual intensity term. Here $1$ denotes one
raw degree day and is used to address the presence of zeros in some state--years. The residential and commercial climate slopes ($\alpha$'s and $\beta$'s) are estimated
on the 51-jurisdiction panel using state and year effects; this preferred
normalization uses within-state weather variation after removing persistent
state levels and common annual shocks.\footnote{Table~\ref{tab:climate_bridge} reports
the associated regression estimates.}

The LMDI-I procedure translates this identity into a decomposition of the
\emph{relative} change in $E_{\text{res,pc}}$ between years $0$ and $T$ into
three multiplicative effects. For each factor (activity, intensity, climate), it
computes the log-change of that factor between $0$ and $T$ for each state, then
aggregates these log-changes across states using weights that reflect the
importance of each state in total residential energy use, averaging information
from the initial and final years by means of the logarithmic mean. 

The same
logic applies to the other sectors, with activity measured by commercial floor
area per capita in the commercial sector, by industrial value added per capita
in the industrial sector, and by vehicle miles traveled per capita in the
transport sector. Climate enters only for residential and commercial buildings.
A key advantage of LMDI-I is that the decomposition is \emph{exact}: by
construction, the product of the aggregated activity, intensity, and climate
effects reproduces the observed change in $E_{k,\text{pc}}$ without leaving a
residual term, conditional on the fitted climate slopes and observed inputs.
Exact accounting therefore does not imply that these inputs are estimated
without uncertainty.

% \subsubsection{Sectoral choices and treatment of climate}

% For the industrial and transport sectors, we do not introduce a
% separate climate term, as annual state-level energy use in these sectors is only
% indirectly affected by heating and cooling needs.  At the sectoral level, activity indicators are defined as follows: residential
% and commercial activity are measured by floor area per capita; industrial
% activity by industrial value added per capita; and transport activity by vehicle
% miles traveled (VMT) per capita. 

% For the residential and commercial sectors, climate is handled explicitly to
% avoid conflating weather-driven fluctuations with efficiency changes. Following
% \citet{Reuter2019} and \citet{SerranoPuente2021}, we construct a climate factor
% based on heating degree days (HDD) and cooling degree days (CDD). In a first
% step, we estimate the elasticity of energy use per unit of floor area with
% respect to HDD and CDD using two-way fixed effects panel regressions at the
% state level. In a second step, we use the estimated elasticities to define
% sector-specific climate factors $\mathrm{CLI}_{\text{res}}$ and
% $\mathrm{CLI}_{\text{com}}$, which are then extracted from the intensity term in
% the LMDI identity. The full details of this procedure are reported in
% \ref{app:lmdi_climate}. In the main text, we treat climate as a
% separate effect whose contribution can be compared directly with those of
% activity and intensity.

\subsection{National evolution of per capita energy use}

We apply the LMDI decomposition to national aggregates in three ways:
(i) a discrete decomposition between 2006 and 2022, (ii) a continuous
decomposition that traces annual contributions over the full period, and (iii) a cross-state heterogeneity analysis.

\paragraph{Discrete decomposition: 2006--2022}

\begin{figure}[htbp]
  \centering
  \caption{Discrete LMDI decomposition of U.S.\ per capita energy consumption, 2006--2022.}
  \includegraphics[width=0.80\textwidth]{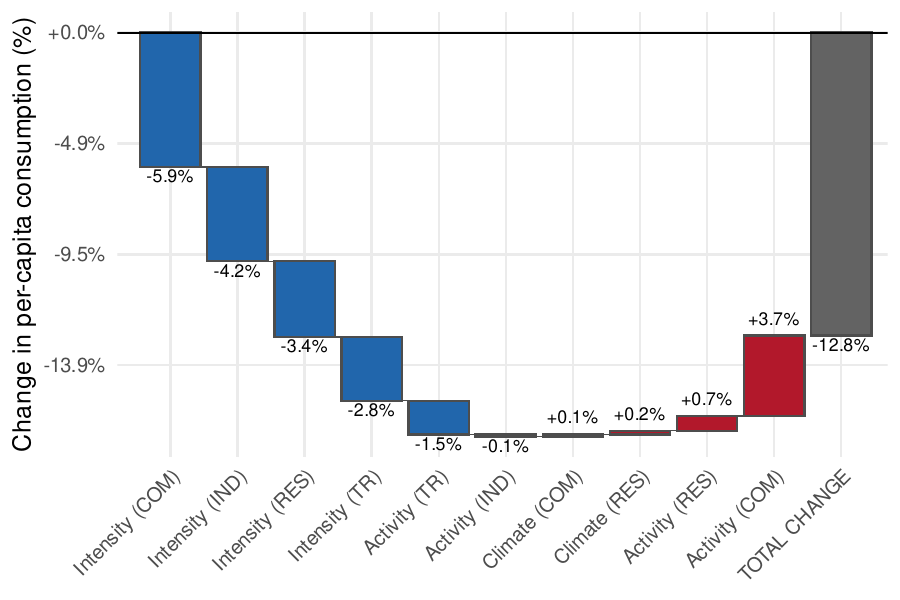}
  \label{fig:LMDI_US_discrete}
\end{figure}

Total per capita energy consumption in the United States declined by 12.8\%
between 2006 and 2022. The discrete LMDI decomposition in
Figure~\ref{fig:LMDI_US_discrete} shows that this decline is overwhelmingly
driven by reductions in energy intensity across all four end-use sectors. The
intensity effects in the commercial, industrial, residential, and transport
sectors contribute about $-5.9$, $-4.2$, $-3.4$, and $-2.8$ p.p.,
respectively, so that aggregate intensity improvements account for roughly
16 p.p. of the change in per capita energy use.

In contrast, changes along the activity margin mainly play an offsetting role. Transport activity contributes a modest reduction of about $-1.5$ p.p., and industrial activity a
negligible $-0.1$ p.p. In the building sectors, however,
increases in residential and commercial floor area per capita raise energy use
by approximately $+0.7$ and $+3.7$ p.p., respectively, thereby
partially offsetting the efficiency gains. Temperature variations, captured by
the climate factors in the residential and commercial sectors, have a
negligible impact on the overall change, with combined climate effects close to
zero over the full period, which is unsurprising given that HDD and CDD have remained relatively stable.

Therefore, the aggregate 12.8\% decline in per capita energy use is more than fully accounted for by within-sector intensity improvements, while changes in activity and climate largely cancel out, providing little net contribution to the observed reduction. These findings are consistent with earlier decomposition studies, which also identify improvements in energy intensity as the dominant driver of lower energy use in advanced economies \citep{Reuter2019,SerranoPuente2021,Jiang2016,Jiang2019}. As with any index-based decomposition, however, LMDI attributes all observed changes to a small set of pre-defined effects without a residual term, so its scope for economic interpretation is inherently limited.

\paragraph{Continuous decomposition: 2006--2022}

\begin{figure}[htbp]
  \centering
  \caption{Continuous LMDI decomposition of U.S.\ per capita energy consumption, 2006--2022.}
  \includegraphics[width=0.80\textwidth]{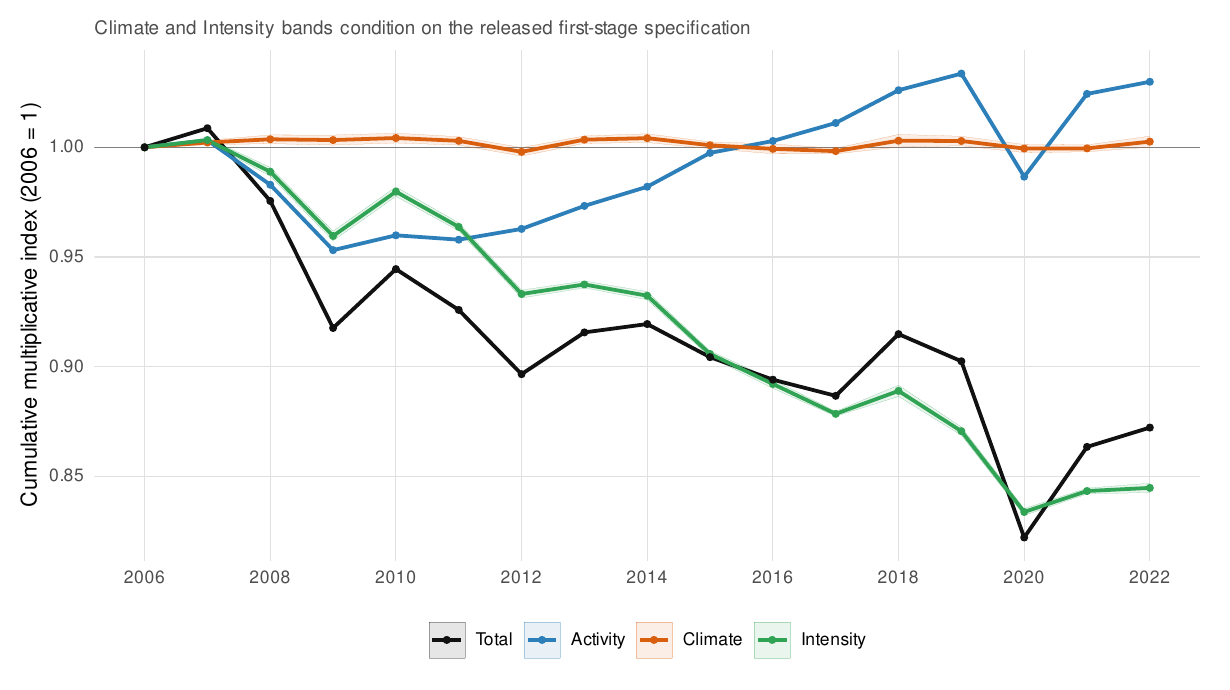}
  \label{fig:LMDI_US_continuous}
\end{figure}

Turning to the continuous decomposition, Figure~\ref{fig:LMDI_US_continuous} shows the
cumulative contributions of activity, intensity, and climate to the change in
per capita energy use relative to 2006 (index = 1).

The climate index remains essentially flat over the entire period, which confirms that year-to-year variation in
heating and cooling degree days has a negligible effect on aggregate per capita
energy use at the national level. We also plot the 95\% confidence interval since the climate coefficients are estimated from regressions. By contrast, the intensity index exhibits a
marked downward trend. It gradually falls from 1 in 2006 to close to 0.84 in 2020. The total energy-use index closely tracks this profile, which indicates that cumulative
changes in intensity account for most of the observed fall in per capita energy
use.

The activity index displays a very different pattern. It drops modestly during
the 2008--2009 financial crisis, then increases almost
monotonically, exceeding its initial level by the mid-2010s and peaking at
around 1.03 on the eve of the COVID-19 pandemic. The pandemic year is
associated with a sharp but temporary decline in activity, followed by a rapid rebound.\footnote{It is
notable that per capita energy consumption continues to decline over much of the
period when the activity index is recovering or above 1: in the LMDI
decomposition, this divergence is attributed to sustained improvements in energy
intensity rather than to lower underlying activity.}

\paragraph{Cross-state heterogeneity}

National aggregates, however, conceal substantial heterogeneity across states.
We therefore repeat the LMDI decomposition at the state level for the discrete
period 2006--2022 and map the resulting changes in per capita energy use and in
the activity and intensity components.

\begin{figure}[htbp]
  \centering
  \caption{State-level activity and intensity effects in per capita energy use, 2006--2022.}
  \begin{minipage}[t]{0.49\textwidth}
  \caption*{(a) Activity effect}
    \includegraphics[width=\linewidth]{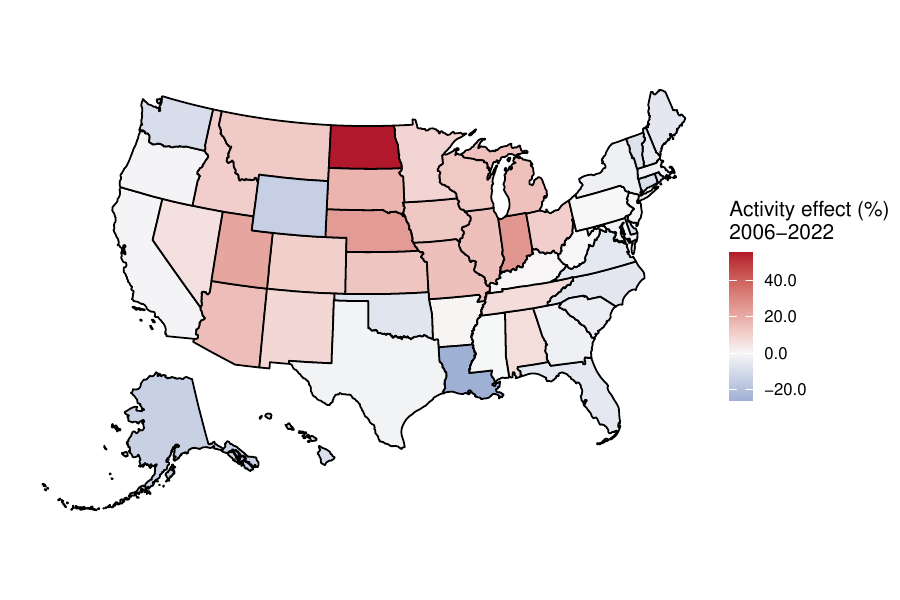}
  \end{minipage}
  \hfill
  \begin{minipage}[t]{0.49\textwidth}
  \caption*{(b) Intensity effect}
    \includegraphics[width=\linewidth]{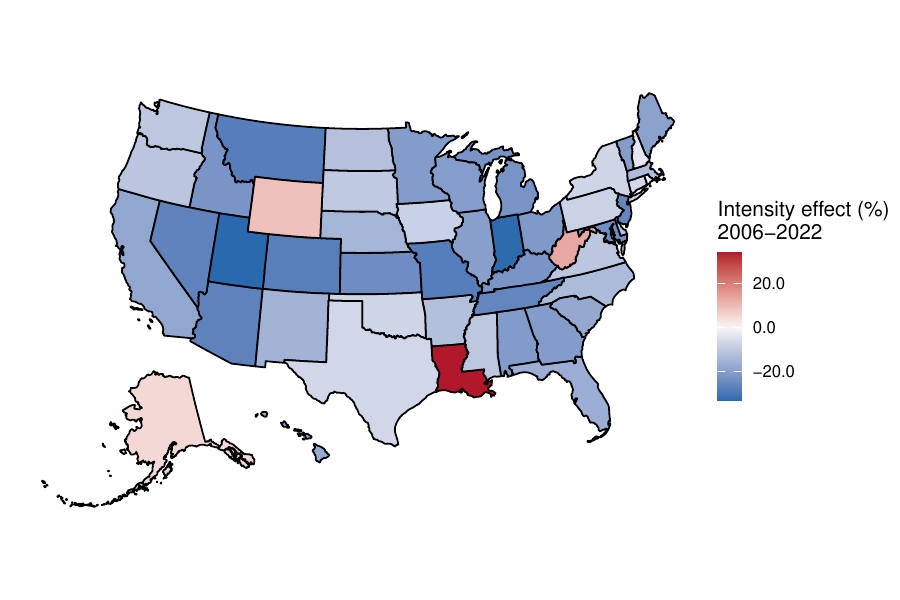}
  \end{minipage}
  \label{fig:map_intensity_activity}
\end{figure}

The state-level activity and intensity effects from the decomposition are shown
in Figure~\ref{fig:map_intensity_activity}. In most states the intensity effect
is clearly negative, often on the order of 10--20\%, whereas activity effects are
smaller in magnitude and more mixed in sign. In other words, efficiency gains
typically reduce per capita energy use even where underlying activity is
expanding. At the same time, states with similar net changes in per capita
energy consumption often get there through very different mixes of the two
components: some combine strong activity growth with large intensity
improvements, while others show little change in activity but only modest gains
in intensity.

Texas and Louisiana both end up with little net change in per capita energy use, but for opposite reasons: in Louisiana, a sharp contraction in (mainly industrial) activity is offset by a strong rise in industrial intensity, whereas in Texas both activity and intensity move little. North Dakota shows the mirror image of Louisiana, with a large surge in (especially industrial) activity and only modest intensity declines, while Wyoming combines falling industrial activity with a noticeable increase in intensity. Among the states with the largest reductions in per capita energy use, a common pattern is stable activity and a pronounced decline in intensity, as in Maine, New Jersey, and Maryland, whereas states such as Utah and Indiana achieve similar reductions despite strong activity growth because efficiency gains more than offset the expansion in demand.

These descriptive results have two main implications for the rest of the paper. First, they show that the post-2006 decline in U.S.\ per capita energy use is driven almost entirely by improvements in energy intensity, with activity broadly returning to pre-crisis levels and climate playing a negligible role, while similar net outcomes across states often reflect very different combinations of activity and intensity changes. Second, they highlight the need for a complementary approach to explain \emph{why} intensity improves in some states and worsens in others by attributing these changes to structural fundamentals or policy levers. This motivates the remainder of the paper, where we turn to a stochastic frontier framework to interpret cross-state differences in energy use.

\section{Economic framework}\label{sec:eco_frame}

Let us decompose observed per capita energy use in state $s$ as
\begin{equation}\label{eq:decomp}
  q_s \;=\; q_s^*\;\! I_s,
\end{equation}
where $q_s^*$ is a conditional energy-demand frontier and $I_s \ge 1$ is an inefficiency factor.\footnote{As a theoretical contribution, \ref{app:microfoundation} provides a formal
microfoundation for this economic framework. Starting from heterogeneous
agents' service needs, capital--energy choices, technology adoption,
policy compliance, and service use, it derives the aggregate relationships
$q^{e}=q^{*}I$ and $q^{*}=S/a$, computes the associated equilibrium quantities,
and establishes sufficient conditions for the comparative statics stated
here. The model thereby provides the theoretical basis for the empirical 
predictions.} The frontier $q_s^*$ represents the minimum energy input consistent with delivering a \emph{baseline} level of energy services. We define it as the level that would be demanded by agents with reference comfort preferences, given the state's income, climate, and structural characteristics. Unlike a thermodynamic engineering limit, this is a \emph{conditional economic} frontier: it depends on the cost-minimizing technology mix available to agents in a given state at given prices. The inefficiency factor $I_s$ captures the wedge between observed and frontier energy use. This wedge reflects not only technological and allocative inefficiency but also excess consumption driven by comfort preferences above the baseline and by rebound responses to efficiency improvements.

The frontier depends on two forces operating in opposite directions. On the demand side, baseline energy-service needs $S_s$ are increasing in per capita income $Y_s$ and in structural and climatic characteristics $X_s$: richer states demand more comfort and mobility, colder or hotter states require more heating or cooling, and states with larger transport or industrial sectors have higher baseline service requirements. On the supply side, best-practice energy efficiency $a_s$ (the maximum ratio of services to energy input attainable given current technology) is increasing in the energy price $P_s$ and in the stringency of energy-efficiency policies $\textit{Eff}_s$. Higher energy prices raise the cost of wasteful technologies and shift the cost-minimizing input mix toward more efficient alternatives. In addition, stricter efficiency policies restrict the technology set by tightening minimum standards and accelerate adoption through subsidies and information \rev{programs}. The frontier is the ratio of these two forces:
\begin{equation}\label{eq:frontier_theory}
  q_s^* \;=\; \frac{S(Y_s, X_s)}{a(P_s, \textit{Eff}_s, X_s)}.
\end{equation}
Three sign predictions follow directly: (1) Higher income raises service needs and therefore the frontier ($\partial q^*/\partial Y > 0$); (2) Higher energy prices improve best-practice efficiency, lowering the frontier ($\partial q^*/\partial P < 0$); and (3) Stricter efficiency policies tighten the technology set, also lowering the frontier ($\partial q^*/\partial \textit{Eff} < 0$).

In practice, not all agents operate at the frontier. The gap between observed and frontier energy use arises through three channels that are difficult to separate empirically: 
\begin{enumerate}
    \item[(i)]  Adoption frictions: information asymmetries, credit constraints, misaligned incentives between landlords and tenants, and behavioral biases prevent some agents from investing in the most efficient available technologies, even when the net present value of doing so is positive;
\item[(ii)]  Comfort-driven usage: agents may consume energy services beyond baseline needs for reasons of convenience, habit, or preference, particularly in high-income states where the marginal utility of energy expenditure is low relative to the value of comfort;
\item[(iii)]  Rebound effects: improvements in the frontier that lower the per unit cost of energy services can induce agents to consume more services, partially offsetting the efficiency gain. 
\end{enumerate}
Rather than attempting to structurally identify each channel---which would require exogenous variation that our cross-state panel does not provide---we treat $I_s$ as a composite energy-intensity gap.\footnote{Our efficiency concept is input-specific: it measures the contraction of energy use to the frontier holding service output constant, corresponding to a non-radial measure common in the energy-efficiency literature.}  This composite interpretation has an important normative implication: $I_s$ should not be read as a welfare loss, since some of the excess consumption it captures reflects rational comfort choices rather than market failures. By the same logic, policy-induced reductions in $I_s$ may mix genuine efficiency gains with reductions in comfort-driven consumption, so the welfare content of closing the intensity gap is ambiguous without further structural assumptions.

The sign predictions for the inefficiency factor mirror and complement those for the frontier. Efficiency policies reduce the intensity gap through both the extensive margin, by raising adoption rates with \rev{subsidies} for  efficient equipment and mandating minimum standards, and the intensive margin, by discouraging wasteful usage through \rev{labeling}, information \rev{programs}, and utility obligations ($\partial I/\partial \textit{Eff} \le 0$). Higher energy prices similarly reduce the gap by raising the private return to efficient investment and making high-comfort consumption more costly ($\partial I/\partial P \le 0$). The effect of income on inefficiency is theoretically ambiguous: higher income relaxes credit constraints and information barriers, facilitating adoption, but it also raises the desired level of comfort and lowers the salience of energy costs. The net sign of $\partial I/\partial Y$ is therefore an empirical matter.

A distinctive feature of this framework is that policy-relevant variables, i.e. prices and efficiency regulation, enter both the frontier and the inefficiency term, operating through economically distinct channels. In the frontier, they shift the best-practice technology available to all agents. In the inefficiency term, they affect the rate at which agents adopt that technology and the extent to which they use energy services beyond baseline needs. Empirically, this dual-channel structure motivates the SFA specification of Section~\ref{sec:empirical}, in which the policy variables appear in both the frontier equation and the inefficiency variance equation, allowing the data to determine the relative importance of each channel.

Conditional on year effects, the pooled SFA combines within-state changes and persistent between-state differences. Because several covariates are slow-moving and state fixed effects are not included, the pooled coefficients may be dominated by cross-state associations and may partly reflect time-invariant state characteristics correlated with the regressors.\footnote{We assess this concern in the frontier using a Mundlak specification that adds each state’s time average of all time-varying frontier covariates. This separates within-state effects, \(\beta\), from total between-state effects, \(\beta+\delta\).}

Taking logs of \eqref{eq:decomp} and defining $u_s \equiv \ln I_s \ge 0$ yields the additive decomposition
\begin{equation}\label{eq:log_decomp}
  \ln q_s \;=\; \ln q_s^*(Y_s, P_s, \textit{Eff}_s, X_s) \;+\; u_s,
\end{equation}
which corresponds to a stochastic frontier model. Adding a time dimension $t$, specifying the frontier as log-linear in its arguments, including year fixed effects to absorb common temporal shocks, and appending a symmetric noise term $v_{st}$ for measurement error and model approximation yields the empirical specification developed in Section~\ref{sec:empirical}. The six sign predictions derived above, i.e. three for the frontier  and three for the inefficiency, provide the testable structure for our empirical analysis.

\section{Data}\label{sec:data}
\subsection{Main data sources}
\label{sec:data_sources}

Our empirical analysis is based on a panel of the 50 U.S.\ states and the District of Columbia, observed annually from 2006 to 2022. The unit of analysis is the state--year, and variables that naturally scale with population (such as energy use or vehicle travel) are expressed in per capita terms whenever relevant.\footnote{Details on variable construction are provided in \ref{def}.} We construct a balanced panel of 867 observations \rev{$(50\text{ states}+\text{DC} \text{ over } 17\text{ years}$)} for the frontier analysis, and sectoral panels for the LMDI decompositions.

Our primary energy variables come from the U.S. Energy Information Administration (EIA) State Energy Data System (SEDS), which provides state-level series on energy consumption and energy prices. We complement these data with information from the EIA on electricity generation by fuel and on fossil fuel production. Macroeconomic variables are taken from the Bureau of Economic Analysis (BEA), which reports state-level gross domestic product and industry value added. Transportation activity is measured using vehicle-miles traveled (VMT) from the U.S. Department of Transportation (U.S. DOT).

Climate conditions are summarized by annual heating and cooling degree days (HDD and CDD), also obtained from EIA's SEDS. To capture differences in the size and composition of the building stock, we use data on residential and commercial floor area from the Commercial Buildings Energy Consumption Survey (CBECS), as well as from the Residential Energy Consumption Survey (RECS) and the American Housing Survey (AHS). Finally, we use the American Council for an Energy-Efficient Economy (ACEEE) State Energy Efficiency Scorecard to summarize state-level energy efficiency policies.

\subsection{Energy consumption: outcome and sectoral decomposition}
\label{subsec:energy_consumption}

%\subsubsection{Total energy use (main outcome)}
%\label{subsubsec:total_energy_use}

The main dependent variable is total end-use-sector energy consumption per capita, measured in MMBtu per person. Following SEDS conventions, it includes allocated electricity-system losses and therefore approximates a primary-energy-equivalent measure. This measure is the outcome in the main stochastic frontier specification. For comparison, we re-estimate the model using two component outcomes: direct primary energy use excluding retail electricity and retail electricity sold to end users, both expressed per capita.

%\subsubsection{Sectoral split for decomposition (LMDI)}
%\label{subsubsec:sectoral_split}

We distinguish four end-use sectors: residential, commercial, industrial, and transport. Except during COVID-19, changes in the share of each of these four sectors in total consumption are very small, with the largest variation remaining below 2\%. In 2022, the residential sector accounted for 20.6\% of total consumption, the commercial sector for 17.4\%, transportation for 29.1\%, and the industrial sector for 32.8\%.\footnote{Figure \ref{fig:US_sect_lev} in the Appendix presents the evolution of each sector's consumption share across time.} Sectoral energy use is employed exclusively for the LMDI decomposition, whereas the frontier analysis relies on total per capita energy consumption.

%\subsubsection{per capita transformation and deflators}
%\label{subsubsec:pc_deflators}

\rev{Variables that scale with population} are normalized by state population, and we use the suffix ``/pc'' to denote per capita variables. \rev{GDP is expressed in chained 2017 dollars, whereas the energy-price series is in current dollars. Year effects absorb common annual price movements.}\footnote{\ref{sec:des_un} provides detailed information on units, conversion factors, and the construction of per capita and real variables.} 

\subsection{Policy and price variables}
\label{subsec:policy_prices}

Our price variable summarizes the average cost of end-use energy in each
state and year. It combines expenditures on retail electricity and primary
fuels relative to the associated end-use energy quantity. A higher value
therefore means that, across the state's overall energy basket, end users pay
more per unit of energy.

This measure is broader than a gasoline price, a household electricity
tariff, or the price of any single fuel. It also reflects the composition of
energy use: states can differ because their component fuel and electricity
prices differ, because they use different mixtures of fuels and sectors, or
because of differences in market conditions, retail-rate structures, taxes, and regulation.\footnote{The measure is the EIA SEDS total-energy
average price, $TETCD_{st}$, in current dollars per million Btu, where
$TETCD_{st}=1000\,TETCV_{st}/TNSCB_{st}$ and
$TETCV_{st}=ESTCV_{st}+PESSV_{st}$. The denominator is SEDS
expenditure-adjusted end-use consumption, not our
primary-energy-equivalent outcome. Tax coverage is not uniform across the
underlying SEDS series, so we do not describe the measure as uniformly
``after tax.'' Figure~\ref{fig:pen_s_ts} shows both common movements over time and
persistent differences in price levels across states.}

The ACEEE State Energy Efficiency Scorecard is a recurring comparison of
how states organize, support, and pursue energy efficiency
\citep{Berg2020}. Our $\textit{Eff}_{st}$ variable summarizes five parts of
that environment: utility and public-benefit \rev{programs}, transportation,
building codes and policies, state-government initiatives, and appliance
standards.

Crucially, the score measures more than whether a policy merely exists.
Depending on the domain and Scorecard edition, points reflect \rev{program}
spending and reported prior-year utility savings; the adoption and strength
of building, transportation, and appliance policies; some compliance or
enforcement criteria; dedicated institutions; lead-by-example \rev{programs};
financial incentives; and selected transportation outcomes.

A higher value of $\textit{Eff}_{st}$ indicates a stronger and more
comprehensive state energy-efficiency environment on the dimensions assessed
by ACEEE. The score reflects the breadth and strength of policies and
\rev{programs}, the resources devoted to them, supporting institutions, and
selected indicators of performance; it therefore conveys more than a simple
count of adopted policies. It should not, however, be interpreted as a direct
measure of implementation quality or realized energy savings. Moreover, a
one-point difference does not necessarily represent the same substantive
policy change across states or years, since the same total can result from
different combinations of domains.

We construct $\textit{Eff}_{st}$ as a harmonized composite
rather than using ACEEE's published annual total or state rank.\footnote{For each
Scorecard edition, the retained domain scores are rescaled from their
edition-specific maxima to fixed 2020-reference maxima of 20 points for
utility and public-benefit programs, 12 for transportation, 9 for building
codes and policies, 6 for state-government initiatives, and 3 for appliance
standards, and are then summed. Signed penalties are preserved, while missing
comparable scores for 2007 and 2021 are interpolated component by component. \ref{sec:oth_dt} reports all details.} 
Although the reference maxima sum to 50, the resulting series is not a
literal common $[0,50]$ index because rubric content, point allocations, and
reference periods vary across editions. For estimation, the contemporaneous
score is pooled-mean-\rev{centered} and enters in levels in both the frontier and
gap-log-variance equations.

\subsection{Other controls and background variables}
\label{subsec:other_controls}

The set of additional controls includes real GDP per capita, denoted \textit{GDP/pc}. We account for sectoral structure through the share of energy-intensive industries excluding fossil fuel extraction (\textit{E-Int Ind}) and fossil fuel extraction per capita (\textit{FF/pc}). This distinction was motivated by the fact that among the five largest per capita energy consumers in the US are states that exploit fossil resources, as shown in Figure \ref{fig:FF_pc_ts} in the Appendix. 

Climate and geographic conditions are captured by heating and cooling degree days (\textit{HDD} and \textit{CDD}). Transport activity is measured by vehicle miles traveled per capita (\textit{VMT/pc}). \rev{We also control for fossil-fuel electricity generation per capita, a production-location measure that may include interstate exports and is not a direct measure of conversion losses}. 

In LMDI decomposition, the activity indicators for the commercial, residential, and transportation sectors are \rev{\textit{ComSurf/pc}, \textit{ResSurf/pc}}, and \textit{VMT/pc}, respectively; and for the industrial sector, we use industrial GDP ($GDP_{pc}^{indus}$).

Finally, residential surface area per capita (\textit{ResSurf/pc}) and commercial surface area per capita (\textit{ComSurf/pc}) are used in both the LMDI decomposition and the stochastic frontier analysis.\footnote{Full definitions, units, data sources, motivation, and justification for the inclusion of these variables are provided in \ref{def}.} The residential series combines raw 2020 RECS state
anchors with AHS Census-division trajectories \rev{using an additive bridge}. For commercial floor area, annual CBECS division totals are constructed
from the 2003, 2012, and 2018 benchmarks and their documented growth paths,
then allocated to states using contemporaneous annual population shares. \rev{Commercial floor area per capita is common to states in the same division--year.}

\begin{table}[H]
\centering
\caption{Unweighted summary statistics (analysis panel, 2006--2022)}
\label{tab:sumstats}
\begin{threeparttable}
\begin{tabular}{lrrrr}
\toprule
Variable & Mean & SD & Min & Max \\
\midrule
\textit{Energy use/pc} (MMBtu)       & 343.3  & 171.6  & 151.2  & 1093.3 \\
\textit{Price} (\$/MMBtu)            & 20.0   & 4.2    & 8.5    & 44.7   \\
\textit{GDP/pc} (\$2017, thousands)  & 58.4   & 22.3   & 35.6   & 214.7  \\
\textit{Eff} (harmonized points)     & 18.2   & 10.7   & 0.0    & 47.6 \\
\textit{VMT/pc} (thousands mi.)      & 10.2   & 2.0    & 4.5    & 19.1   \\
\textit{ComSurf/pc} (sq ft)          & 285.9  & 52.5   & 207.6  & 472.8  \\
\textit{ResSurf/pc} (sq ft)          & 705.3  & 82.6   & 475.5  & 856.6  \\
\textit{E-Int Ind} (\%)              & 6.1    & 3.5    & 0.1    & 23.1   \\
\textit{FF/pc} (billion Btu/person) & 0.6 & 2.2 & 0.0 & 19.9 \\
\textit{FFelec/pc} (MWh/person) & 11.0 & 11.7 & $<0.01$ & 83.1 \\
\textit{HDD} (thousands)             & 5.1    & 2.3    & 0.0    & 11.7   \\
\textit{CDD} (thousands)             & 1.2    & 1.0    & 0.0    & 5.1    \\
\bottomrule
\end{tabular}
\begin{tablenotes}
\footnotesize
\item \textit{Notes:} Statistics based on $N=867$ jurisdiction--year observations with equal weighting. See \ref{def} for complete definitions and construction rules.
\end{tablenotes}
\end{threeparttable}
\end{table}

\section{Empirical framework}
\label{sec:empirical}
\subsection{Empirical model}
The decomposition $q_{st}=q_{st}^{*}I_{st}$ derived in
Section~\ref{sec:eco_frame} describes the systematic component of energy use.
Because observed state energy use may be measured with error and the frontier
is necessarily an approximation to the underlying relationship, we allow the
observed outcome to differ by introducing a multiplicative
error into $q_{st}^{\mathrm{obs}}=q_{st}^{*}I_{st}\exp(v_{st})$. Taking logarithms then gives
\begin{equation}
\label{eq:log_sfa_decomposition}
  \ln q_{st}^{\mathrm{obs}}
  =\ln q_{st}^{*}+v_{st}+u_{st},
\end{equation}
where $u_{st}\equiv\ln I_{st}\geq0$  is the one-sided composite energy-use gap and $v_{st}$ is
symmetric noise capturing measurement error and model approximation.

We specify the stochastic frontier model as
\begin{equation}
\label{eq:sf_deterministic}
  \ln q_{st}
  \;=\;
  \beta_0
  \;+\; \beta_Y \ln Y_{st}
  \;+\; \beta_P \ln P_{st}
  \;+\; \beta_\textit{Eff} \textit{Eff}_{st}
  \;+\; X_{st}'\beta_X
  \;+\; \lambda_t
  \;+\; v_{st} \;+\; u_{st}, 
\end{equation}

where $Y_{st}$ is real GDP per capita, $P_{st}$ is the average \rev{end-use}\old{after-tax} energy price per million BTU, $\textit{Eff}_{st}$ is the ACEEE energy-efficiency policy index. The vector $X_{st}$ includes other fundamentals, including structural and climatic controls: \textit{VMT/pc}, commercial floor area per capita \rev{(\textit{ComSurf/pc})}, residential floor area per capita \rev{(\textit{ResSurf/pc})}, \textit{E-Int Ind}, \textit{FF/pc}, \textit{HDD}, \textit{CDD}, and \textit{FFelec/pc}. The terms $\lambda_t$ are year effects capturing common shocks across states (e.g.\ macroeconomic conditions, federal policies, fuel price shocks). The parameters $\beta_Y$, $\beta_P$, $\beta_\textit{Eff}$ and $\beta_X$ capture how best-practice per capita energy use responds to changes in income, prices, policies, and other fundamentals. Consistent with our theoretical framework, we expect $\beta_Y>0$ (higher income raises service needs), while higher energy prices and stricter efficiency policies should reduce the frontier, suggesting $\beta_P<0$ and $\beta_\textit{Eff}<0$.

The composite error $\varepsilon_{st} = v_{st} + u_{st}$ combines symmetric noise with the one-sided inefficiency term. Under the input-requirement
frontier convention, $u_{st}\geq0$ shifts energy use upward relative to the
conditional frontier, whereas $v_{st}$ permits observations on either side of
it. Our main specification models $u_{st}$ as conditionally
heteroskedastic half-normal 
\begin{equation}
\label{eq:u_dist}
  u_{st}\mid X_{st},Z_{st},\lambda_t
  \sim
  \mathcal{N}^{+}\!\left(0,\sigma_{u,st}^{2}\right),
\end{equation}
where $\mathcal{N}^{+}$ denotes a zero-mean normal distribution truncated at
zero. \rev{The variance  of the underlying untruncated normal distribution is specified as}
\begin{equation}
\label{eq:u_var}
  \ln\sigma_{u,st}^{2}
  =
  \gamma_0
  +\gamma_Y\ln Y_{st}
  +\gamma_P\ln P_{st}
  +\gamma_{\textit{Eff}}\textit{Eff}_{st}.
\end{equation}
\rev{The actual conditional variance of the half-normal gap is $(1-2/\pi)\sigma_{u,st}^{2}$. Thus, $Z_{st}$ contains exactly}
$(\ln Y_{st},\ln P_{st},\textit{Eff}_{st})$. For the half-normal model,
the conditional mean of the inefficiency term is 
  $\mathbb E[u_{st}\mid X_{st},Z_{st},\lambda_t]
  =
  \sigma_{u,st}\sqrt{\frac{2}{\pi}}$. The covariates in $Z_{st}$ therefore determine both the dispersion and the
expected magnitude of the composite gap. In particular, a negative $\gamma_{\textit{Eff}}$ would indicate that stronger energy-efficiency policies are associated with a smaller distance from the conditional energy-demand frontier.\footnote{To assess sensitivity to the assumed half-normal distribution of $u_{st}$,
we re-estimate the model using an exponential
one-sided distribution and a truncated-normal location-and-scale model in \ref{sec:robust}.}

In this model, the efficiency policy index $\textit{Eff}_{st}$ affects energy use through two distinct channels. First, through $\beta_{\textit{Eff}}$ in the frontier equation~\eqref{eq:sf_deterministic}, more ambitious energy-efficiency policies shift the conditional demand frontier $q_{st}^*$ inward, for example by tightening minimum standards or otherwise steering the technology set toward high-efficiency options. Second, through $\gamma_{\textit{Eff}}$ in the inefficiency equation~\eqref{eq:u_var}, these policies reduce the expected distance from the frontier by increasing the private return to efficient technologies and discouraging wasteful usage. The structural framework implies $\gamma_{\textit{Eff}} \le 0$, and it likewise predicts that higher energy prices decrease expected inefficiency, $\gamma_P \le 0$, because they both raise the value of energy savings and make high comfort levels more costly. By contrast, the effect of income on $u_{st}$ is theoretically ambiguous: higher income relaxes financial constraints on the adoption of efficient technologies but also raises desired comfort and usage, so the sign of $\gamma_Y$ is ultimately an empirical matter.

Figure~\ref{fig:sfa_schematic} illustrates the empirical model and this dual-channel structure
  schematically. The downward-sloping frontier captures the first channel:
  states with stronger efficiency policies reach a lower conditional minimum
  of energy use ($\beta_{\textit{Eff}} < 0$). The narrowing of the shaded
  region from left to right captures the second channel: stronger policies
  compress the distribution of inefficiency above the frontier
  ($\gamma_{\textit{Eff}} \le 0$). Each observed value $y_i$ deviates from
  the frontier by symmetric noise $v_i$ and one-sided inefficiency
  $u_i \geq 0$; only the latter is amenable to policy.

  % sfa_schematic.tex — Stochastic energy-demand frontier schematic
%
% Required in preamble:
%   \usepackage{tikz}
%   \usetikzlibrary{arrows.meta, decorations.pathreplacing}
%   \usepackage{amsmath}

\begin{figure}[htbp]
\centering
\begin{tikzpicture}[>=Stealth, font=\small]

% ── Colours ──────────────────────────────────────────────────────────
\definecolor{cfr}{HTML}{2166AC}   % frontier blue
\definecolor{cin}{HTML}{CB4335}   % inefficiency red
\definecolor{cno}{HTML}{1B9E77}   % noise teal
\definecolor{cdt}{HTML}{808080}   % data points grey
\definecolor{cpo}{HTML}{7B2D8E}   % dual-channel purple
\definecolor{csh}{HTML}{FADBD8}   % inefficiency shading

% ── Layout constants ────────────────────────────────────────────────
%   y-axis starts at 1.0 (not 0) — this is log energy, no natural zero.
%   Frontier fills most of the vertical range.
\def\ybot{1.0}

% ── Axes ─────────────────────────────────────────────────────────────
\draw[->, thick] (-0.3, \ybot) -- (12.5, \ybot);
\draw[->, thick] (0, {\ybot - 0.2}) -- (0, 7.8);

\node[below, font=\normalsize] at (6.0, {\ybot - 0.5})
  {Energy-efficiency policy score\; ($\textit{Eff}_{st}$)};
\node[rotate=90, anchor=south, font=\normalsize] at (-0.9, 4.2)
  {Log per-capita energy use\; ($\ln q_{st}$)};
\node[below, gray, font=\footnotesize] at (1.0, \ybot) {\emph{Low}};
\node[below, gray, font=\footnotesize] at (11.0, \ybot) {\emph{High}};

% ── Frontier function ────────────────────────────────────────────────
%   f(z) = 5.0 − 0.20z − 0.008z²    (concave, fills y ≈ 1.8 to 4.9)
%   Upper envelope: f(z) + 2.2 − 0.12z  (narrows: y ≈ 2.7 to 7.0)

% Shaded inefficiency region (large, fills most of the plot)
\fill[csh, opacity=0.30]
  plot[smooth, samples=50, domain=0.5:11.0]
    (\x, {5.0 - 0.20*\x - 0.008*\x*\x})
  -- plot[smooth, samples=50, domain=11.0:0.5]
    (\x, {5.0 - 0.20*\x - 0.008*\x*\x + 2.2 - 0.17*\x})
  -- cycle;

% Frontier curve
\draw[cfr, line width=1.8pt, smooth, samples=60, domain=0.5:11.0]
  plot (\x, {5.0 - 0.20*\x - 0.008*\x*\x});

% Upper envelope (dashed)
\draw[cin!40, dashed, thin, smooth, samples=50, domain=0.5:11.0]
  plot (\x, {5.0 - 0.20*\x - 0.008*\x*\x + 2.2 - 0.17*\x});

% Frontier label (below frontier at right end)
\node[cfr, font=\footnotesize\bfseries, below right] at
  (10.8, {5.0 - 0.20*11 - 0.008*121 - 0.1})
  {$\ln q_{st}^*$};

% ── Data points ──────────────────────────────────────────────────────
% Points ABOVE frontier (majority)
\foreach \px/\dy in {%
  0.8/1.80, 1.2/1.00, 1.6/1.90, 2.0/0.60, 2.4/1.45,
  2.8/0.45, 3.2/1.20, 3.6/0.80, 4.4/1.10, 4.8/0.40,
  5.3/0.85, 5.7/0.55, 6.1/0.70, 6.5/0.40, 6.9/0.55,
  7.3/0.30, 7.7/0.42, 8.1/0.22, 8.5/0.35,
  9.0/0.20, 9.4/0.26, 9.8/0.15, 10.2/0.12, 10.6/0.20%
}{%
  \fill[cdt, opacity=0.50]
    (\px, {5.0 - 0.20*\px - 0.008*\px*\px + \dy}) circle (1.8pt);
}

% Points NEAR or BELOW frontier (v < 0, small u)
\foreach \px/\dy in {%
  1.8/-0.12, 3.4/-0.08, 5.0/0.02, 6.3/-0.10,
  7.9/0.03, 9.2/-0.05, 10.0/0.02, 10.8/0.01%
}{%
  \fill[cdt, opacity=0.50]
    (\px, {5.0 - 0.20*\px - 0.008*\px*\px + \dy}) circle (1.8pt);
}

% ── Highlighted observation (x = 5.5, prominent) ────────────────────
\pgfmathsetmacro{\xh}{5.5}
\pgfmathsetmacro{\fh}{5.0 - 0.20*\xh - 0.008*\xh*\xh}     % frontier ≈ 3.66
\pgfmathsetmacro{\vh}{0.20}                                   % v (symmetric)
\pgfmathsetmacro{\uh}{1.10}                                   % u > 0
\pgfmathsetmacro{\yh}{\fh + \vh + \uh}                        % observed ≈ 4.96

% Observed point
\fill[black] (\xh, \yh) circle (3pt);

% Drop line from frontier to observation
\draw[gray, densely dashed] (\xh, \fh) -- (\xh, \yh);

% Tick on deterministic frontier
\draw[cfr, thick] ({\xh - 0.20}, \fh) -- ({\xh + 0.20}, \fh);

% ── Small brace: v_st CENTRED on frontier (symmetric noise) ─────────
%   Goes from frontier - |v| to frontier + |v|, centred on frontier line.
\draw[decorate, decoration={brace, amplitude=4pt},
      cno, very thick]
  ({\xh - 0.35}, {\fh + \vh}) -- ({\xh - 0.35}, {\fh - \vh})
  node[midway, left=6pt, cno, font=\footnotesize] {$v_{st}$};
% Small ticks at v boundaries
\draw[cno, thin] ({\xh - 0.12}, {\fh + \vh}) -- ({\xh + 0.12}, {\fh + \vh});
\draw[cno, thin] ({\xh - 0.12}, {\fh - \vh}) -- ({\xh + 0.12}, {\fh - \vh});

% ── Large brace: total gap from frontier to observation ──────────────
%   This is the main visual: ε_st = v_st + u_st
\draw[decorate, decoration={brace, amplitude=8pt, mirror},
      cin, very thick]
  ({\xh + 0.35}, \fh) -- ({\xh + 0.35}, \yh)
  node[midway, right=12pt, cin, font=\footnotesize, align=left]
  {$\varepsilon_{st} = v_{st} + u_{st}$};

% Label: observed value
\node[above right, font=\footnotesize] at ({\xh + 0.15}, {\yh + 0.10})
  {$\ln q_{st}$};

% ── σ_u narrowing ───────────────────────────────────────────────────
% Left: wide (moved right of y-axis to avoid overlap)
\draw[cin!60, <->, semithick]
  (1.2, {5.0 - 0.20*1.2 - 0.008*1.44 + 0.08})
  -- (1.2, {5.0 - 0.20*1.2 - 0.008*1.44 + 1.75});
\node[cin!60, font=\scriptsize, anchor=west] at
  (1.35, {5.0 - 0.20*1.2 - 0.008*1.44 + 0.92})
  {large $\sigma_u$};

% Right: narrow (matches tighter envelope)
\draw[cin!60, <->, semithick]
  (11.3, {5.0 - 0.20*11 - 0.008*121 + 0.02})
  -- (11.3, {5.0 - 0.20*11 - 0.008*121 + 0.33});
\node[cin!60, font=\scriptsize, anchor=west] at
  (11.5, {5.0 - 0.20*11 - 0.008*121 + 0.17})
  {small $\sigma_u$};

% ── Annotation boxes (ABOVE the plot, outside axes) ─────────────────

% Box 1 — Frontier (right, above plot)
\node[draw=cfr!70, fill=cfr!5, rounded corners=3pt,
      text width=4.3cm, inner sep=5pt, font=\footnotesize,
      align=left]
  (bF) at (9.5, 7.6)
  {\textcolor{cfr}{\textbf{Frontier}}\; $\ln q_{st}^*$
   \hfill\textcolor{cfr}{\scriptsize eq.\,\eqref{eq:sf_deterministic}}\\[2pt]
   $\beta_P < 0$\,,\;\; $\beta_Y > 0$\,,\;\;
   $\beta_{\textit{HDD}},\beta_{\textit{CDD}} > 0$\\[1pt]
   \textcolor{cpo}{$\beta_{\textit{Eff}} < 0$}};

% Box 2 — Inefficiency (left, above plot)
\node[draw=cin!70, fill=cin!5, rounded corners=3pt,
      text width=4.3cm, inner sep=5pt, font=\footnotesize,
      align=left]
  (bU) at (3.0, 7.6)
  {\textcolor{cin}{\textbf{Inefficiency}}\;
       $\sigma^2_{u,st}$
   \hfill\textcolor{cin}{\scriptsize eq.\,\eqref{eq:u_var}}\\[2pt]
   $\gamma_P \leq 0$\,,\;\;
   $\gamma_Y$\; ambiguous\\[1pt]
   \textcolor{cpo}{$\gamma_{\textit{Eff}} \leq 0$}};

% Straight arrows from boxes down into plot
\draw[->, cfr!70, thick, shorten >=4pt]
  (bF.south) -- (8.5, {5.0 - 0.20*8.5 - 0.008*72.25 + 0.08});

\draw[->, cin!70, thick, shorten >=4pt]
  (bU.south) -- (3.5, {5.0 - 0.20*3.5 - 0.008*12.25 + 0.90});

\end{tikzpicture}

\caption{Schematic of the stochastic energy-demand frontier.
The frontier $\ln q_{st}^*$ represents minimum energy use
conditional on state characteristics.
Observed energy use $\ln q_{st} = \ln q_{st}^* + v_{st} + u_{st}$
deviates from the frontier by symmetric noise~$v_{st}$
(which can place observations below~$\ln q_{st}^*$) and
one-sided inefficiency $u_{st} \geq 0$.
The shaded region narrows from left to right:
stronger efficiency policies lower the frontier
($\beta_{\textit{Eff}} < 0$) and compress the
inefficiency distribution
($\gamma_{\textit{Eff}} \leq 0$).}
\label{fig:sfa_schematic}
\end{figure}

As discussed in Section~\ref{sec:eco_frame}, \rev{our pooled estimates combine within-state and between-state variation conditional on year effects, with persistent cross-state differences.} This \rev{pooled design} is particularly relevant for prices in the inefficiency equation: states with persistently high prices often differ from low-price states in unobserved, time-invariant ways, e.g. due to older infrastructure or regulatory histories that both raise prices and leave a legacy of inefficient technologies. Such confounders can generate a positive cross-sectional correlation between prices and inefficiency even if the causal effect is negative. The price coefficients in the inefficiency term should therefore be interpreted as potentially biased associations.

\subsection{Results}

The theoretical framework of Section~\ref{sec:eco_frame} generates six testable predictions,\footnote{In the frontier equation, we expect $\beta_Y > 0$ (higher income raises service needs and therefore the frontier), $\beta_P < 0$ (higher energy prices drive adoption of more efficient technologies, lowering the frontier), and $\beta_{\textit{Eff}} < 0$ (stricter efficiency policies tighten the technology set). In the inefficiency equation, we expect $\gamma_{\textit{Eff}} \le 0$ (policies reduce the distance to the frontier), $\gamma_P \le 0$ (higher prices lower the intensity gap), and we note that $\gamma_Y$ is theoretically ambiguous: higher income relaxes adoption constraints but also raises comfort-driven usage.} which we now confront with the results reported in Table \ref{tab:mundlak}.\footnote{The model is estimated by maximum likelihood using the \texttt{npsf} package in R. Standard errors are heteroskedasticity-robust. Additional robustness checks (truncated-normal specification, subsample sensitivity, etc.) are in \ref{sec:robust}.} We report both the baseline, i.e. a pooled SFA specification, and the Mundlak specification.

The pooled SFA specification identifies frontier parameters from both cross-sectional and time-series variation. Since a number of the regressors---energy prices, GDP, building stocks---exhibit substantial persistent differences across states, the estimated coefficients primarily capture long-run cross-state associations rather than within-state responses to changes over time. To complement these results, we re-estimate the frontier with a Mundlak adjustment.\footnote{This involves augmenting the frontier equation with the within-state time averages of all time-varying regressors:
%\begin{equation}
  $\ln q_{st} = \beta_0 + X_{st}'\beta + \bar{X}_s'\delta + \lambda_t + u_{st} + v_{st}$,
%\end{equation}
where $\bar{X}_s = T^{-1}\sum_t X_{st}$.} The original coefficients $\beta$ now capture within-state (time-series) variation, while the Mundlak terms $\beta+\delta$ absorb persistent cross-state differences. %If $\delta = 0$, the pooled specification is consistent and the correlated random effects are unnecessary.

\begin{table}[h!]
\centering
\caption{Pooled SFA and Mundlak regression results}
\label{tab:mundlak}
\begin{threeparttable}
{\setstretch{1} \small
\begin{tabular}{lccc}
\toprule
& \multicolumn{1}{c}{Main} & \multicolumn{2}{c}{Mundlak} \\
\cmidrule(lr){2-2} \cmidrule(lr){3-4}
& Estimate & Within $\beta$ & Total between $\beta+\delta$ \\
\midrule
\multicolumn{4}{l}{\textit{Frontier equation}} \\[4pt]
\textit{Price} & ${-}0.7524^{***}$ & ${-}0.1708^{**}$ & ${-}0.7793^{***}$ \\
  & (0.0459) & (0.0853) & (0.0466) \\[2pt]
\textit{GDP/pc} & $0.1931^{***}$ & $0.2254^{***}$ & $0.2440^{***}$ \\
  & (0.0322) & (0.0841) & (0.0323) \\[2pt]
\textit{Eff} & ${-}0.0017^{***}$ & $0.0043^{***}$ & ${-}0.0027^{***}$ \\
  & (0.0007) & (0.0011) & (0.0010) \\[2pt]
\textit{VMT/pc} & $0.3933^{***}$ & $0.1356$ & $0.3853^{***}$ \\
  & (0.0337) & (0.0902) & (0.0346) \\[2pt]
\textit{ComSurf/pc} & ${-}0.0136$ & ${-}0.0013$ & ${-}0.0060$ \\
  & (0.0246) & (0.0273) & (0.0643) \\[2pt]
\textit{ResSurf/pc} & $0.0141$ & $0.0039$ & ${-}0.0133$ \\
  & (0.0534) & (0.2334) & (0.0603) \\[2pt]
\textit{E-Int Ind} & $0.1077^{***}$ & $0.0045$ & $0.1071^{***}$ \\
  & (0.0099) & (0.0275) & (0.0111) \\[2pt]
\textit{FF/pc} & $0.0364^{***}$ & ${-}0.0102$ & $0.0450^{***}$ \\
  & (0.0024) & (0.0076) & (0.0028) \\[2pt]
\textit{FFelec/pc} & $0.0094^{***}$ & $0.0454^{***}$ & $0.0030$ \\
  & (0.0031) & (0.0162) & (0.0029) \\[2pt]
\textit{HDD} & $0.0456^{***}$ & $0.0345^{***}$ & $0.0447^{***}$ \\
  & (0.0044) & (0.0110) & (0.0062) \\[2pt]
\textit{CDD} & $0.1007^{***}$ & $0.0285$ & $0.1013^{***}$ \\
  & (0.0090) & (0.0286) & (0.0132) \\[2pt]
\midrule
\multicolumn{4}{l}{\textit{Composite-gap log-variance equation}
  $(\ln \sigma_{u,st}^{2})$} \\[3pt]
& \textit{Main estimate}
& \multicolumn{2}{c}{\textit{Mundlak-model estimate}} \\[3pt]

\textit{Eff}
  & ${-}0.1555^{***}$
  & \multicolumn{2}{c}{${-}0.1591^{***}$} \\
  & (0.0096) & \multicolumn{2}{c}{(0.0092)} \\[2pt]

\textit{Price}
  & $0.3107$
  & \multicolumn{2}{c}{$0.0172$} \\
  & (0.3447) & \multicolumn{2}{c}{(0.3692)} \\[2pt]

\textit{GDP/pc}
  & $2.6264^{***}$
  & \multicolumn{2}{c}{$2.4708^{***}$} \\
  & (0.2873) & \multicolumn{2}{c}{(0.2910)} \\[2pt]

$\overline{\mathbb{E}[u\mid Z]}$
  & $0.1935$
  & \multicolumn{2}{c}{$0.1831$} \\
\bottomrule
\end{tabular}
\begin{tablenotes}
\footnotesize
 \item \textit{Notes:} $^{***}$ $p<0.01$, $^{**}$ $p<0.05$,
  $^{*}$ $p<0.10$. Standard errors in parentheses are based on the
  inverse Hessian.
\end{tablenotes}}
\end{threeparttable}
\end{table}

Let us first focus on the baseline specification. \rev{Most} frontier coefficients are statistically significant and carry signs consistent with the theoretical framework of Section~\ref{sec:eco_frame}. The estimated price elasticity of the frontier is $-0.75$, implying that \rev{a 10\% increase in average energy prices is associated with approximately a 6.9\% reduction} in best-practice per capita energy use. This estimate is larger in absolute value than the residential demand elasticities reported by \citet{FilippiniHunt2012}, who find elasticities in the range of $-0.07$ to $-0.12$. The difference likely reflects two factors: our frontier covers total energy demand (not residential only), and the sample period differs.\footnote{We estimate this specification using retail
electricity consumption per capita and the corresponding retail electricity
price and obtain a frontier price elasticity of $-0.46$ (Table~\ref*{tab:sfa_electricity_current}).}

\rev{The GDP per capita elasticity of $0.19$ is within the range of $0.17$--$0.39$ reported across the residential SFA specifications of \citet{FilippiniHunt2012}.} The ACEEE policy coefficient $\textit{Eff}_{st}$ ($-0.0017$, $p < 0.01$) implies that a 15-point increase in the Scorecard---approximately the interquartile range---is associated with a 2.5\% reduction in frontier energy use.

Climate variables enter with expected positive signs: heating degree days ($0.046$) and cooling degree days ($0.101$) both raise the frontier, with cooling having a larger effect per unit. Vehicle-miles traveled per capita and fossil fuel variables also raise the frontier, consistent with structural energy intensity.

The two surface area variables are not statistically significant, which may reflect the lack of sufficient cross-state variation in the underlying data series.

Turning to the inefficiency equation, stronger efficiency policies are associated with substantially reduced inefficiency ($\gamma_{\textit{Eff}} = -0.16$, $p < 0.001$), indicating that policies work through both channels: tightening the frontier and reducing the gap to it. GDP per capita has a large positive coefficient (\rev{$2.63$}, $p < 0.001$), consistent with the theoretical prediction that higher income raises desired comfort and usage, thereby increasing the energy-intensity gap. The price coefficient in the inefficiency equation is positive but statistically insignificant \rev{($0.31$, $p = 0.37$)}, consistent with the endogeneity concern discussed above: states with persistently high prices may also have structural features that raise inefficiency, biasing the cross-sectional correlation upward.%\footnote{Although all coefficients appear statistically significant in the frontier, standard errors should be interpreted cautiously since they do not account for within-state serial correlation.} 

The three frontier predictions are hence confirmed ($\hat\beta_Y > 0$, $\hat\beta_P < 0$, $\hat\beta_{\textit{Eff}} < 0$), and the efficiency-policy prediction for inefficiency holds ($\hat\gamma_{\textit{Eff}} < 0$). Of the two remaining predictions, $\gamma_Y$ was theoretically ambiguous and the positive estimate ($\hat\gamma_Y > 0$) suggests that comfort and rebound channels dominate the adoption channel. The prediction $\gamma_P \le 0$ is not confirmed: the positive but insignificant estimate likely reflects endogeneity bias rather than a true positive causal effect of prices on inefficiency.

A likelihood-ratio test strongly rejects the pooled-frontier restriction
that the Mundlak state-mean coefficients are jointly zero
($\chi^2(11)=145.76$, $p=1.09\times10^{-25}$). However, the Mundlak correction leads to broadly similar coefficients to the pooled SFA, indicating that the pooled frontier largely reflects persistent between-state relationships. The decomposition reveals a striking pattern. The within-state price elasticity is $-0.17$, while the between-state coefficient is $-0.78$. This means that most of the estimated price effect in the pooled model reflects persistent cross-state differences in price levels rather than within-state responses to price changes over time. The within-state elasticity of $-0.17$ is more comparable to previous short-run residential demand estimates.

Several structural variables---VMT, commercial surface area, energy-intensive industry share--- lose significance in the within-state specification, confirming that these slow-moving variables primarily capture cross-state heterogeneity.

The policy index ($\textit{Eff}_{st}$) exhibits an even sharper
within--between contrast. Conditional on year effects and the other
covariates, the Mundlak within-state slope is positive ($0.0043$,
$p<0.001$), whereas the total-between slope is negative ($-0.0027$,
$p=0.0085$). Thus, state-years in which a state scores above its own
2006--2022 average are associated with a higher fitted frontier, while states
with persistently higher average scores have lower fitted frontiers.

The positive contemporaneous within-state association is consistent with
dynamic endogeneity. States may strengthen scored policies and \rev{programs} in
response to rising or anticipated energy-use pressures, whereas reductions in
energy use may materialize only after implementation and the gradual turnover
of buildings, vehicles, and equipment. Changes in the total score may also
reflect shifts in its domain composition. These feedback and timing mechanisms
could produce a positive contemporaneous association without implying that
stronger efficiency policies raise energy use. Such mechanisms are less directly relevant to the comparison of long-run
state averages. Nevertheless, the negative total-between association is not
necessarily causal either: it may combine cumulative policy environments with
persistent differences in institutions, infrastructure, preferences, and
other state characteristics. The model cannot distinguish among these
mechanisms, so both coefficients are interpreted as conditional associations
rather than causal policy effects.

The key finding is that the inefficiency equation is robust to the Mundlak adjustment. The $\textit{Eff}_{st}$ coefficient remains negative and highly significant $\gamma_{\textit{Eff}} = -0.155$ vs.\ $-0.159$ in the baseline, \rev{while the GDP coefficient remains positive and highly significant ($\gamma_Y = 2.63$ vs.\ $2.47$)}. The cross-observation mean inefficiency decreases slightly ($\overline{E[u]} = 0.19$ vs.\ $0.18$), consistent with the Mundlak terms absorbing persistent cross-state variation that was partly attributed to inefficiency in the pooled model. This robustness is reassuring: the conclusion that stronger efficiency policies are associated with lower inefficiency does not depend on whether the frontier is identified from cross-sectional or within-state variation.\footnote{Subsample analyses excluding fossil-fuel-producing states, crisis years, industrial states, and early/recent sub-periods (\ref{sec:subset_robust}) confirm the stability of the key coefficients.}

The robustness analysis comprises 34 alternative regressions, with 20
for the main outcome variable (total energy use), and seven each for
direct primary energy and retail electricity. For total energy, the principal results are stable when we exclude
fossil-producing or energy-intensive states and crisis years, restrict the
sample to 2014--2022, lag the ACEEE score by one to three years, replace the
aggregate score with policy blocks or domains, change the one-sided
distribution, alter the energy and price definitions, add investor-owned
utility context, omit the two floor-area controls, or suppress their endpoint
extrapolation. \rev{The frontier price and gap-policy associations are generally stable, but not every sign or significance level is preserved; the outcome-specific exceptions are described below and in \ref{sec:robust}.} An extra
2006--2013 specification did not converge.\footnote{Complete results
are reported in Tables~\ref{tab:subset_robust},
\ref{tab:sfa_total_timing_composition},
\ref{tab:measurement_distribution}, \ref{tab:iou_context}, and
\ref{tab:floor_sensitivity}, together with the floor-omission continuation in
Table~\ref{tab:mundlak_no_surfaces}.}

Changing the energy outcome and its corresponding price measure leads to estimating different
energy-demand margins rather than minor variants of the same estimand. The
frontier price coefficient remains negative and precisely estimated for
delivered total energy, direct primary energy, and retail electricity. Other
coefficients are more outcome-dependent. The GDP frontier coefficient is weak
or negative outside the main total-energy specification, while the ACEEE
frontier coefficient is positive for delivered and direct primary energy but
negative for retail electricity. The ACEEE inefficiency coefficient remains
negative in every outcome--price pairing, although it is imprecisely estimated
for retail electricity, and the GDP inefficiency coefficient remains positive and
precise throughout. By contrast, the price coefficient in the gap equation is
not stable in sign or magnitude.
Finally, the policy-domain regressions reveal substantial
heterogeneity hidden by the aggregate ACEEE score.\footnote{These comparisons are
reported in Appendix Tables~\ref{tab:measurement_matched} and
\ref{tab:measurement_no_ff_matched}; the complete timing and domain results
appear in Tables~\ref{tab:sfa_electricity_current},
\ref{tab:sfa_electricity_timing_composition},
\ref{tab:sfa_primary_current}, and
\ref{tab:sfa_primary_timing_composition}. Tables~
\ref{tab:price_policy_sensitivity}--\ref{tab:aceee_gap_policy_sensitivity}
provide a compact summary of the focal coefficient sensitivities.
}

 We turn to quantifying the relative contribution of each variable using a variable-importance analysis.
\subsection{Contribution of each variable}

We construct contribution measures that decompose panel variations
  \rev{in the predicted log frontier $\ln \widehat{q}_{st}^*$ and the predicted inefficiency term $\hat{u}_{st}$. Common year effects are removed for the frontier attribution only.}
  Formally, we first partial out year fixed effects from both the predicted
  frontier and all regressors, so that the importance measures capture the
  share of \emph{within-year cross-state} variation explained by each
  variable. We then attribute the residual explained variance to blocks of
  regressors.

  Our principal attribution method is the exact LMG decomposition of an
auxiliary regression's $R^2$ \citep{Groemping2006,Gromping2007}. LMG averages
each variable's incremental contribution over every possible entry order,
thereby producing an additive allocation that is invariant to variable
ordering. Because the target is itself the prediction 
$X\widehat\beta$, the full auxiliary $R^2$ equals one by construction and the
frontier allocations sum to one. For the inefficiency term, we evaluate all
 subsets of $\textit{Eff}$, price, and GDP; those allocations sum to
the auxiliary regression's explained $R^2$, and the unexplained remainder is
retained explicitly.\footnote{As a frontier-only predictive sensitivity, we also fit
Random Forests to the same year-demeaned frontier design and report raw
marginal and conditional permutation-importance scores
\citep{strobl2008,Gromping2009}. These scores provide a non-additive ranking
of predictive relevance: they are not normalized to sum to one and are not
combined numerically with either LMG allocation.}

To combine the attribution exercises on a common descriptive scale, we
decompose the sample variance of
$y_{st}\equiv\ln q_{st}^{\mathrm{obs}}$. For the half-normal model,
the model-implied variance of the one-sided composite gap is
$\widehat V_U=N^{-1}\sum_{st}\widehat{\sigma}_{u,st}^{\,2}
-(2/\pi)(N^{-1}\sum_{st}\widehat{\sigma}_{u,st})^2$. The covariance
residual is hence 
$C_{FU}=\Var(y)-\Var(\widehat F)-\widehat V_U-\widehat{\sigma}_v^{\,2}$.
Under the maintained variance identity, $C_{FU}$ corresponds to the total
covariance contribution $2\Cov(\widehat F,U)$. Because its allocation is not
unique, we divide it equally between the frontier and the one-sided gap. We
therefore define the frontier share as
$s_F=[\Var(\widehat F)+\tfrac{1}{2}C_{FU}]/\Var(y)$, the gap share as
$s_U=[\widehat V_U+\tfrac{1}{2}C_{FU}]/\Var(y)$, and the noise share as
$s_V=\widehat{\sigma}_v^{\,2}/\Var(y)$. For the main specification, these
shares are 56.35\%, 42.20\%, and 1.45\%, respectively. Under this
convention, the frontier is assigned somewhat
more than half of total log-outcome variation, the inefficiency gap
slightly more than two fifths, and symmetric noise only a small remainder.
These shares describe variation across all 867 state--year observations; the
calendar-year demeaning described above applies specifically to the frontier
LMG exercise.

  We then combine the two channels into a single contribution measure.
  Each variable's share of frontier variation is rescaled by~$s_F$, and
  variables that also enter the inefficiency variance equation receive an
  additional term equal to~$s_U$ times their \emph{raw} (un-normalized)
  LMG in a secondary decomposition of $\hat{u}_{st}$ on Eff, $\ln P$, $\ln\text{GDP/pc}$. The three contributions sum to the
auxiliary regression's $R_G^2=0.35$, so the observed gap covariates jointly
account for $s_UR_G^2=14.84$ percentage points of
$\Var(\ln q^{\mathrm{obs}})$. The remaining
$s_U(1-R_G^2)=27.36$ percentage points are retained explicitly as unexplained
variation.\footnote{\ref{sec:imp} reports the target definitions, implementation, and  normalization.} We also report confidence intervals  obtained via a
  state-level block bootstrap ($B = 500$), holding $s_F$ and $s_U$ at
  their point estimates so that the intervals reflect LMG uncertainty
  alone.

\begin{figure}[H]
  \centering
  \caption{Sunburst decomposition of $\Var(\ln q)$: inner ring shows the
  three-way variance split (frontier, inefficiency, noise); outer ring
  disaggregates into individual drivers with matching contour colors.}
\includegraphics[width=0.7\textwidth,trim={85 85 85 85}, clip]{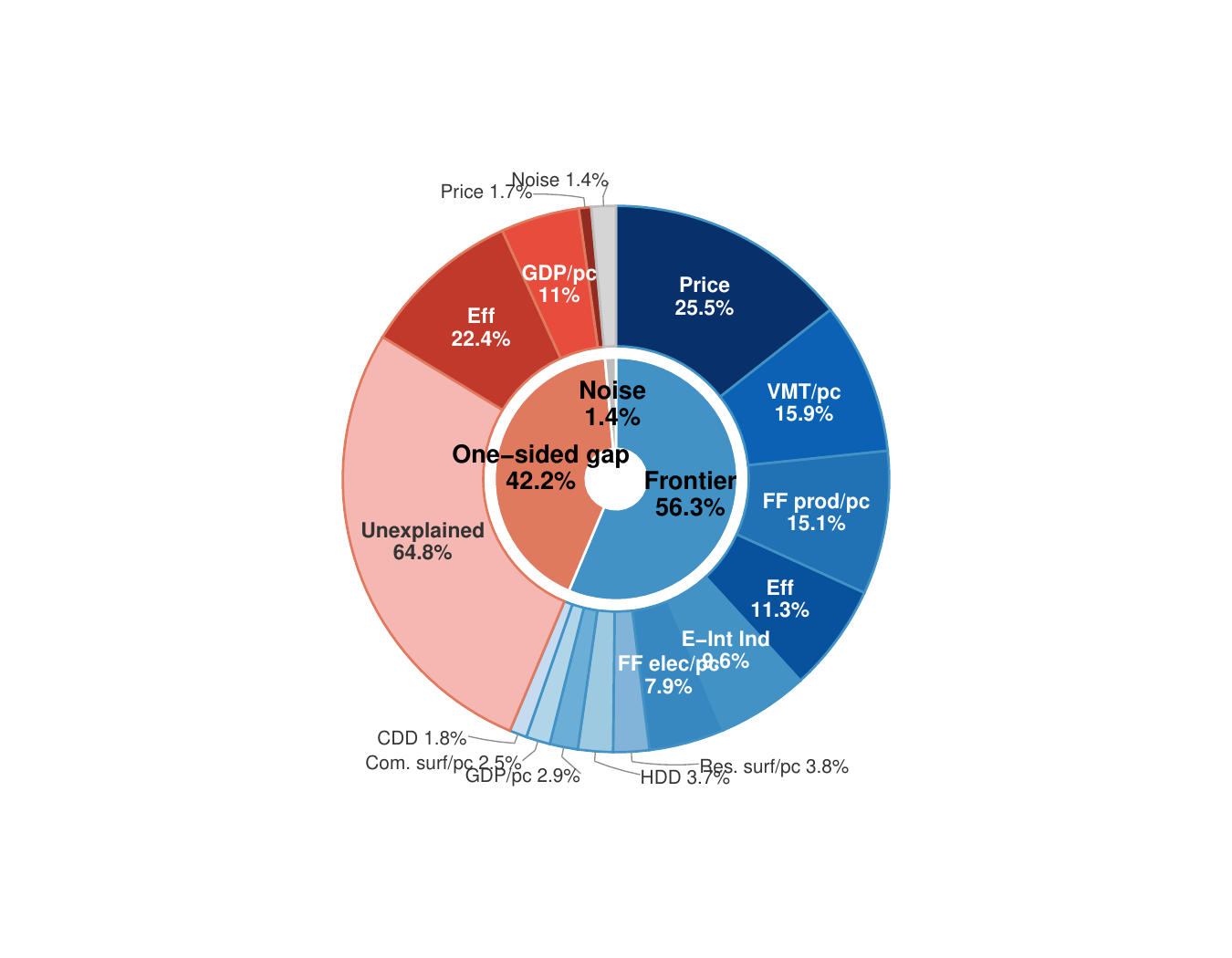}
  \label{fig:donut}
\end{figure}

Figure~\ref{fig:donut} summarizes these results in a sunburst decomposition of
variance. The inner ring shows the three-way variance split: fitted frontier
\rev{($s_F=56.35\%$)}, model-based one-sided composite gap \rev{($s_U=42.20\%$)}, and
symmetric noise \rev{($s_V=1.45\%$)}. The outer ring disaggregates these components
into eleven frontier variables (blue), three observed gap covariates plus an
unexplained remainder (red), and a single noise slice (grey).

In this decomposition, energy prices are the most influential frontier
determinant, accounting for roughly 14.4\% of total variation through the
frontier channel alone (25.5\% of fitted-frontier variation). State
energy-efficiency policy (\textit{Eff}) ranks fourth within the frontier,
accounting for about 6.4\% of total variation (11.3\% of frontier variation),
but gains a further 9.5 percentage points through the composite-gap channel.
Its combined contribution is therefore about 15.8\%, making it the largest
contributor, slightly ahead of the 15.1\% combined contribution
of price. Together, price and \textit{Eff} account for about 30.9\% of
$\Var(\ln q)$ when both channels are included.

GDP per capita is a minor frontier driver, accounting for about 1.7\% of total
variation through that channel, but gains a further 4.7 percentage points
through the composite-gap channel. Its combined contribution is therefore
approximately 6.3\%, making it the fifth most important observed variable
overall.

The two climate variables together account for about 3.1\% of total variation,
with heating degree days contributing more than cooling degree days. Their
relatively modest allocation is obtained jointly with the other correlated
predictors included in the LMG decomposition.\footnote{Year-centered pairwise
correlations are $\rho(\text{price},\text{HDD})=-0.19$ and
$\rho(\text{price},\text{CDD})=0.14$.} Residential and commercial floor area
jointly account for approximately 3.6\%, indicating that building-stock size
is a second-tier predictor in this attribution.\footnote{National-level time
series for the four most important frontier variables are reported in
Figure~\ref{fig:var_us_ts} in the Appendix.}

The matched outcome--price specifications preserve the importance of price
and \textit{Eff}, although their ordering changes with the energy measure. For
retail electricity, price ranks first with a combined contribution of 20.0\%,
followed by GDP per capita (16.5\%), VMT per capita (12.6\%), and
\textit{Eff} (12.5\%). The inner split assigns 71.0\% to the frontier, 20.4\%
to the one-sided gap, and 8.6\% to noise. For direct primary energy,
\textit{Eff} ranks first (12.4\%), followed by fossil production (9.3\%),
the matched non-electric energy price (8.0\%), and VMT per capita (7.8\%);
the corresponding inner shares are 45.0\%, 54.0\%, and 1.0\%.\footnote{Because each
decomposition is scaled by its own outcome variance, these comparisons concern
the composition and ranking of variation rather than absolute variance
levels; complete results are reported in
Table~\ref{tab:sunburst_matched_outcomes_contributions}.}
  
%The decomposition reveals an important asymmetry: the frontier is well explained
%by observables, but more than
%half of the inefficiency variation lies beyond the three modelled determinants.

\subsubsection{Uncertainty}

\begin{figure}[H]
  \centering
  \caption{State-block bootstrap uncertainty in main-model LMG contributions}
  \includegraphics[width=0.98\textwidth]
  {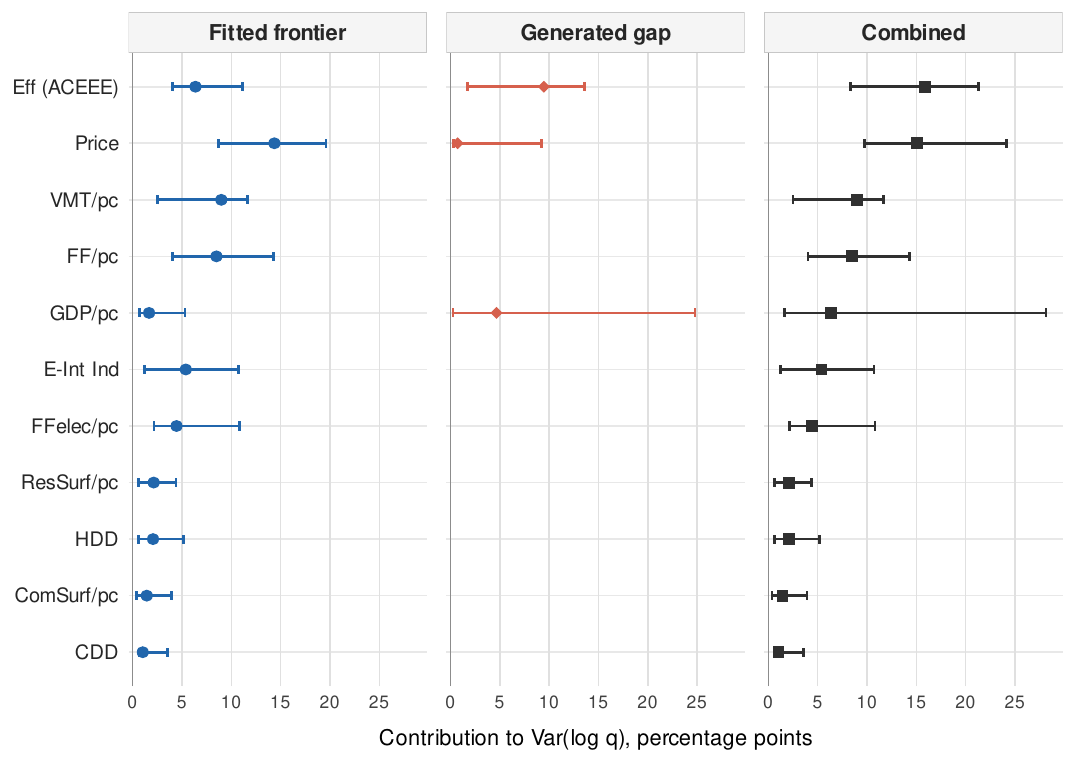}
  \label{fig:lmg_contribution_bootstrap}
  \vspace{-0.4em}
  \begin{minipage}{0.98\textwidth}
  \footnotesize
  \textit{Notes:} The left panel reports the fitted-frontier contribution; the middle panel reports the gap
  contribution; and the right panel reports their draw-level sum. Whiskers are
  type-7 2.5th--97.5th percentile intervals from 500 resamples. Complete jurisdiction histories are resampled and the main
  half-normal SFA is re-estimated.
  \end{minipage}
\end{figure}

Figure~\ref{fig:lmg_contribution_bootstrap} holds the main specification fixed
and uses a state block bootstrap to summarize sampling uncertainty in
the LMG attribution. The combined 95\% intervals are 8.30--21.32 percentage points of total log-outcome variance for \textit{Eff} policy and
9.74--24.15 points for energy price, respectively, and overlap substantially. The intervals for
VMT and fossil production are 2.50--11.67 and 4.03--14.30 points. GDP has a
particularly wide 1.67--28.14 interval because its generated-gap allocation is
right-skewed across state resamples. The bootstrap therefore supports the broad
importance of the leading variables.

As a separate frontier-only predictive sensitivity, we compare marginal and
Strobl-style conditional Random Forest permutation importance \citep{strobl2008}. Price remains the
leading marginal predictor and ranks within the top two conditionally, whereas
\textit{Eff} is approximately fifth marginally but its conditional rank is
implementation-sensitive. The RF evidence therefore reinforces the predictive
prominence of price but gives a less stable assessment of \textit{Eff}'s frontier contribution. These raw signed scores are non-additive and are not
variance shares.\footnote{Appendix Table~\ref{tab:rf_conditional_importance}
reports the complete implementation comparison, and
Tables~\ref{tab:rf_total_variants}--\ref{tab:rf_domain_detail_variants} report
the expanded outcome, timing, and policy sensitivities.}

\subsubsection{Robustness}

\begin{figure}[H]
  \centering
  \caption{Rank stability of total contributions across primary-equivalent-total specifications}
  \includegraphics[width=0.98\textwidth]{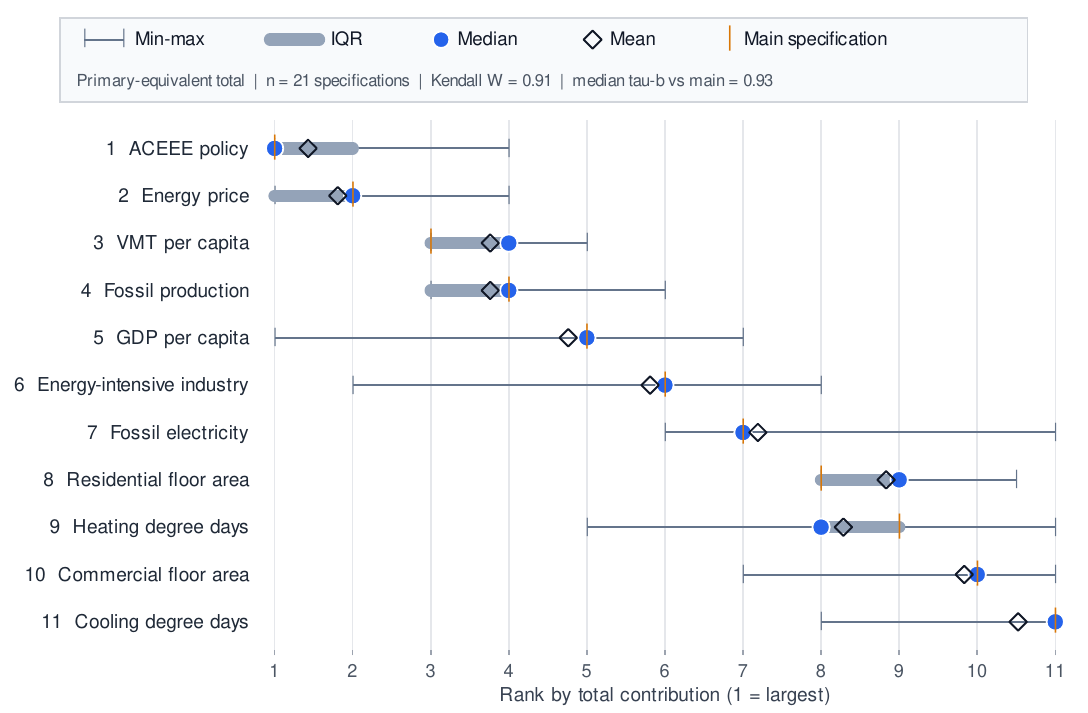}
  \label{fig:robustness_rank_total}
  \vspace{-0.4em}
  \begin{minipage}{0.98\textwidth}
  \footnotesize
  \textit{Notes:} Rank 1 denotes the largest combined contribution. The thin capped line is the full minimum--maximum
  range across specifications; the thick line is the interquartile range
  (IQR), hence 50\% of model specs; the filled circle and open diamond are the median and mean; and the
  gold tick is the main specification. Table~\ref{tab:robustness_stability_scope} inventories the specifications.
  \end{minipage}
\end{figure}

\begin{figure}[H]
  \centering
  \caption{Stability of total contribution shares across primary-equivalent-total specifications}
  \includegraphics[width=0.98\textwidth]{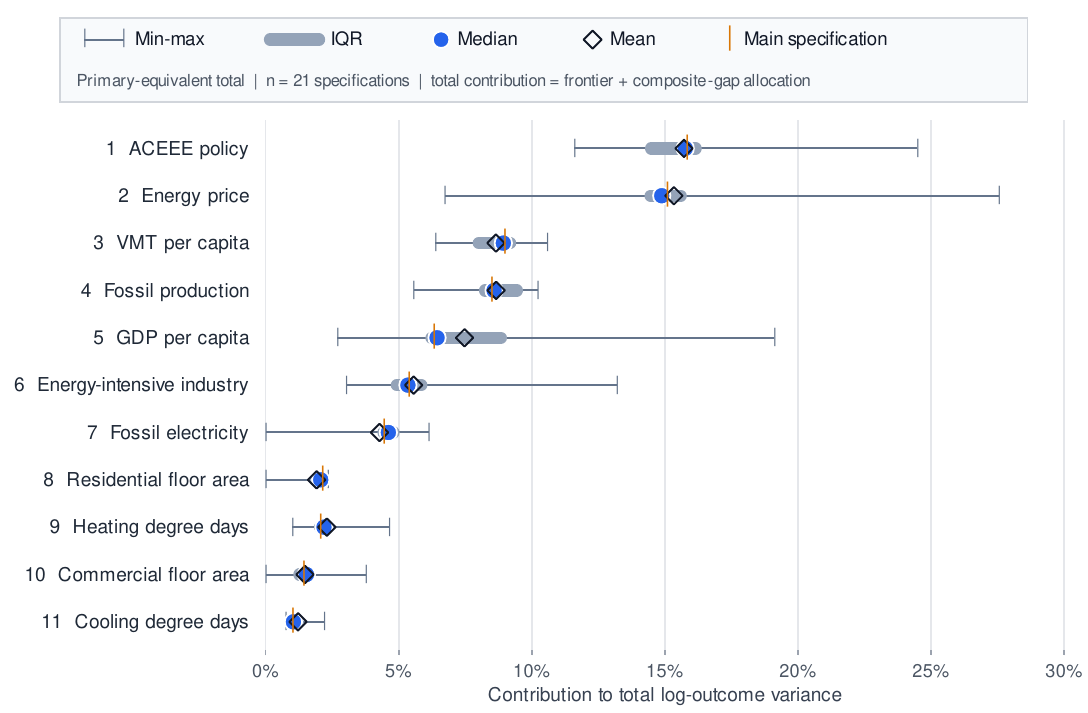}
  \label{fig:robustness_contribution_total}
  \vspace{-0.4em}
  \begin{minipage}{0.98\textwidth}
  \footnotesize
  \textit{Notes:} Marks have the same meaning as in
  Figure~\ref{fig:robustness_rank_total}. A total contribution is the sum of a
  driver's fitted-frontier and generated-composite-gap allocations, expressed
  in percentage points of that specification's own $\Var(\log q)$.
  \end{minipage}
\end{figure}

Figures~\ref{fig:robustness_rank_total}
and~\ref{fig:robustness_contribution_total} report rank and contribution stability statistics across specifications. This approach describes dispersion over a finite, dependent set
of design perturbations and does not produce confidence intervals. Instead, we hold the attribution method
fixed and vary the pooled SFA specification. For each specification,
we add the contributions from the frontier and gap, after scaling both by their
shares of that model's own empirical variance. We then rank the covariates by this combined contribution. The main specification determines the row order and is shown by the
gold tick.

The main-model ordering is substantively stable even though the magnitudes are
not invariant. ACEEE policy is first in the main model and has median rank 1,
an IQR of 1--2, and a full range of 1--4; energy price is second in the main
model and has median rank 2, the same 1--2 IQR, and the same 1--4 range. Their
main-specification total contributions are 15.83 and 15.09 percentage points,
respectively. Across the 21 specifications, the corresponding medians are
15.76 and 14.88 points and the IQRs are 14.48--16.15 and 14.46--15.59 points.
VMT and fossil production form the next stable pair, with median contributions
of 8.93 and 8.58 points. GDP has a wider full range than its IQR: a few design
changes alter its allocation even though its typical rank remains near fifth.
This ordering remains descriptive, not causal.

Corresponding summaries for direct primary energy and retail electricity are
reported in Figures~\ref{fig:robustness_rank_primary}--
\ref{fig:robustness_contribution_retail}. Direct primary energy has especially
stable ranks ($W=0.98$): ACEEE policy and fossil production rank first and
second in all seven specifications. Retail electricity also shows strong
agreement ($W=0.95$), although the magnitudes allocated to GDP and ACEEE are
more specification-sensitive.

\subsubsection{State-level results}

The SFA yields the observation-level Battese--Coelli proximity index
$\widehat e^{\,BC}_{st}
=E[\exp(-u_{st})\mid\widehat\varepsilon_{st}]$
\citep{battese_coelli1988}. Values closer to one indicate greater proximity
to the frontier. We also report the JLMS composite log-gap predictor, for which smaller values indicate a
smaller residual-conditional gap. 
Averaging over 2006--2022, Oregon, Utah, California, North Carolina, and
New York are closest to the fitted frontier. Texas, North Dakota, DC,
Louisiana, and Alaska are farthest. The mean Battese--Coelli proximity index
across all state--years is 0.844. This average should not be interpreted
directly as a percentage of avoidable energy use: it is a nonlinear,
residual-conditional model diagnostic rather than an observed energy-use
ratio. The rankings are established from the specification. Table~\ref{tab:eff_rankings} reports the statistics and selected
descriptive characteristics for the five closest and five farthest
jurisdictions.

\begin{table}[H]
\centering
\caption{Selected state-level model diagnostics and descriptive context}
\label{tab:eff_rankings}
\begin{threeparttable}
\small
\begin{tabular}{lccccc}
\toprule
State & BC proximity & JLMS gap & $\textit{Eff}$ & Energy/pc
      & $\Delta$Energy \\
\midrule
\multicolumn{6}{l}{\textit{Closest to the fitted frontier}} \\[2pt]
Oregon         & 0.974 & 0.027 & 35.0 & 202 & $-$11.6\% \\
Utah           & 0.974 & 0.027 & 18.7 & 269 & $-$18.2\% \\
California     & 0.969 & 0.032 & 42.6 & 189 & $-$19.0\% \\
North Carolina & 0.968 & 0.033 & 14.8 & 259 & $-$18.0\% \\
New York       & 0.968 & 0.033 & 34.9 & 183 & $-$9.2\%  \\
\midrule
\multicolumn{6}{l}{\textit{Farthest from the fitted frontier}} \\[2pt]
Texas          & 0.694 & 0.368 & 13.9 & 461 & $-$7.8\%  \\
North Dakota   & 0.607 & 0.505 &  3.1 & 756 & $+$36.6\% \\
DC             & 0.542 & 0.619 & 23.3 & 258 & $-$34.5\% \\
Louisiana      & 0.470 & 0.763 &  8.1 & 898 & $-$0.5\%  \\
Alaska         & 0.295 & 1.226 &  8.1 & 881 & $-$9.1\%  \\
\bottomrule
\end{tabular}
\begin{tablenotes}
\footnotesize
\item \textit{Notes:} Jurisdictions are selected and ordered using their
2006--2022 mean Battese--Coelli (BC) frontier-proximity index. The JLMS
column reports the mean residual-conditional composite log gap. Energy/pc is mean primary-equivalent total
energy consumption in MMBtu per capita. $\Delta$Energy is the percentage
change between 2006 and 2022.
\end{tablenotes}
\end{threeparttable}
\end{table}

\begin{figure}[htbp]
  \centering
  \caption{Mean frontier proximity and ACEEE Scorecard by jurisdiction}
  \label{fig:eff_aceee}
  \includegraphics[width=0.80\textwidth]
    {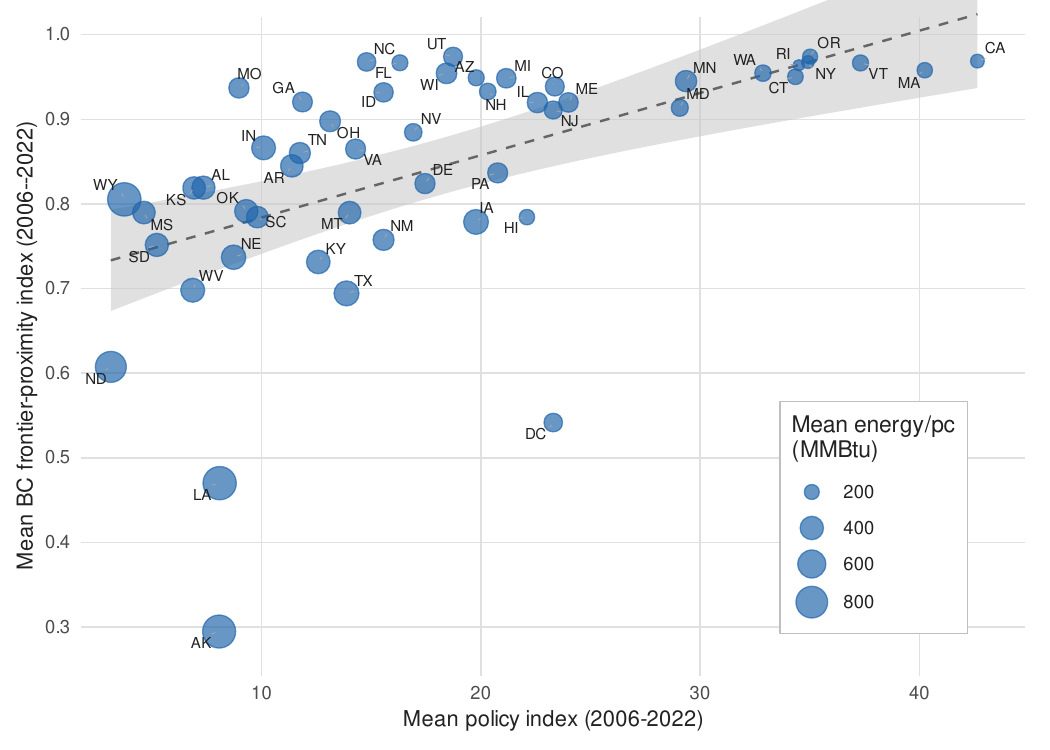}

  \begin{minipage}{0.80\textwidth}
  \footnotesize
  \textit{Notes:} Both axes report jurisdiction means over 2006--2022. The dashed line and shaded region show the OLS fit
  and its 95\% confidence band.
  \end{minipage}
\end{figure}

\begin{figure}[htbp]
  \centering
  \caption{State-level efficiency policy scores}
  \label{fig:map_eff}
  \includegraphics[width=0.95\textwidth]{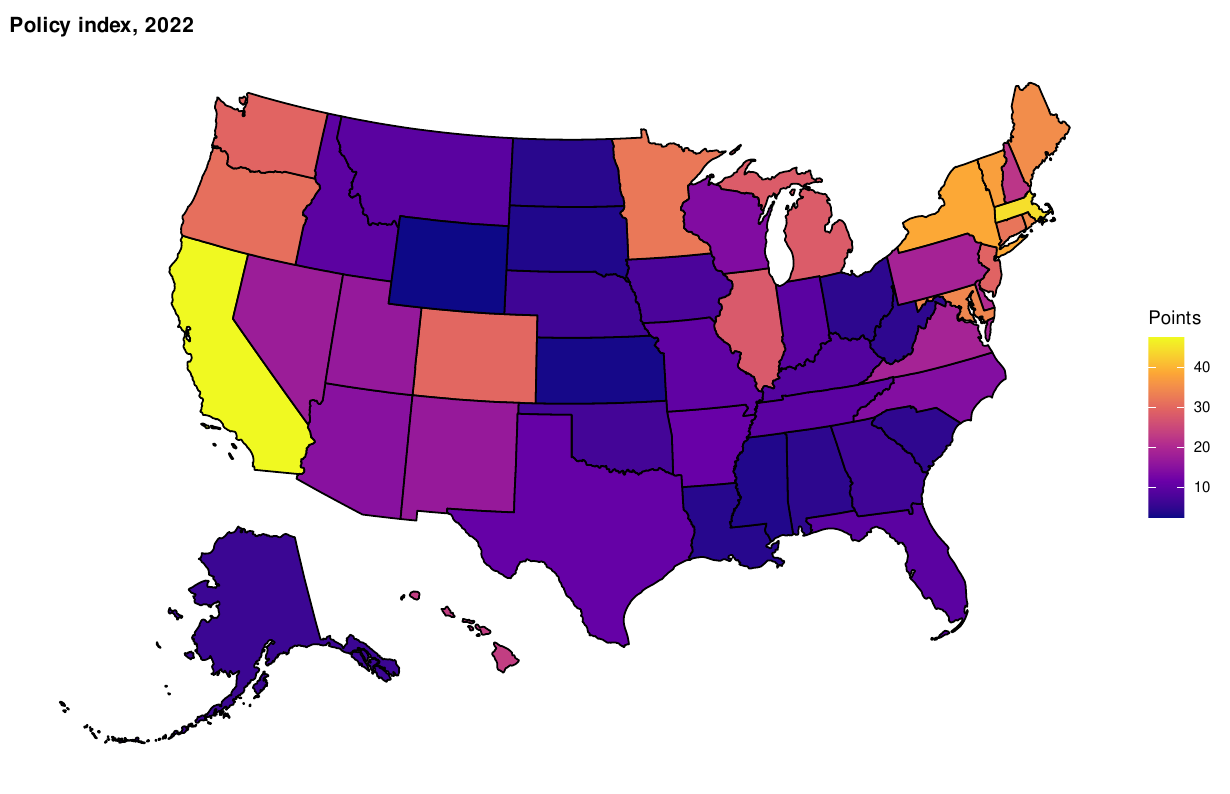}

\end{figure}

Figure~\ref{fig:eff_aceee} plots average frontier proximity against the
 average ACEEE score by state, with point size proportional to mean per capita
energy use. The positive correlation ($r=0.54$) indicates that states with
higher average scores also tend to lie closer to their fitted frontiers. Figure~\ref{fig:map_eff} presents the geographic distribution.

\section{Policy Implications}\label{sec:policy}

The results and comparative evidence point to several contributors to high
U.S.\ energy use: comparatively low end-user prices in several major energy
markets, automobile-intensive mobility, fossil-fuel production,
energy-intensive activities, and uneven energy-efficiency policy
environments. These findings offer several insights for the design of energy demand-side
policy. We organize the discussion around three themes: the role of energy prices, the design of energy-efficiency \rev{programs}, and equity considerations.

\paragraph{Energy prices} Energy prices receive approximately 25.5\% of the
LMG allocation of frontier variation, associated with an elasticity of $-0.75$, while
the Mundlak specification yields a total-between elasticity of $-0.78$ and a
much smaller within-state elasticity of $-0.17$. 
Using the exact exponential mapping, a 10\% price difference is associated
with a 6.9\% lower frontier under the pooled estimate, compared with
approximately 1.6\% using the within-state estimate. The negative price association also remains
precisely estimated when total energy is replaced by delivered energy, direct
primary energy, or retail electricity.
The results point to persistent differences in state price
environments as an important structural feature of U.S.\ energy demand.

The U.S.\ state comparison illustrates the potential scale of this price
margin. In 2022, the twelve lowest-price jurisdictions had an average price of
\$21.71/MMBtu, compared with a median of \$25.51/MMBtu across all
jurisdictions. In an illustrative counterfactual, we raise the price in each
of these twelve jurisdictions to the national median, holding the other
covariates constant. The resulting changes imply a
frontier that is 13.6\% lower across the affected jurisdictions.\footnote{The
frontier change is calculated separately for each jurisdiction using its own
initial price and the pooled price coefficient. The twelve changes are then
weighted by each jurisdiction's 2022 energy use.}

The international comparison strengthens this interpretation. Using all seven
IEA end-user price series available for the United States, Canada, France, and
Germany, we construct a simple relative-price index. The resulting index
places end-user prices in France and Germany at 2.03 times the
United States--Canada level; equivalently, the North American price index is
approximately one-half of the Franco-German index. In the same IEA data,
2019 total energy supply averaged 295~MMBtu per person in the United States and
Canada, compared with 141~MMBtu in France and Germany---approximately twice as
much energy per person where prices were approximately half as high
\citep{IEAKeyWorldEnergy2021}.\footnote{The price comparison uses seven energy products: household light
fuel oil, automotive diesel, premium gasoline, industrial and household
natural gas, and industrial and household electricity. For each product, we
first average the U.S.\ and Canadian prices and separately average the French
and German prices. We then divide the French--German average by the
U.S.--Canadian average. Finally, we take the geometric mean, giving every product equal weight. This produces an
overall price ratio of 2.03. Oil-product prices refer to the first quarter of
2021; natural-gas and electricity prices refer to 2020. For the energy comparison, we convert the IEA figures to MMBtu per person.
The values are 267.3 for the United States and 322.5 for Canada, producing an
average of 294.9; the corresponding values are 142.6 for France and 140.6 for
Germany, producing an average of 141.6. Dividing 294.9 by 141.6 gives an
energy-supply ratio of 2.08. Weighting countries by population instead gives
a similar price ratio of 2.03 and an energy-supply ratio of 1.93.} Taken together, the U.S.\ state
results and the wider evidence on price responsiveness indicate
that comparatively low end-user prices are part of the explanation for high
U.S.\ energy use and plausibly contribute to Canada's high-use pattern.

The policy implication is that a credible and predictable marginal price
signal should be a central component of long-run demand-side policy. Depending
on the objective and sector, this may involve carbon or energy charges,
motor-fuel taxation, or electricity and gas tariff reform. If emissions are
the primary concern, the price signal should reflect carbon content; if
reducing energy demand or system costs is the objective, a broader energy-price
signal may be appropriate. Such signals must be sufficiently durable to shape
decisions about vehicles, equipment, buildings, and retrofits. Price reform should therefore
complement efficiency standards, retrofit finance, public transport, and
infrastructure that expand the alternatives available to energy users.
Affordability can be maintained without suppressing marginal incentives by
recycling revenues through lump-sum or income-linked transfers.

\paragraph{Efficiency policies}
Low prices are only one part of the explanation for high U.S.\ energy use.
The results indicate that uneven efficiency-policy environments are another.
In the main specification, the ACEEE policy index is negatively
associated with both the frontier
($\beta_{\textit{Eff}}=-0.0017$, $p=0.0096$) and the scale of the 
one-sided energy-use gap
($\gamma_{\textit{Eff}}=-0.1555$, $p<0.001$). Using the exact exponential
mapping, a 10-point higher score is associated with a 1.7\% lower 
frontier. The policy index receives 11.3\% of the LMG allocation of frontier
variation. Once the frontier and gap channels are combined, it receives
15.83 percentage points of $\Var(\ln q)$. Across the 21 
total-energy specifications, its median contribution rank remains first, with
an interquartile range of one to two. Differences in state policy environments
are therefore among
the most consistently important observed characteristics that predict state energy use.

In
2022, the five highest-scoring states---California, Massachusetts, New York,
Vermont, and Rhode Island---averaged 40.7 points in the policy index,
compared with 15.4 points among the other 46 jurisdictions. In an
illustrative counterfactual, we evaluate each of these 46 jurisdictions at
the high-five mean, holding the other covariates constant. The resulting
changes imply a frontier that is 4.43\% lower across
the affected jurisdictions.\footnote{The frontier change is calculated
separately for each jurisdiction using its own initial score and the common
target of 40.74 points; the resulting changes are then weighted by 2022 energy
use. The gap coefficient is not added as further savings because it
describes the scale of the entire composite gap, and equal index scores can
represent different combinations of policies.}

The international comparison points in the same direction. Following \citet{ACEEEInternational2022}'s international ACEEE comparison in 2022, France ranked first with 74.5 points
out of 100 and Germany third with 71.5, whereas the U.S. ranked
tenth with 54 points and Canada thirteenth with 49.5. Transportation was the weakest category for
the United States and one of the weakest for Canada. The same four-country
comparison therefore reveals two reinforcing differences: France and Germany
had substantially higher end-user energy prices and stronger assessed
efficiency-policy environments, while using approximately half as much
primary energy per person. The international scores are not identical to the
state index used in our regressions, but they reinforce the conclusion that
high U.S.\ and Canadian energy use reflects not only comparatively low prices,
but also weaker efficiency policy environments.

The policy implication is not only to add more \rev{programs} where absent, but to make
efficiency policy more coherent and durable across the country. At the federal level, that includes having nationwide
performance standards for appliances, equipment, and vehicles. Common rules for measuring and verifying energy savings,
together with predictable multi-year funding, would also make \rev{program}
results easier to compare and sustain. States and local governments can then
adapt this framework to local conditions through enforceable building codes,
requirements that utilities deliver verified energy savings, retrofit
\rev{programs}, industrial energy-management measures, and transport policies. This division of responsibilities combines national consistency with local
implementation. It also makes efficiency policy a natural complement to
energy-price reform. 

The broader OECD evidence reinforces this emphasis on coherent policy
governance rather than simply increasing the number of \rev{programs}. Using
stochastic-frontier models for 29 OECD countries,
\citet{BarreraSantana2022EJ} find that stronger energy-efficiency governance
is associated with higher  efficiency. Within their governance
index, measurable policy targets and extensive evaluation are the elements
most strongly associated with efficiency. This supports the view that policy
effectiveness depends not only on adopting individual measures, but also on
embedding them in a durable framework with clear objectives, coordinated
implementation, and systematic evaluation.

This governance perspective is particularly relevant to Canada, where
responsibility for energy efficiency is divided between federal and
provincial governments. The corresponding priority is therefore a coordinated
national direction combined with provincial
implementation adapted to local energy systems and conditions
\citep{IEACanada2022}.

\paragraph{The energy-intensity gap} 
Prices, policy environments, and observable structural differences are not the sole determinants for high U.S.\ energy use. At the mean values of the
gap covariates, the main model estimates
$\sigma_u=0.184$ and
$\mathbb{E}[u_{st}\mid Z_{st}]=0.146$ log points. Because the gap variance is
heteroskedastic, averaging the  predicted values across all state--years
gives a larger value,
$\overline{\mathbb{E}[u\mid Z]}=0.194$ log points. 

The model therefore points to a third component of the U.S.\
high-energy-use pattern: comparatively low energy efficiency, reflected
in a sizeable \rev{modeled} gap above the conditional frontier. This state-level
finding is consistent with earlier OECD frontier evidence placing both the
United States and Canada toward the lower-efficiency end of the comparison
\citep{FilippiniHunt2011}.

To illustrate its scale, we mechanically halve each jurisdiction's expected
composite gap in 2022 while holding its frontier unchanged. This
produces an aggregate energy-use level that is 11\% lower,
equivalent to 30.6~MMBtu per person or 10.2 quadrillion Btu.\footnote{The expected log gap is
calculated separately for each jurisdiction as
$\mathbb{E}[u\mid Z]=\sigma_u(Z)\sqrt{2/\pi}$. The calculation halves this
quantity, applies the exact exponential transformation, and aggregates the
resulting changes using observed 2022 energy use.} The composite gap is not
entirely wasted energy: it may combine incomplete technology adoption,
financing and information constraints, split incentives, operating practices,
service demand, and rebound. Nevertheless, its scale suggests that incomplete
adoption and implementation are potentially important parts of the North American 
energy-use puzzle. 
The policy priority is therefore to diagnose this margin sector by sector and
address its remediable components. Comparable provincial analysis
could help determine how much of Canada's high use reflects a similar margin,
as opposed to climate, distance, and resource-intensive production.

\paragraph{Equity and differentiated responsibilities}
Another dimension to the explanation for high U.S.\
energy use is related to income.
GDP per capita is positively associated with both the frontier and the scale of the composite inefficiency. Specifically, a 10\% higher GDP
per capita is associated with a 1.9\% higher frontier and a 13.3\% larger
expected gap, conditional on the other covariates. GDP per capita
accounts for 11.0\% of inefficiency variation \rev{(4.7 percentage points of total log-energy variance)}, making it the second most
important predictor \rev{in the gap channel}. This suggests that high U.S.\ energy use
is not solely a consequence of prices, policies, climate, or production structure. It
is also associated with the higher demand for energy services, comfort, and
mobility as incomes rise. This pattern is consistent with comfort-driven and rebound-related usage in high-income states, and it implies that a large share of the aggregate efficiency gap is concentrated where economic slack for policy action is greatest. This supports a differentiated approach in which high-income, high-inefficiency states would contribute disproportionately to demand reductions, while lower-income states receive capacity-building support and compensatory transfers.

The same principle applies to Canada through differentiated federal--provincial
burden sharing. Better-resourced provinces can assume greater responsibility
for investment and implementation, while federal standards and financing can
prevent differences in provincial capacity from producing an uneven
transition. Because climate, distance, and resource-intensive production
create different energy requirements across provinces, differentiation should
reflect both capacity and the scope for reducing avoidable energy use rather
than imposing identical per capita reductions everywhere.

%\textit{Limitations.} Our analysis is descriptive rather than causal. The cross-sectional identification does not rule out confounding by time-invariant unobserved heterogeneity, and the price elasticity may be upward-biased in absolute terms due to the omission of state fixed effects. The ACEEE Scorecard measures policy ambition rather than policy outcomes, and its endogeneity to state energy use cannot be fully addressed in our framework. The inefficiency term captures a composite energy-intensity gap---not pure technical inefficiency---and we do not decompose it into rebound, adoption frictions, and behavioral components. Finally, the analytical standard errors from maximum likelihood do not account for within-state serial correlation; the state-level block bootstrap confidence intervals reported above address this concern for the key policy parameters, but significance levels for individual coefficients should be interpreted with caution. A Mundlak adjustment (Section~\ref{sec:mundlak}) confirms that the frontier coefficients primarily reflect cross-sectional variation, while the inefficiency estimates are robust. 

\section{Conclusion}\label{sec:conclusion}
This paper combines LMDI decomposition, stochastic frontier analysis,
   and variable-importance methods to study cross-state variation in
  U.S.\ per capita energy consumption over 2006--2022. The main finding is that \rev{energy prices and efficiency policies are the two leading observed contributors when the frontier and composite-gap channels are combined.} When the inefficiency channel is included, prices and
policies together account for roughly \rev{31\%} of total variation, while the majority of the inefficiency variation remains unexplained,
pointing to unobserved institutional and behavioral heterogeneity. This finding provides direct support for energy and climate policies using price signals and energy efficiency regulation to promote lower energy consumption. The energy-intensity gap above the frontier is best
  interpreted as a composite of technological inefficiency, adoption
  frictions, behavioral responses, and rebound effects rather than a
  single distortion.

  Two extensions would strengthen the analysis. First, decomposing the
  intensity gap into its constituent channels, such as rebound, adoption
  frictions, and preference-driven usage, would clarify the mechanisms
  behind the aggregate inefficiency measure. \rev{Second, building on our five-domain disaggregation of the ACEEE Scorecard, linking domain-specific policy instruments to sector-level energy demand with richer longitudinal implementation data could better identify which regulatory mechanisms drive realized energy savings.}

\section*{Acknowledgements}

 We acknowledge funding from the Social Sciences and Humanities Research Council (SSHRC) Grant
 430-2026-00414, and the funds made available by the Chair in Energy Sector Management (HEC Montréal) to support this research.

\section*{CRediT authorship contribution statement}

\textbf{David Benatia:} Conceptualization, Software, Methodology, Formal analysis, Writing -- original draft, review \& editing, Visualization, Funding acquisition, Supervision. \textbf{Rémy Molinié:} Data curation, Software, Formal analysis, Writing -- original draft. \textbf{Pierre-Olivier Pineau:} Conceptualization, Funding acquisition, Supervision.

\section*{Declaration of competing interest}

The authors declare that they have no known competing financial interests or personal relationships that could have appeared to influence the work reported in this paper.

\bibliographystyle{elsarticle-harv}

\bibliography{references_energy_efficiency}

\pagebreak

\appendix                     
\setcounter{page}{1}

\begin{center}
    {\Large\bfseries Appendices}\\
    \vspace{1em}
    {\emph{What Drives Energy Use? Prices, Policy, and Inefficiency} \\
    \vspace{0.5em}
David Benatia, Rémy Molinié \&
Pierre-Olivier Pineau}
\end{center}

\noindent
These appendices provide the theoretical, data, methodological, and
robustness details supporting the main analysis. For ease of navigation,
the material is organized as follows:

\begin{itemize}[
    leftmargin=*,
    itemsep=0.2em,
    topsep=0.5em
]

\item
\hyperref[app:data]{\textbf{\ref*{app:data}: Data and variable construction.}}
Sources, definitions, descriptive statistics, and variable construction.

\item
\hyperref[app:energy-series]{\textbf{\ref*{app:energy-series}: Energy-consumption time series.}}
State-level energy consumption over 2006--2022.

\item
\hyperref[ts]{\textbf{\ref*{ts}: Other state time series.}}
Fossil production, electricity generation, prices, and related variables.

\item
\hyperref[app:lmdi]{\textbf{\ref*{app:lmdi}: LMDI decomposition.}}
Sectoral identities, aggregation weights, climate adjustment, and uncertainty.

\item
\hyperref[sec:imp]{\textbf{\ref*{sec:imp}: Variable-importance methods.}}
Implementation and interpretation of the LMG and Random Forest analyses.

\item
\hyperref[sec:des_un]{\textbf{\ref*{sec:des_un}: Units and transformations.}}
Units, logarithmic transformations, centering, and model-ready variables.

\item
\hyperref[sec:oth_dt]{\textbf{\ref*{sec:oth_dt}: ACEEE Scorecard.}}
Construction, harmonization, timing, composition, and econometric limitations.

\item
\hyperref[sec:add_fig]{\textbf{\ref*{sec:add_fig}: Additional figures.}}
Supporting state and national time-series figures.

\item
\hyperref[sec:robust]{\textbf{\ref*{sec:robust}: Robustness checks.}}
Alternative samples, outcomes, distributions, controls, and contribution results.

\item
\hyperref[sec:policy_diagnostics]{\textbf{\ref*{sec:policy_diagnostics}: Policy and state-context diagnostics.}}
Coefficient sensitivities, scale illustrations, and descriptive state contexts.

\item
\hyperref[app:microfoundation]{\textbf{\ref*{app:microfoundation}: Economic foundation.}}
Microfoundation of the frontier and composite gap and its mapping to the SFA.

\end{itemize}

%% Literature review section removed per author request.
%% \section{Literature review}\label{lr}

\pagebreak

\section{Data}\label{app:data}

This appendix provides more detailed information about our dataset.

\subsection{Additional descriptive statistics}\label{app:descriptive-statistics}

Table \ref{tab:stats-desc} and Table \ref{tab:variable-units} present additional summary statistics and information about the key variables.

\begin{table}[h!]
\centering
\scriptsize
\setlength{\tabcolsep}{2pt}
\begin{tabularx}{\textwidth}{l*{8}{Y}}
\toprule
\textbf{Variable} & \textbf{n} & \textbf{mean} & \textbf{sd} & \textbf{median} & \textbf{Q0.25} & \textbf{Q0.75} & \textbf{min} & \textbf{max} \\
\midrule
CDD             & 867 & 1.22   & 0.96  & 0.97   & 0.54   & 1.70   & 0.00   & 5.13 \\
HDD             & 867 & 5.11   & 2.29  & 5.24   & 3.48   & 6.74   & 0.00   & 11.70 \\
\rev{Energy use/pc}     & 867 & 343.28 & 171.59& 302.20 & 232.90 & 383.05 & 151.20 & 1093.30 \\
Price           & 867 & 19.99  & 4.22  & 19.54  & 17.02  & 22.21  & 8.47   & 44.71 \\
Pop             & 867 & 6229   & 7009  & 4376   & 1763   & 7111   & 523    & 39503 \\
GDP/pc         & 867 & 58.44  & 22.34 & 54.48  & 48.20  & 62.17  & 35.61  & 214.65 \\
\rev{Eff}             & 867 & 18.20  & 10.71 & 16.14  & 10.15  & 24.69  & 0.00   & 47.58 \\
VMT/pc          & 867 & 10201  & 2036  & 10242  & 8838   & 11243  & 4516   & 19132 \\
ComSurf/pc      & 867 & 285.93 & 52.53 & 278.94 & 260.78 & 295.26 & 207.60 & 472.78 \\
ResSurf/pc      & 867 & 705.30 & 82.58 & 723.17 & 647.44 & 769.72 & 475.53 & 856.63 \\
E-Int Ind       & 867 & 6.11   & 3.45  & 5.70   & 3.60   & 7.94  & 0.09   & 23.07 \\
FFelec/pc       & 867 & 11.04  & 11.70 & 8.15   & 5.00   & 13.00  & $<0.01$ & 83.06 \\
FF/pc & 867 & 0.60   & 2.25  & 0.01   & 0.00   & 0.25  & 0.00   & 19.89 \\
\bottomrule
\end{tabularx}
\begin{minipage}{\textwidth}
\vspace{2pt}\footnotesize
\rev{\textit{Notes:} Variable units match Table~\ref{tab:variable-units}. In this table, \textit{VMT/pc} is reported in unscaled vehicle-miles per person (versus thousands of vehicle-miles per person in Table~\ref{tab:sumstats}), and \textit{Pop} is in thousands of residents.}
\end{minipage}
\caption{Descriptive statistics of the variables}\label{tab:stats-desc}
\end{table}

\begin{table}[htbp]
\centering
\caption{Variable units and definitions}
\label{tab:variable-units}
\small
\begin{tabularx}{\textwidth}{lX}
\toprule
\textbf{Variable} & \textbf{Unit or definition} \\
\midrule
CDD/HDD
    & Thousand degree days \\

\rev{GDP/pc}
    & Thousands of chained 2017 U.S.\ dollars per person \\

Pop
    & Thousands of residents in Table~\ref{tab:stats-desc} \\

\rev{Energy use/pc}
    & Million Btu (MMBtu) per person \\

Price
    & Current U.S.\ dollars per MMBtu \\

VMT/pc
    & Vehicle-miles traveled per person \rev{in Table~\ref{tab:stats-desc} (reported in thousands in Table~\ref{tab:sumstats})} \\

\rev{Eff}
    & Author-harmonized 2020-reference points; \rev{reference domain maxima sum to 50 points; signed scores are retained} \\

ComSurf/pc
    & Square feet per resident at the Census-division--year level, assigned to member states \\

E-Int Ind
    & Percentage of economic activity in energy-intensive industries \\

ResSurf/pc
    & Square feet per person \\

FF/pc
    & Billion Btu per person, equivalent to 1,000 MMBtu per person \\

FFelec/pc
    & Megawatt-hours per person \\
\bottomrule
\end{tabularx}
\end{table}

\subsection{Sources and variable construction}\label{def}

\noindent
Table~\ref{tab:variable-units} summarizes the measurement units. This
subsection documents the source, construction, interpretation, and model
treatment of each variable.

\subsubsection*{Energy, climate, and economic variables}

\noindent
The \href{https://www.eia.gov/state/seds/}{EIA State Energy Data System
(SEDS)} provides the principal energy, climate, expenditure, and population
series. Coverage varies by series; the estimation panel covers 2006--2022.

\begin{vardictionary}

\item
\varheading{Total primary energy consumption per capita (\rev{\textit{Energy use/pc}})}
{SFA outcome; million Btu (MMBtu) per person}

\textit{Construction.}
The baseline outcome is SEDS total energy consumption per resident:
\[
TETPB_{st}=\frac{TETCB_{st}}{TPOPP_{st}}.
\]
Here, $TETCB$ is economy-wide total consumption on a primary-energy
equivalent basis, including electricity-system inputs and
allocated losses. $TPOPP$ is July 1 resident population, measured in
thousands and including members of the Armed Forces.

\textit{Model treatment.}
The SFA response is $\ln(TETPB_{st})$ and is not mean-centered. It is distinct
from delivered end-use consumption.

\item
\varheading{Cooling and heating degree days (\textit{CDD}, \textit{HDD})}
{Temperature exposure; thousands of degree days}

Annual cooling and heating degree days compare mean outdoor temperature with
a $65^{\circ}\mathrm{F}$ reference. CDD measure temperatures above the
reference point, while HDD measure temperatures below it. Larger values
indicate greater potential demand for cooling or heating.

\item
\varheading{Real GDP per capita (\textit{GDP/pc})}
{Thousands of chained 2017 U.S. dollars per resident}

Real state GDP is divided by resident population. The variable captures the
scale of economic activity and income-related energy-service demand.

\item
\varheading{Population (\textit{Pop})}
{July 1 resident population}

Population is used to construct per-capita variables. It is displayed in
thousands in Table~\ref{tab:stats-desc} but is not a standalone frontier
regressor.

\item
\varheading{Aggregate energy price (\textit{Price})}
{SFA baseline; current U.S. dollars per MMBtu}

\textit{Construction.}
The baseline price measure is
\[
TETCD_{st}
  =1000\,\frac{TETCV_{st}}{TNSCB_{st}},
\qquad
TETCV_{st}=ESTCV_{st}+PESSV_{st}.
\]
It is EIA's aggregate end-user expenditure index divided by
expenditure-adjusted end-use consumption, $TNSCB$. The denominator is neither
primary-equivalent consumption, $TETCB$, nor unadjusted delivered energy use.

\textit{Model treatment and interpretation.}\par
The model uses pooled-mean-centered $\ln(TETCD)$. Its frontier coefficient is
an aggregate conditional elasticity, not a sector-specific causal own-price
elasticity. Its gap-equation coefficient is an association with
$\ln(\sigma_u^2)$. The \hyperref[sec:robust]{robustness appendix} reports same-vintage checks using
delivered consumption and an upstream price measure; neither check is a
sector model.

\item
\varheading{Retail electricity consumption and price}
{Robustness outcome and matched price}

The retail-electricity outcome is electricity sold to ultimate customers per
resident,
\[
\ln\!\left(ESTCB_{st}/TPOPP_{st}\right).
\]
Its matched price is pooled-mean-centered $\ln(ESTCD_{st})$, EIA's average
retail electricity price expressed on an energy-equivalent dollars-per-MMBtu
basis. This is an aggregate retail-electricity association across all end-use
sectors, not a sectoral demand elasticity or a causal price effect.

\item
\varheading{Direct primary energy excluding retail electricity}
{Robustness outcome and matched price}

The direct-primary outcome is
\[
\ln\!\left((TNTCB_{st}-ESTCB_{st})/TPOPP_{st}\right).
\]
Its matched price is pooled-mean-centered $\ln(PETXD_{st})$, a unit value for
direct end-use primary fuels weighted by current consumption. The
physical quantity retains some process or no-direct-cost energy that is absent
from the price accounting, so outcome and price are conceptually aligned but
not denominator-identical. The two component outcomes close in levels:
\[
TNTCB=ESTCB+(TNTCB-ESTCB).
\]

\item
\varheading{Fossil-fuel production per capita (\textit{FF/pc})}
{Billion Btu per resident}

\textit{Construction.}
SEDS coal production, marketed natural-gas production, and crude-oil
production including lease condensate are summed and divided by resident
population.

\textit{Interpretation.}
The variable measures fossil energy produced within a state, not fossil
energy consumed by its residents.

\end{vardictionary}

\subsubsection*{Policy, mobility, industry, and built environment}

\begin{vardictionary}

\item
\varheading{Energy-efficiency policy score (\textit{Eff})}
{Author-harmonized ACEEE composite; 2020-reference points}

The model uses a signed five-domain composite derived from the ACEEE State
Energy Efficiency Scorecard. It is neither ACEEE's official annual total nor
an enforcement index. The retained domains are utility and public-benefit
programs (UPB), transportation, buildings, state-government initiatives, and
appliances. Values for 2007 and 2021 are interpolated component by component;
all other panel years use audited official editions.

For official source edition $t$, the signed raw score in domain $d$ is
normalized by that edition's audited maximum and rescaled to the corresponding
2020-reference maximum:
\[
\begin{aligned}
\textit{Eff}_{st}
&=
\sum_{d\in\{UPB,Tr,Building,SGI,Appliances\}}
\frac{Raw_{dst}}{Max^{source}_{dt}}Max^{2020}_{d},\\
\bigl(Max^{2020}_{d}\bigr)&=(20,12,9,6,3).
\end{aligned}
\]
Early labels are mapped to these five domains, omitted categories are not
folded into the composite, signed penalties are retained, and interpolated
years are calculated component by component. The
\hyperref[sec:oth_dt]{ACEEE Scorecard appendix}
reports the source audit and limitations.

\item
\varheading{Vehicle-miles traveled per capita (\textit{VMT/pc})}
{Vehicle-miles per resident}

The \href{https://data.transportation.gov/Roadways-and-Bridges/Highway-Statistics-Annual-Vehicle-Miles-of-Travel-/b9i2-d7ii}{U.S. Department of Transportation}
reports annual vehicle-miles traveled by vehicle type and urban or rural
location. Mileage is summed across the reported vehicle classes and both
location categories, then divided by resident population.

\item
\varheading{Fossil electricity generation per capita (\textit{FFelec/pc})}
{Megawatt-hours per resident}

Annual net generation from coal, petroleum, and natural gas by the Total
Electric Power Industry is summed and divided by resident population. The
variable measures generation at the production location rather than
residents' electricity consumption, fossil-fuel input, or conversion losses;
it may therefore include electricity exported to other states. The source is
the \href{https://www.eia.gov/electricity/data.php#generation}{EIA electricity
data portal}.

\item
\varheading{Energy-intensive industry share (\textit{E-Int Ind})}
{Percentage of total state GDP}

Using \href{https://www.bea.gov/data/gdp/gdp-industry}{BEA GDP-by-industry
data}, the variable equals 100 times the sum of state GDP in eight
energy-intensive manufacturing categories divided by total state GDP.
Following \citet{Gerres2019}, the retained BEA categories are chemical; food,
beverage, and tobacco; fabricated metal; nonmetallic mineral; paper;
petroleum and coal products; plastics and rubber; and primary metal
manufacturing. Mining, quarrying, and oil and gas extraction are excluded.

\item
\varheading{Commercial floor area per capita (\textit{ComSurf/pc})}
{Constructed Census-division--year benchmark; square feet per resident}

\textit{Construction.}
The \href{https://www.eia.gov/consumption/commercial/}{Commercial Buildings
Energy Consumption Survey} provides total commercial floor-area benchmarks
$F_{d,t}$ for the nine Census divisions in 2003, 2012, and 2018. The annual
division path compounds the 2003--2012 division growth rate over the first
interval and the 2012--2018 rate over the second. After 2018, the 2012--2018
compound annual growth rate is extended through 2022.

Let $P_{s,t}$ denote annual BEA population for state $s$ in division $d(s)$,
and let $P_{d,t}=\sum_{j\in d}P_{j,t}$. Each division-year total is allocated
to its member states in proportion to contemporaneous population:
\[
\widehat F_{s,t}
  =F_{d(s),t}\frac{P_{s,t}}{P_{d(s),t}},
\qquad
\operatorname{ComSurf/pc}_{s,t}
  =\frac{\widehat F_{s,t}}{P_{s,t}}
  =\frac{F_{d(s),t}}{P_{d(s),t}}.
\]

\textit{Interpretation.}
The allocation preserves each CBECS division-year total and uses no state
commercial-building-inventory shares. Because population cancels in the
per-capita expression, all states in the same division receive the same value
in a given year. The variable is a constructed division-year benchmark
assigned to member states, not an observed annual state building stock or a
source of within-division cross-state variation.

\item
\varheading{Residential floor area per capita (\textit{ResSurf/pc})}
{Constructed state-year series; square feet per resident}

\textit{Construction.}
The \href{https://www.eia.gov/consumption/residential/data/2020/}{2020
Residential Energy Consumption Survey} provides the state anchor: total
square feet in primary residences divided by BEA state population. Let
$r^{RECS}_{s,2020}$ denote this anchor and $m^{AHS}_{d,t}$ the
\href{https://www2.census.gov/programs-surveys/ahs/2011/2011\%20AHS\%20Definitions.pdf}{American Housing Survey}
Census-division median square feet per person. The additive bridge is
\[
r_{st}
=r^{RECS}_{s,2020}
+\left(m^{AHS}_{d(s),t}-m^{AHS}_{d(s),2020}\right).
\]
Direct AHS division inputs are used for 2011, 2013, 2015, 2017, 2019, and
2021. Intervening years, including the 2020 index, are linearly interpolated.
The 2011 division index is carried backward unchanged over 2006--2010, and
the 2022 endpoint is the midpoint of the 2021 and 2023 division
inputs.

\textit{Interpretation.}
The additive formula reproduces each state's raw RECS anchor in 2020. Every
state-year observation nevertheless combines a state anchor with a
division-level trajectory; the series is not a directly observed annual state
panel.

\end{vardictionary}

\subsubsection*{Econometric treatment}

\begin{itemize}[leftmargin=1.5em,itemsep=0.3em,topsep=0.35em]
  \item Variables permitted to equal zero---\textit{FF/pc}, \textit{Eff},
  \textit{HDD}, and \textit{CDD}---enter in levels.

  \item All remaining positive covariates, including \textit{ComSurf/pc},
  \textit{ResSurf/pc}, and \textit{FFelec/pc}, enter in logarithms.

  \item Every regressor is centered on its pooled 867-observation mean. The
  logged energy consumption outcome is not centered.
\end{itemize}

\section{Energy consumption - State time series}\label{app:energy-series}

\begin{figure}[H]
  \centering
  \caption{Total energy consumption per capita and state, 2006-2022}
\includegraphics[width=0.80\textwidth]{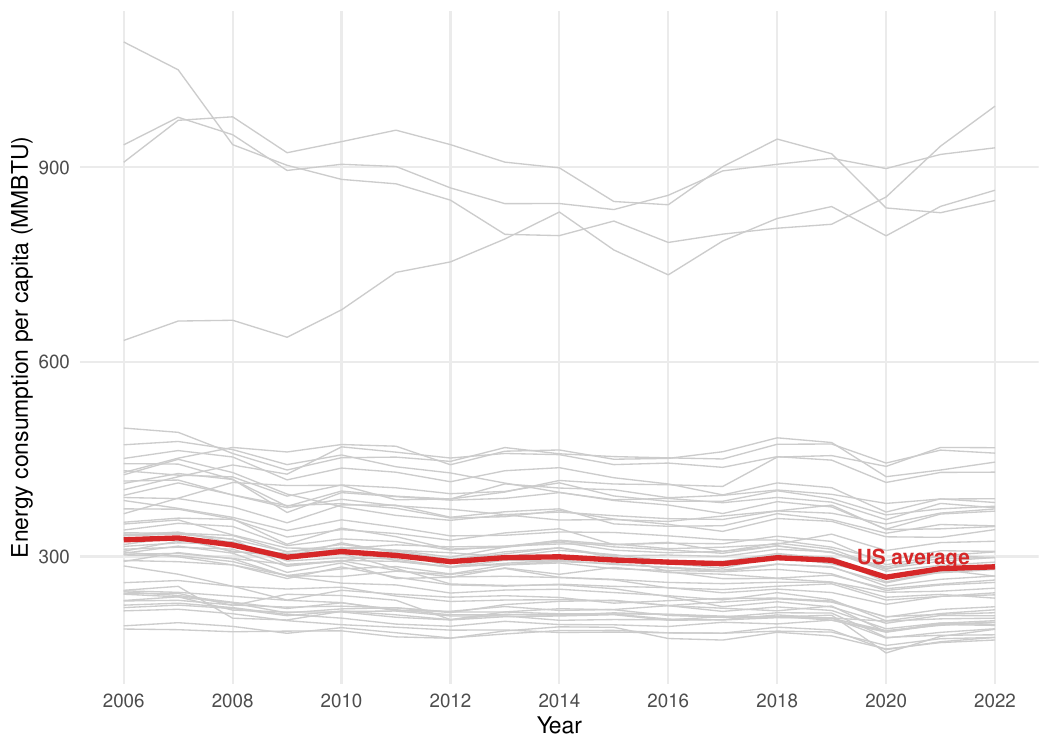}
\label{fig:en_cons_tot_ts}
\end{figure}

\section{Other time series}\label{ts}

\begin{figure}[H]
  \centering
  \caption{Fossil fuel extraction per capita and state, 2006-2022}
\includegraphics[width=0.80\textwidth]{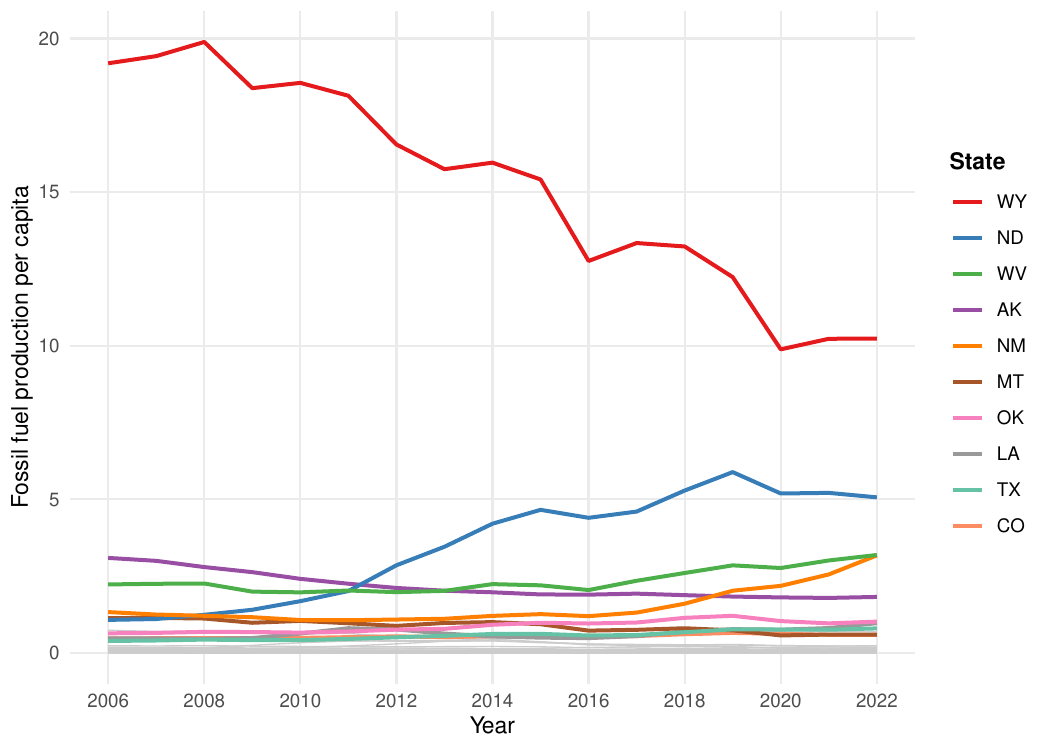}

\label{fig:FF_pc_ts}
\end{figure}

\begin{figure}[H]
  \centering
\includegraphics[width=0.80\textwidth]{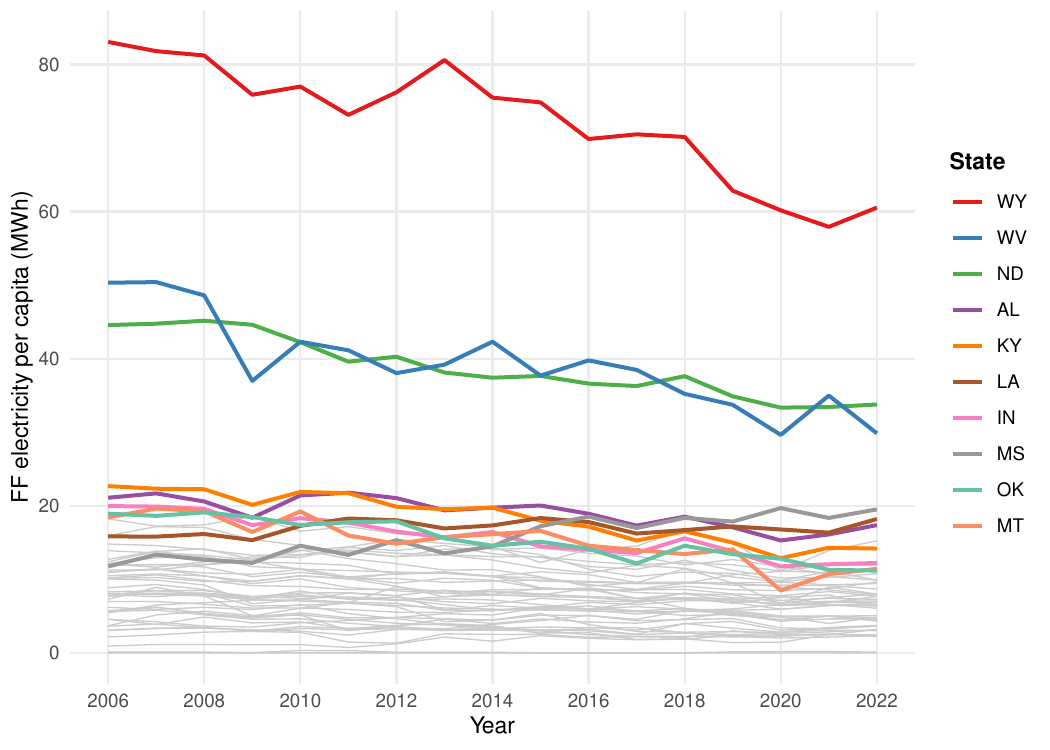}
\caption{Generation of electricity using fossil fuel per capita and state, 2006-2022}
\label{fig:gen_ff_ts}
\end{figure}

\begin{figure}[H]
  \centering
\includegraphics[width=0.80\textwidth]{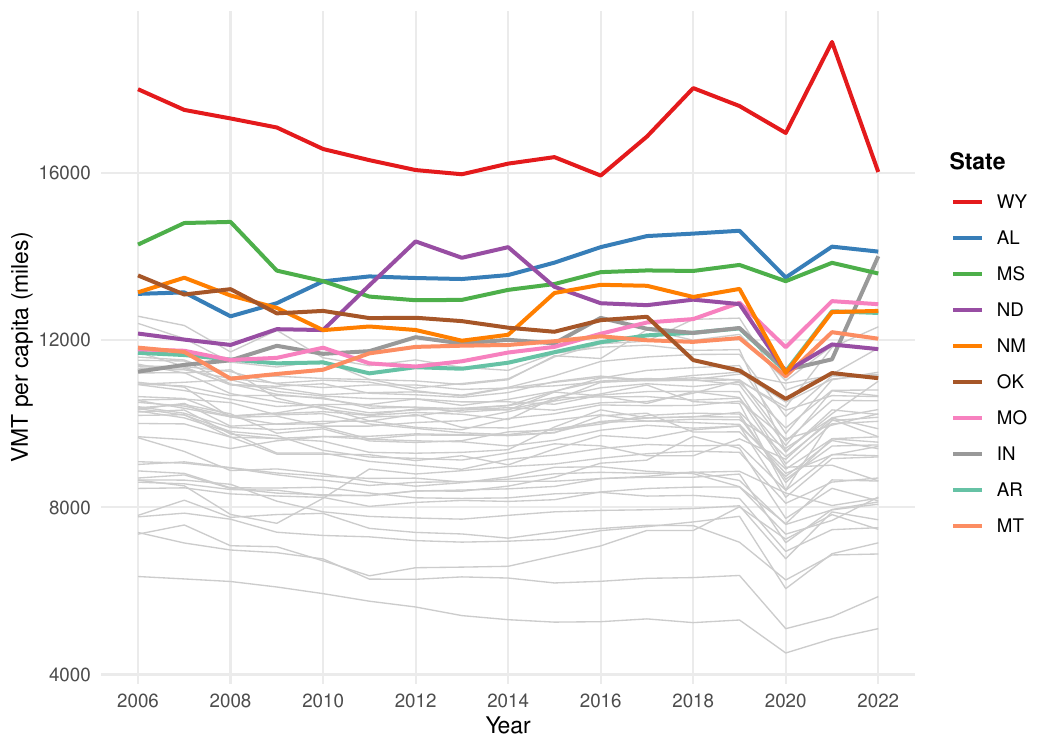}
\caption{Vehicle miles traveled per capita and state, 2006-2022}
\label{fig:VMT_ts}
\end{figure}

\begin{figure}[H]
  \centering
\includegraphics[width=0.80\textwidth]{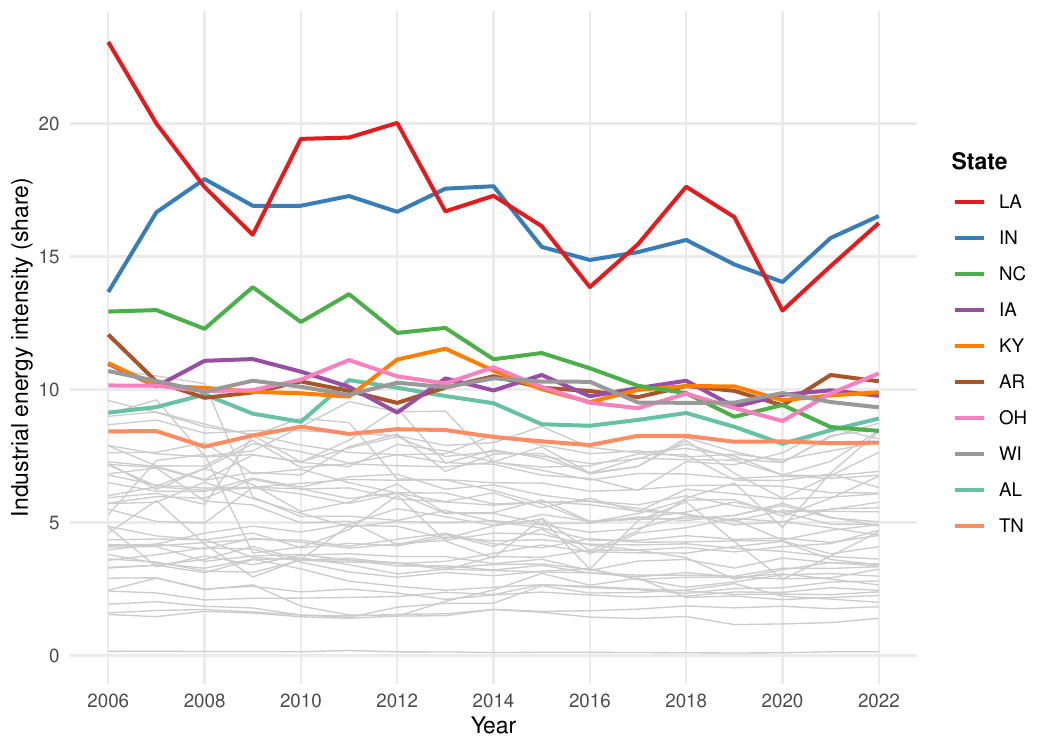}
\caption{Share of energy intensive industry by state, 2006-2022}
\label{fig:pct_ener_ts}
\end{figure}

\clearpage

\section{LMDI decomposition methodology}\label{app:lmdi}

We apply the multiplicative LMDI-I decomposition of \citetA{ANG2005,Ang2015}. Total per-capita energy consumption is the sum over four end-use sectors $k \in \{\text{RES}, \text{COM}, \text{IND}, \text{TR}\}$:
\begin{equation}
  E_{\text{tot,pc}} = \sum_{k} E_{k,\text{pc}} = \sum_{k} A_{k,\text{pc}} \times \mathrm{INT}_k \times \mathrm{CLI}_k,
\end{equation}
where $A_{k,\text{pc}}$ is a per-capita activity indicator, $\mathrm{INT}_k$ is the energy intensity net of climate, and $\mathrm{CLI}_k$ is a climate factor ($\mathrm{CLI}_k = 1$ for the industrial and transport sectors).

\paragraph{Sectoral activity indicators.}
\begin{itemize}[nosep]
  \item \textbf{Residential:} residential floor area per capita,
  $A_{\text{res,pc}}=\rev{\textit{ResSurf/pc}}$ (square feet per person).
  \item \textbf{Commercial:} $A_{\text{com,pc}} = \rev{\textit{ComSurf/pc}}$ (CBECS Census-division commercial floor area per resident assigned to member states, sqft/person).
  \item \textbf{Industrial:} $A_{\text{ind,pc}} = \text{GDP}_{\text{pc}}^{\text{indus}}$ (industrial real GDP per capita).
  \item \textbf{Transport:} $A_{\text{tr,pc}} = \textit{VMT/pc}$ (vehicle miles traveled per capita).
\end{itemize}

\paragraph{Climate factors.}
For the residential and commercial sectors, climate factors are constructed from heating and cooling degree days (HDD, CDD) relative to a reference value:
\begin{equation}
  \mathrm{CLI}_k = \left(\frac{\text{HDD}+1}{\text{HDD}^{\text{ref}}+1}\right)^{\!\alpha_k} \left(\frac{\text{CDD}+1}{\text{CDD}^{\text{ref}}+1}\right)^{\!\beta_k}, \quad k \in \{\text{res}, \text{com}\},
\end{equation}
where $\text{HDD}^{\text{ref}}$ and $\text{CDD}^{\text{ref}}$ are the 2006--2022 means of the annual population-weighted U.S. raw degree-day series. The continuity correction $1$ is one raw degree day; because the stored SFA variables are in thousands, its stored-unit equivalent is $0.001$. On the 51-jurisdiction panel, excluding any synthetic U.S. unit, $\alpha_k$ and $\beta_k$ are estimated from
\begin{equation}
  \log(E_{k,st}/A_{k,st}) = \alpha_k \log\!\left(\frac{\mathrm{HDD}_{st}+1}{\mathrm{HDD}^{\mathrm{ref}}+1}\right) + \beta_k \log\!\left(\frac{\mathrm{CDD}_{st}+1}{\mathrm{CDD}^{\mathrm{ref}}+1}\right) + \mu_s + \lambda_t + e_{k,st}.
\end{equation}
The state and year effects use within-state weather variation after removing persistent state levels and common annual shocks. Table~\ref{tab:climate_bridge} reports the four slopes with state-clustered HC1 standard errors. These slopes are accounting-normalization inputs, not causal weather effects.

\paragraph{LMDI-I weights.}
The multiplicative LMDI-I decomposition uses the logarithmic mean as a weighting function. For the change between years $0$ and $T$, the contribution of each factor is computed using the \texttt{LMDIR} package \citep{Heun_LMDIR_0_1_14_2024}. The decomposition satisfies the following properties:
\begin{itemize}[nosep]
  \item \emph{Exactness:} $D_{\text{tot}} = D_{\text{act}} \times D_{\text{int}} \times D_{\text{cli}}$, where $D$ denotes the multiplicative effect. There is no accounting residual conditional on the fitted climate slopes and observed inputs.
  \item \emph{Symmetry:} The decomposition is invariant to the choice of reference year.
  \item \emph{Aggregation consistency:} Sector-level effects aggregate to the economy-wide effect.
\end{itemize}

For the discrete decomposition, we compare $t = 2006$ and $t = 2022$. For the continuous decomposition, we chain year-on-year effects from 2006 through 2022. State-level decompositions use the same framework applied to each state individually.

Uncertainty from the four climate slopes is propagated with exactly 9,999 draws shared across the residential and commercial equations. Each draw holds observed energy and activity paths and Total fixed, reallocating Climate one-for-one to residual Intensity in log space. The resulting ranges are conditional first-stage uncertainty, not full-input LMDI sampling intervals. %Displayed $100(D-1)$ contributions are multiplicative; exact additive closure is in $100\log(D)$ points.

\section{Variable importance: implementation details}\label{sec:imp}

Let $\widehat F_{st}=X_{st}\widehat\beta$ be the fitted log frontier from the
main specification and let $X_{st}^{(j)}$ denote its 11 non-year predictors.
For both the frontier target and predictors we subtract the relevant
calendar-year mean,
\begin{equation}
  \widetilde F_{st}=\widehat F_{st}-\overline F_{\cdot t},
  \qquad
  \widetilde X_{st}^{(j)}=X_{st}^{(j)}-\overline X_{\cdot t}^{(j)}.
\end{equation}
No predictor is dropped, perturbed, or replaced. We compute exact LMG increments over all $2^{11}=2{,}048$ subsets. The auxiliary regression of $\widetilde F$ on all 11 columns of $\widetilde X$ has $R^2=1$ because its target is the linear fitted frontier. The allocation is therefore an additive, sample-specific description of fitted-frontier variation, not observed-energy variance, a coefficient, an effect, or a mechanism decomposition.

\begin{table}[H]
\centering
\caption{Additive LMG attribution of the fitted frontier}
\label{tab:frontier_lmg_all11_app}
\begin{threeparttable}
\small
\begin{tabular}{rlrr}
\toprule
Rank & Variable & \shortstack{LMG\\allocation} & \shortstack{Share of full-\\model $R^2$} \\
\midrule
1 & Price & 0.254971 & 0.254971 \\
2 & VMT per capita & 0.159450 & 0.159450 \\
3 & Fossil production per capita & 0.150703 & 0.150703 \\
4 & ACEEE total score & 0.113001 & 0.113001 \\
5 & Energy-intensive industry share & 0.095580 & 0.095580 \\
6 & Fossil electricity generation per capita & 0.078890 & 0.078890 \\
7 & Residential floor area per capita & 0.037825 & 0.037825 \\
8 & HDD (thousand degree days) & 0.036626 & 0.036626 \\
9 & GDP per capita & 0.029482 & 0.029482 \\
10 & Commercial floor area per capita & 0.025345 & 0.025345 \\
11 & CDD (thousand degree days) & 0.018128 & 0.018128 \\
\bottomrule
\end{tabular}
\begin{tablenotes}
\footnotesize
\item \textit{Notes:} The full-model $R^2$ and the LMG sum are 1.000000; the closure error is $\rev{-2.998\times10^{-15}}$. The last two columns consequently coincide up to rounding.
\end{tablenotes}
\end{threeparttable}
\end{table}

The second LMG target is the residual-conditional composite gap, $-\log(\mathrm{JLMS})$. We regress this generated target on ACEEE, price, and GDP and compute exact LMG over all $2^3=8$ subsets. The fitted $R^2$ and LMG sum are both 0.351732; retaining the 0.648268 remainder closes the total at one.

\begin{table}[H]
\centering
\caption{LMG attribution of the generated composite energy-use gap}
\label{tab:generated_gap_lmg_app}
\begin{threeparttable}
\small
\begin{tabular}{lrrr}
\toprule
Component & \shortstack{Total generated-\\gap share} & \shortstack{Share of\\explained $R^2$} & \shortstack{Predictor\\rank} \\
\midrule
ACEEE total score & 0.224208 & 0.637440 & 1 \\
GDP per capita & 0.110378 & 0.313812 & 2 \\
Price & 0.017146 & 0.048748 & 3 \\
Unexplained remainder & 0.648268 & -- & -- \\
\bottomrule
\end{tabular}
\begin{tablenotes}
\footnotesize
\item \textit{Notes:} The target and allocation are generated and model-dependent. This is not a decomposition of latent mechanisms. Random forests are not applied to the gap and are not combined with either LMG allocation.
\end{tablenotes}
\end{threeparttable}
\end{table}

For the frontier-only Random Forest sensitivity, the primary implementation uses R~4.5.2 and \texttt{party}~1.3-21 on the same $N=867$ year-demeaned design. Each fixed-seed forest (101, 202, and 303) uses 500 trees and $mtry=3$. Within each seed, marginal importance and conditional importance at $1-p>.2$ and $1-p>.95$ are computed from the same fitted forest. The thresholds apply to the conditional-inference association statistic $1-p$, not absolute correlation. Direct legacy raw importance is the all-tree mean per-tree increase in out-of-bag squared error, with zero assigned when a predictor is absent from a tree. Panel B repeats the same design, tree count, $mtry$, and seeds under \texttt{partykit}~1.2-29; its default regression-risk and averaging definition differs, so raw magnitudes are not comparable across panels.

\begin{table}[H]
\centering
\caption{Raw marginal and conditional Random Forest permutation importance}
\label{tab:rf_conditional_importance}
\scriptsize
\setlength{\tabcolsep}{3pt}
\begin{threeparttable}
\begin{tabular}{lrrrrrr}
\toprule
& \multicolumn{2}{c}{Marginal} & \multicolumn{2}{c}{Conditional, $1-p>.2$} & \multicolumn{2}{c}{Conditional, $1-p>.95$} \\
\cmidrule(lr){2-3} \cmidrule(lr){4-5} \cmidrule(lr){6-7}
Variable & Raw VI & Rank & Raw VI & Rank & Raw VI & Rank \\
\midrule
\multicolumn{7}{l}{\textit{Panel A: direct legacy \texttt{party} implementation}} \\
\addlinespace
Price & 3.459e-02 & 1 & \rev{4.418e-04} & 1 & 4.264e-04 & 2 \\
VMT per capita & 1.527e-02 & 2 & 1.066e-04 & 2 & 1.854e-04 & 3 \\
Fossil production per capita & 1.229e-02 & 3 & 5.523e-05 & 3 [3--4] & 5.508e-04 & 1 \\
Fossil electricity generation per capita & 1.145e-02 & 4 [4--5] & 3.631e-05 & 6 [3--7] & \rev{3.485e-05} & \rev{9 [9--10]} \\
ACEEE total score & 1.135e-02 & 5 [4--5] & 4.526e-05 & 5 [4--5] & \rev{4.348e-05} & 7 [5--8] \\
Energy-intensive industry share & 8.063e-03 & 6 & 4.248e-05 & 5 [4--6] & 1.307e-04 & 4 \\
Residential floor area per capita & 3.459e-03 & 7 & 1.186e-05 & 9 [8--11] & 2.657e-05 & 10 [9--10] \\
GDP per capita & 2.796e-03 & 8 & 2.346e-05 & 7 [6--8] & 3.674e-05 & \rev{7 [5--8]} \\
HDD (thousand degree days) & 2.483e-03 & 9 & 1.085e-05 & 10 [9--11] & 4.544e-05 & 6 [5--8] \\
Commercial floor area per capita & 1.658e-03 & 10 & 1.174e-05 & 9 [7--10] & 4.638e-05 & 6 [6--7] \\
CDD (thousand degree days) & 1.024e-03 & 11 & 1.059e-05 & 10 [8--11] & 1.803e-05 & 11 \\
\addlinespace
\midrule
\addlinespace
\multicolumn{7}{l}{\textit{Panel B: modern \texttt{partykit} implementation sensitivity}} \\
\addlinespace
Price & 2.962e-02 & 1 & 3.866e-04 & 1 & 3.817e-04 & 2 [1--2] \\
VMT per capita & 1.443e-02 & 2 & 7.974e-06 & 3 [2--5] & 8.042e-05 & 3 \\
Fossil production per capita & 1.229e-02 & 3 & -6.207e-05 & 9 [5--10] & 4.351e-04 & 1 [1--2] \\
Fossil electricity generation per capita & 1.150e-02 & 4 & -4.828e-05 & 10 [9--10] & \rev{-4.005e-05} & 10 [9--10] \\
ACEEE total score & 9.947e-03 & 5 & \rev{-9.186e-05} & 11 & \rev{-8.506e-05} & 11 \\
Energy-intensive industry share & 7.399e-03 & 6 & -1.986e-05 & 6 [5--7] & \rev{4.004e-05} & 4 \\
Residential floor area per capita & 3.478e-03 & 7 & 1.219e-05 & 2 [2--3] & 1.446e-05 & 6 [5--6] \\
GDP per capita & 2.646e-03 & 8 [8--9] & -4.190e-05 & 8 & \rev{-1.547e-05} & 8 [8--10] \\
HDD (thousand degree days) & 2.471e-03 & 9 [8--9] & -3.169e-05 & 6 [6--9] & 3.364e-06 & 7 [6--7] \\
Commercial floor area per capita & 1.690e-03 & 10 & -9.247e-06 & 4 [3--7] & 1.091e-05 & 5 [5--7] \\
CDD (thousand degree days) & 1.281e-03 & 11 & \rev{-1.586e-05} & 4 [4--7] & -1.832e-05 & 9 [8--9] \\
\bottomrule
\end{tabular}
\begin{tablenotes}
\footnotesize
\item \textit{Notes:} Raw VI is the median signed score across draws; rank is the median with the seed minimum--maximum in brackets. Scores are non-additive predictive diagnostics, not variance shares, coefficients, effects, significance tests, or causal estimates. Fixed-seed ranges describe algorithmic stability, not sampling uncertainty or confidence intervals.
\end{tablenotes}
\end{threeparttable}
\end{table}

In the primary direct implementation, price ranks first marginally and
at the $1-p>.2$ conditional cutoff, and second at the $1-p>.95$ cutoff; across
all nine seed--method comparisons it remains in the top two. ACEEE has median
rank 5 marginally (seed range 4--5), median rank 5 at the .2 cutoff (4--5), and
median rank 7 at the .95 cutoff (5--8), with positive scores in all six
conditional seed--threshold runs. Under the modern implementation, ACEEE ranks
fifth marginally and eleventh under both conditional rules, with negative
scores in all six conditional runs. The latter denotes no positive unique
predictive signal under that implementation, not a negative coefficient,
causal null, or contradiction of the SFA association.

\section{Description of units and transformations}\label{sec:des_un}

Table~\ref{tab:units} summarizes the units and transformations applied to each variable in the SFA frontier equation.

\begin{table}[H]
\centering
\small
\caption{Variable units and transformations in the SFA}\label{tab:units}
\begin{tabular}{@{}p{2.3cm}p{2.85cm}p{2.75cm}p{4.4cm}@{}}
\toprule
\textbf{Variable} & \textbf{Raw unit} & \textbf{Transform} & \textbf{SFA term} \\
\midrule
\rev{Energy use/pc} & million Btu/person & natural log; not centered & $\ln(TETPB_{st})$ (response) \\
\rev{Price} & current \$/million Btu & natural log; pooled-mean center & centered $\ln(TETCD_{st})$ \\
\rev{GDP/pc} & thousand chained 2017 \$/person & natural log; pooled-mean center & centered log GDP/person \\
\rev{VMT/pc} & vehicle-miles/person & natural log; pooled-mean center & centered log VMT/person \\
\rev{ComSurf/pc} & square feet/person & natural log; pooled-mean center & centered log CBECS division floor/person assigned to member states \\
\rev{ResSurf/pc} & square feet/person & natural log; pooled-mean center & centered log constructed residential floor/person \\
\rev{E-Int Ind} & \rev{percent of state GDP} & natural log; pooled-mean center & centered log industry share \\
\rev{FFelec/pc} & MWh/person & natural log; pooled-mean center & centered log generation/person \\
\rev{FF/pc} & billion Btu/person & level; pooled-mean center & centered production/person \\
\rev{Eff} & signed 2020-reference points & level; pooled-mean center & $\textit{Eff}_{st}-\overline{\textit{Eff}}$ \\
HDD & thousand degree days & level; pooled-mean center & centered HDD \\
CDD & thousand degree days & level; pooled-mean center & centered CDD \\
\bottomrule
\end{tabular}
\end{table}

\noindent Mean-centering subtracts the pooled sample mean from each variable, so the SFA intercept is interpretable as \rev{the log frontier evaluated at the pooled means of the transformed continuous regressors in the reference year, 2006}. Variables with zero values (FF/pc, Eff, HDD, CDD) are entered in levels rather than logs to avoid undefined log-transforms. Year dummies $y_{07}$ through $y_{22}$ capture common time effects (reference: 2006).

\section{ACEEE Scorecard: construction and econometric considerations}\label{sec:oth_dt}

\subsection{Harmonized score}

The ACEEE index variable retains five signed domains and places them on the fixed 2020 nominal maxima: UPB 20, Transportation 12, Building 9, State Government Initiatives 6, and Appliances 3 points. For each official ACEEE edition, the construction rescales the signed source-domain score by its source-year maximum and then sums the five components. The changing early categories are mapped explicitly; separate CHP, 2008 RD\&D, and 2022 Industry categories are omitted rather than reassigned. The 2007 and 2021 components are linear interpolations between adjacent audited editions, accounting for 102 state--year rows. The stored total equals the sum of the five signed components to numerical precision.

The measure is deliberately described as a policy index including dimensions relative to existing \rev{programs}, institutional aspects, and performance measures. UPB can include reported prior-year savings and \rev{program} spending; Building includes codes and some compliance or enforcement criteria; Transportation mixes policies and adoption with selected observed outcomes; State Government combines institutions, lead-by-example activity, and incentives; and Appliances is sparse in many years. Rubric content, metric definitions, reference periods, and content change across editions. Consequently, the composite is neither an implementation or enforcement index nor a directly observed measure of realized energy savings.

\subsection{Timing, composition, and endogeneity}

The score enters both the frontier and gap equations in levels. Its frontier coefficient is therefore an aggregate conditional semi-elasticity per harmonized point. Policy selection, persistent state differences, embedded performance measures, feedback, and reverse causality can still remain, and the net direction of reverse-causality bias is not signed a priori in this joint nonlinear model.

Lagging the score is a timing sensitivity. On
the common 2009--2022 sample, all four contemporaneous-through-three-year-lag
frontier coefficients are negative and distinguishable from zero at 5\%,
ranging from $-0.0021$ to $-0.0019$. Across the four timings, the
gap-log-variance coefficient remains negative, from $-0.1643$ to $-0.1566$.
Alternative specifications likewise help gauge
composition or context. The two-block policy model rejects a common score slope
($\chi^2(2)=39.12$, $p<.001$), and the five-domain model further improves on
the two-block restriction ($\chi^2(6)=20.65$, $p=.0021$). This suggests that disaggregating the policy index helps improve predictions.

\subsection{Investor-owned retail-provider context sensitivity}

The IOU context variable is identified investor-owned customer accounts at retail providers or suppliers divided by EIA-861's published state total of retail customer accounts. It measures provider composition. Behind-the-meter accounts are excluded from the identified provider numerator and remain in the signed unallocated residual. The official-account identity is verified as
\[
\begin{aligned}
&\text{IOU accounts}+\text{other identified provider accounts}\\
&\qquad+\text{signed unallocated residual}=\text{published EIA total accounts}
\end{aligned}
\]
to numerical precision.

As a robustness check, we add each state's 2006 IOU share as controls, repeated through 2022, to the full $N=867$ baseline rows. Because 2006 is the first outcome and policy year, this measure is fixed within state but not pre-period or predetermined. The second check uses the contemporaneous annual share and a strict quality rule, $|\text{signed unallocated share}|\leq 0.01$, for both the baseline and augmented model on exactly the same ordered $N=755$ rows. This support is unbalanced: 39--50 jurisdictions appear per year, all 51 appear at least once, Massachusetts contributes two years, and Maine and Ohio five years each. The annual share can be post-treatment. Both checks are bounded context sensitivities.

\section{Additional figures}\label{sec:add_fig}

\begin{figure}[H]
  \centering
  \caption{Energy price by state, 2006--2022.}
  \includegraphics[width=0.80\textwidth]{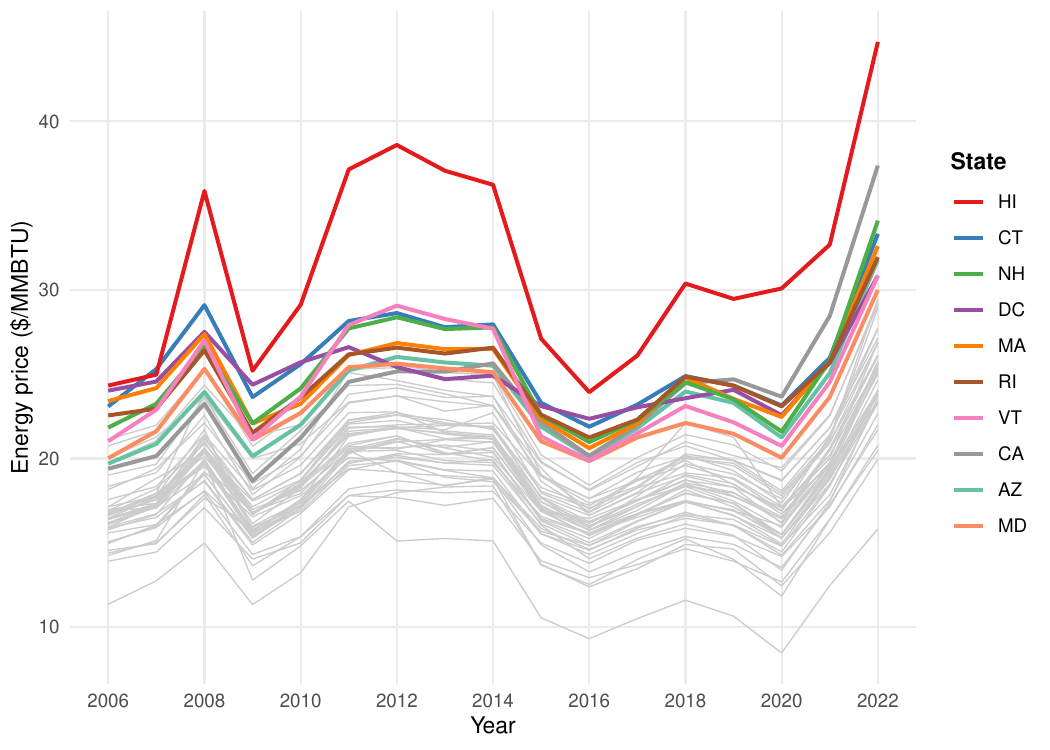}
  \label{fig:pen_s_ts}
\end{figure}

\begin{figure}[H]
  \centering
  \caption{National-level time series of the four most important frontier variables (index, 2006 = 100). \rev{The ACEEE series uses the harmonized scores employed in estimation.}}
  \includegraphics[width=0.80\textwidth]{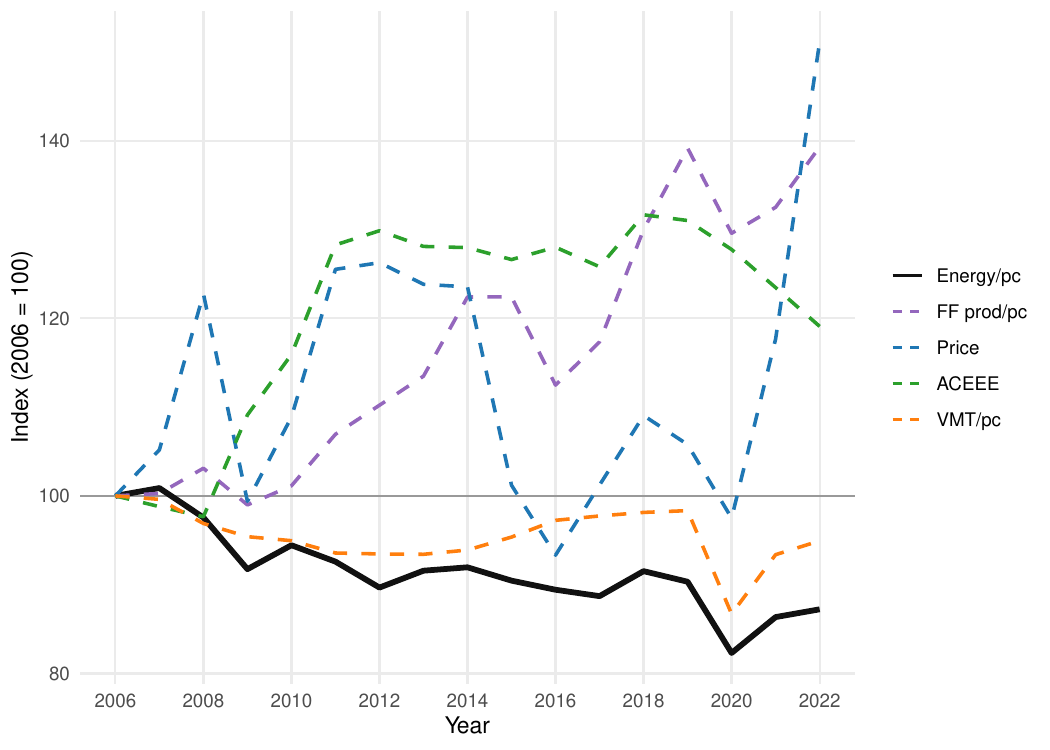}
  \label{fig:var_us_ts}
\end{figure}

\begin{figure}[H]
  \centering
  \caption{Share of each sector in energy consumption per capita, 2006--2022.}
  \includegraphics[width=0.80\textwidth]{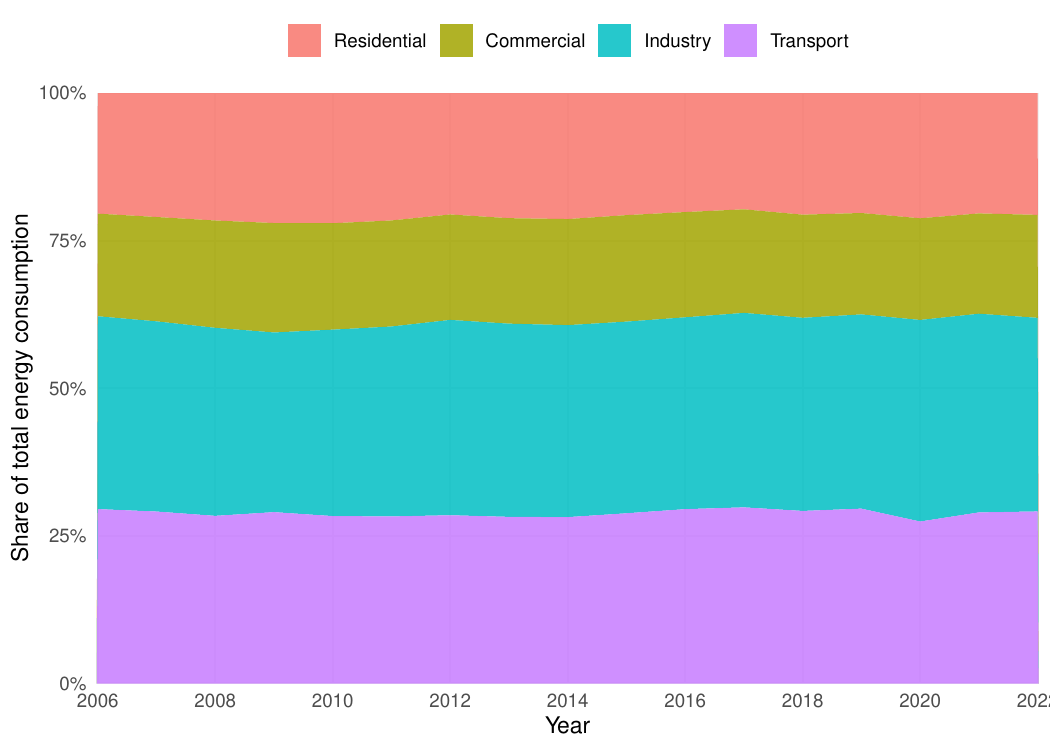}
  \label{fig:US_sect_lev}
\end{figure}

\section{Robustness checks}\label{sec:robust}

\subsection{Measurement and one-sided-distribution sensitivity}

Table~\ref{tab:measurement_distribution_summary} separates changes in the
outcome/price concept from changes in the one-sided distribution. The
half-normal is the predeclared parsimonious reference, not the empirically
best-fitting distribution: the truncated-normal model has lower AIC and BIC.
The negative ACEEE gap association is present in the three measurement pairs
and three distributions, whereas the frontier magnitude and conventional
precision vary. These specifications do not share a single price estimand, and
the distribution-specific gap parameters are not pooled.

\begin{table}[H]
\centering
\caption{Measurement and one-sided-distribution sensitivity}
\label{tab:measurement_distribution_summary}
\scriptsize
\resizebox{\linewidth}{!}{%
\begin{tabular}{lrrrrrr}
\toprule
Specification & Price frontier & ACEEE frontier & ACEEE gap & Log-lik. & AIC & BIC \\
\midrule
\multicolumn{7}{l}{\textit{Panel A: outcome/price concepts}}\\
Preferred TETPB/TETCD & $-0.7524$ $(0.0459)$ & $-0.0017$ $(0.0007)$ & $-0.1555$ $(0.0096)$ & 588.22 & -- & --\\
Strict TNTCB/TETXD, FFelec omitted & $-0.7930$ $(0.0536)$ & $+0.0015$ $(0.0008)$ & $-0.1368$ $(0.0081)$ & 411.17 & -- & --\\
Upstream TETCB/PETCD & $-0.1673$ $(0.0373)$ & $-0.0041$ $(0.0009)$ & $-0.1390$ $(0.0098)$ & 442.49 & -- & --\\
\midrule
\multicolumn{7}{l}{\textit{Panel B: one-sided distributions}}\\
Half normal & $-0.7524$ $(0.0459)$ & $-0.0017$ $(0.0007)$ & $-0.1555$ $(0.0096)$ & 588.22 & $-1110.44$ & $-953.19$\\
Truncated normal & $-1.2238$ $(0.1198)$ & $-0.0108$ $(0.0016)$ & $-0.1841$ $(0.0102)$ & 629.16 & $-1184.31$ & $-1008.01$\\
Exponential & $-0.7578$ $(0.0639)$ & $-0.0037$ $(0.0008)$ & $-0.1751$ $(0.0158)$ & 583.82 & $-1101.63$ & $-944.39$\\
\bottomrule
\end{tabular}
}
\begin{minipage}[t]{\linewidth}
\footnotesize
\textit{Notes:} Standard errors are
in parentheses. All rows use $N=867$. The preferred, strict-end-use, and
upstream-price rows use different outcome/price concepts and the strict model
omits FFelec by design. The half-normal, truncated-normal, and exponential
average expected-gap changes for one additional ACEEE point are $-7.48\%$,
$-1.54\%$, and $-8.38\%$, respectively.
\end{minipage}
\end{table}

 The following generated
tables provide the complete coefficient vectors used in the 
robustness analysis.

\begin{table}[H]
\centering
\caption{One-sided-distribution sensitivity for the preferred specification}\label{tab:measurement_distribution}
\begin{threeparttable}
{\setstretch{1}\scriptsize
\setlength{\tabcolsep}{2pt}
\begin{tabularx}{\textwidth}{@{}>{\raggedright\arraybackslash}p{.28\textwidth}*{3}{Y}@{}}
\toprule
Variable & \shortstack{Half-\\normal} & \shortstack{Truncated\\normal} & Exponential \\
\midrule
\multicolumn{4}{l}{\textit{Frontier equation}} \\[2pt]
Price & \shortstack[c]{$-0.7524^{***}$\\[-1pt]{\scriptsize (0.0459)}} & \shortstack[c]{$-1.2238^{***}$\\[-1pt]{\scriptsize (0.1198)}} & \shortstack[c]{$-0.7578^{***}$\\[-1pt]{\scriptsize (0.0639)}} \\
GDP/pc & \shortstack[c]{$0.1931^{***}$\\[-1pt]{\scriptsize (0.0322)}} & \shortstack[c]{$0.9218^{***}$\\[-1pt]{\scriptsize (0.0510)}} & \shortstack[c]{$0.1684^{***}$\\[-1pt]{\scriptsize (0.0403)}} \\
Eff & \shortstack[c]{$-0.0017^{***}$\\[-1pt]{\scriptsize (0.0007)}} & \shortstack[c]{$-0.0108^{***}$\\[-1pt]{\scriptsize (0.0016)}} & \shortstack[c]{$-0.0037^{***}$\\[-1pt]{\scriptsize (0.0008)}} \\
VMT/pc & \shortstack[c]{$0.3933^{***}$\\[-1pt]{\scriptsize (0.0337)}} & \shortstack[c]{$0.4406^{***}$\\[-1pt]{\scriptsize (0.0285)}} & \shortstack[c]{$0.4127^{***}$\\[-1pt]{\scriptsize (0.0371)}} \\
ComSurf/pc & \shortstack[c]{$-0.0136$\\[-1pt]{\scriptsize (0.0246)}} & \shortstack[c]{$-0.0321$\\[-1pt]{\scriptsize (0.0233)}} & \shortstack[c]{$-0.0293$\\[-1pt]{\scriptsize (0.0278)}} \\
ResSurf/pc & \shortstack[c]{$0.0141$\\[-1pt]{\scriptsize (0.0534)}} & \shortstack[c]{$-0.0581$\\[-1pt]{\scriptsize (0.0466)}} & \shortstack[c]{$-0.0963$\\[-1pt]{\scriptsize (0.0712)}} \\
E-Int Ind & \shortstack[c]{$0.1077^{***}$\\[-1pt]{\scriptsize (0.0099)}} & \shortstack[c]{$0.1276^{***}$\\[-1pt]{\scriptsize (0.0095)}} & \shortstack[c]{$0.1071^{***}$\\[-1pt]{\scriptsize (0.0109)}} \\
FF/pc & \shortstack[c]{$0.0364^{***}$\\[-1pt]{\scriptsize (0.0024)}} & \shortstack[c]{$0.0151^{***}$\\[-1pt]{\scriptsize (0.0023)}} & \shortstack[c]{$0.0343^{***}$\\[-1pt]{\scriptsize (0.0026)}} \\
FFelec/pc & \shortstack[c]{$0.0094^{***}$\\[-1pt]{\scriptsize (0.0031)}} & \shortstack[c]{$0.0154^{***}$\\[-1pt]{\scriptsize (0.0026)}} & \shortstack[c]{$0.0136^{***}$\\[-1pt]{\scriptsize (0.0043)}} \\
HDD & \shortstack[c]{$0.0456^{***}$\\[-1pt]{\scriptsize (0.0044)}} & \shortstack[c]{$0.0558^{***}$\\[-1pt]{\scriptsize (0.0042)}} & \shortstack[c]{$0.0488^{***}$\\[-1pt]{\scriptsize (0.0053)}} \\
CDD & \shortstack[c]{$0.1007^{***}$\\[-1pt]{\scriptsize (0.0090)}} & \shortstack[c]{$0.1348^{***}$\\[-1pt]{\scriptsize (0.0082)}} & \shortstack[c]{$0.1040^{***}$\\[-1pt]{\scriptsize (0.0105)}} \\
\midrule
\multicolumn{4}{l}{\textit{One-sided log-scale equation}} \\[2pt]
Eff & \shortstack[c]{$-0.1555^{***}$\\[-1pt]{\scriptsize (0.0096)}} & \shortstack[c]{$-0.1841^{***}$\\[-1pt]{\scriptsize (0.0102)}} & \shortstack[c]{$-0.1751^{***}$\\[-1pt]{\scriptsize (0.0158)}} \\
Price & \shortstack[c]{$0.3107$\\[-1pt]{\scriptsize (0.3447)}} & \shortstack[c]{$0.1683$\\[-1pt]{\scriptsize (0.3603)}} & \shortstack[c]{$-0.3621$\\[-1pt]{\scriptsize (0.8626)}} \\
GDP/pc & \shortstack[c]{$2.6264^{***}$\\[-1pt]{\scriptsize (0.2873)}} & \shortstack[c]{$2.1644^{***}$\\[-1pt]{\scriptsize (0.3051)}} & \shortstack[c]{$3.8518^{***}$\\[-1pt]{\scriptsize (0.5870)}} \\
\midrule
\multicolumn{4}{l}{\textit{Pre-truncation mean equation (truncated-normal only)}} \\[2pt]
Eff & \textemdash & \shortstack[c]{$0.0058^{***}$\\[-1pt]{\scriptsize (0.0015)}} & \textemdash \\
Price & \textemdash & \shortstack[c]{$0.5728^{***}$\\[-1pt]{\scriptsize (0.1194)}} & \textemdash \\
GDP/pc & \textemdash & \shortstack[c]{$-0.7287^{***}$\\[-1pt]{\scriptsize (0.0565)}} & \textemdash \\
\midrule
$N$ & 867 & 867 & 867 \\
Log likelihood & 588.22 & 629.16 & 583.82 \\
AIC & -1110.44 & -1184.31 & -1101.63 \\
BIC & -953.19 & -1008.01 & -944.39 \\
\bottomrule
\end{tabularx}
\begin{tablenotes}
\footnotesize
\item \textit{Notes:} Standard errors are in parentheses. $^{***}$ $p<0.01$, $^{**}$ $p<0.05$, $^{*}$ $p<0.10$. These three columns hold the preferred outcome, aggregate price, sample, controls, and year effects fixed; only the one-sided distribution changes. The truncated-normal column additionally reports its pre-truncation location equation. Its scale parameters are not directly comparable to the half-normal and exponential parameterizations.
\end{tablenotes}}
\end{threeparttable}
\end{table}

\subsection{Outcome and price measurement sensitivity}

The matched-price specifications preserve the manuscript controls and pair each physical outcome with exactly one corresponding EIA price construct. No displayed regression includes both retail-electricity and direct-primary prices. The second table repeats the three matched pairs after omitting the fossil-electricity-generation control.

\begin{table}[H]
\centering
\caption{Outcome-specific single-price measurement sensitivity}\label{tab:measurement_matched}
\begin{threeparttable}
{\setstretch{1}\scriptsize
\setlength{\tabcolsep}{2pt}
\begin{tabularx}{\textwidth}{@{}>{\raggedright\arraybackslash}p{.24\textwidth}*{4}{Y}@{}}
\toprule
Variable & \shortstack{Preferred\\TETPB/TETCD} & \shortstack{Delivered total\\TNTCB/TETCD} & \shortstack{Direct primary\\PETXD} & \shortstack{Retail electricity\\ESTCD} \\
\midrule
\multicolumn{5}{l}{\textit{Frontier equation}} \\[2pt]
Price & \shortstack[c]{$-0.7524^{***}$\\[-1pt]{\scriptsize (0.0459)}} & \shortstack[c]{$-0.8469^{***}$\\[-1pt]{\scriptsize (0.0599)}} & \shortstack[c]{$-0.8186^{***}$\\[-1pt]{\scriptsize (0.0671)}} & \shortstack[c]{$-0.4625^{***}$\\[-1pt]{\scriptsize (0.0405)}} \\
GDP/pc & \shortstack[c]{$0.1931^{***}$\\[-1pt]{\scriptsize (0.0322)}} & \shortstack[c]{$0.0547$\\[-1pt]{\scriptsize (0.0367)}} & \shortstack[c]{$-0.1255^{***}$\\[-1pt]{\scriptsize (0.0308)}} & \shortstack[c]{$-0.0362$\\[-1pt]{\scriptsize (0.0343)}} \\
Eff & \shortstack[c]{$-0.0017^{***}$\\[-1pt]{\scriptsize (0.0007)}} & \shortstack[c]{$0.0011$\\[-1pt]{\scriptsize (0.0008)}} & \shortstack[c]{$0.0019^{**}$\\[-1pt]{\scriptsize (0.0008)}} & \shortstack[c]{$-0.0060^{***}$\\[-1pt]{\scriptsize (0.0010)}} \\
VMT/pc & \shortstack[c]{$0.3933^{***}$\\[-1pt]{\scriptsize (0.0337)}} & \shortstack[c]{$0.3694^{***}$\\[-1pt]{\scriptsize (0.0390)}} & \shortstack[c]{$0.5453^{***}$\\[-1pt]{\scriptsize (0.0479)}} & \shortstack[c]{$0.2192^{***}$\\[-1pt]{\scriptsize (0.0347)}} \\
ComSurf/pc & \shortstack[c]{$-0.0136$\\[-1pt]{\scriptsize (0.0246)}} & \shortstack[c]{$-0.0527^{*}$\\[-1pt]{\scriptsize (0.0296)}} & \shortstack[c]{$-0.0529$\\[-1pt]{\scriptsize (0.0349)}} & \shortstack[c]{$-0.0296$\\[-1pt]{\scriptsize (0.0305)}} \\
ResSurf/pc & \shortstack[c]{$0.0141$\\[-1pt]{\scriptsize (0.0534)}} & \shortstack[c]{$-0.2066^{***}$\\[-1pt]{\scriptsize (0.0615)}} & \shortstack[c]{$-0.5937^{***}$\\[-1pt]{\scriptsize (0.0650)}} & \shortstack[c]{$0.1647^{***}$\\[-1pt]{\scriptsize (0.0490)}} \\
E-Int Ind & \shortstack[c]{$0.1077^{***}$\\[-1pt]{\scriptsize (0.0099)}} & \shortstack[c]{$0.0614^{***}$\\[-1pt]{\scriptsize (0.0117)}} & \shortstack[c]{$0.0704^{***}$\\[-1pt]{\scriptsize (0.0150)}} & \shortstack[c]{$0.0947^{***}$\\[-1pt]{\scriptsize (0.0091)}} \\
FF/pc & \shortstack[c]{$0.0364^{***}$\\[-1pt]{\scriptsize (0.0024)}} & \shortstack[c]{$0.0436^{***}$\\[-1pt]{\scriptsize (0.0027)}} & \shortstack[c]{$0.0455^{***}$\\[-1pt]{\scriptsize (0.0030)}} & \shortstack[c]{$0.0332^{***}$\\[-1pt]{\scriptsize (0.0029)}} \\
FFelec/pc & \shortstack[c]{$0.0094^{***}$\\[-1pt]{\scriptsize (0.0031)}} & \shortstack[c]{$-0.0078^{**}$\\[-1pt]{\scriptsize (0.0036)}} & \shortstack[c]{$0.0065^{*}$\\[-1pt]{\scriptsize (0.0038)}} & \shortstack[c]{$-0.0042$\\[-1pt]{\scriptsize (0.0041)}} \\
HDD & \shortstack[c]{$0.0456^{***}$\\[-1pt]{\scriptsize (0.0044)}} & \shortstack[c]{$0.0497^{***}$\\[-1pt]{\scriptsize (0.0054)}} & \shortstack[c]{$0.0749^{***}$\\[-1pt]{\scriptsize (0.0057)}} & \shortstack[c]{$-0.0023$\\[-1pt]{\scriptsize (0.0042)}} \\
CDD & \shortstack[c]{$0.1007^{***}$\\[-1pt]{\scriptsize (0.0090)}} & \shortstack[c]{$0.0412^{***}$\\[-1pt]{\scriptsize (0.0111)}} & \shortstack[c]{$0.0130$\\[-1pt]{\scriptsize (0.0130)}} & \shortstack[c]{$0.0622^{***}$\\[-1pt]{\scriptsize (0.0097)}} \\
\midrule
\multicolumn{5}{l}{\textit{Gap log-variance equation}} \\[2pt]
Eff & \shortstack[c]{$-0.1555^{***}$\\[-1pt]{\scriptsize (0.0096)}} & \shortstack[c]{$-0.1362^{***}$\\[-1pt]{\scriptsize (0.0081)}} & \shortstack[c]{$-0.1206^{***}$\\[-1pt]{\scriptsize (0.0074)}} & \shortstack[c]{$-0.0393$\\[-1pt]{\scriptsize (0.0338)}} \\
Price & \shortstack[c]{$0.3107$\\[-1pt]{\scriptsize (0.3447)}} & \shortstack[c]{$0.1649$\\[-1pt]{\scriptsize (0.2907)}} & \shortstack[c]{$-0.4179^{*}$\\[-1pt]{\scriptsize (0.2532)}} & \shortstack[c]{$-3.9378^{***}$\\[-1pt]{\scriptsize (0.9131)}} \\
GDP/pc & \shortstack[c]{$2.6264^{***}$\\[-1pt]{\scriptsize (0.2873)}} & \shortstack[c]{$1.9342^{***}$\\[-1pt]{\scriptsize (0.3373)}} & \shortstack[c]{$2.4888^{***}$\\[-1pt]{\scriptsize (0.3482)}} & \shortstack[c]{$5.2862^{***}$\\[-1pt]{\scriptsize (0.6130)}} \\
\midrule
$N$ & 867 & 867 & 867 & 867 \\
Log likelihood & 588.22 & 413.53 & 282.57 & 689.54 \\
AIC & -1110.44 & -761.05 & -499.13 & -1313.07 \\
BIC & -953.19 & -603.81 & -341.88 & -1155.83 \\
\bottomrule
\end{tabularx}
\begin{tablenotes}
\footnotesize
\item \textit{Notes:} Standard errors are in parentheses. $^{***}$ $p<0.01$, $^{**}$ $p<0.05$, $^{*}$ $p<0.10$. All models use $N=867$, include 2007--2022 year effects with 2006 omitted, and suppress intercepts, year indicators, and the symmetric-noise constant. The preferred column uses TETPB/TETCD. The delivered-total column uses $TNTCB/TPOPP$ with TETCD; direct primary uses $(TNTCB-ESTCB)/TPOPP$ with PETXD; retail electricity uses $ESTCB/TPOPP$ with ESTCD.
\end{tablenotes}}
\end{threeparttable}
\end{table}

\clearpage
\begin{table}[H]
\centering
\caption{Outcome-specific single-price sensitivity without the fossil-electricity-generation control}\label{tab:measurement_no_ff_matched}
\begin{threeparttable}
{\setstretch{1}\scriptsize
\setlength{\tabcolsep}{2pt}
\begin{tabularx}{\textwidth}{@{}>{\raggedright\arraybackslash}p{.24\textwidth}*{3}{Y}@{}}
\toprule
Variable & \shortstack{Delivered\\ total} & \shortstack{Direct\\ primary} & \shortstack{Retail\\ electricity} \\
\midrule
\multicolumn{4}{l}{\textit{Frontier equation}} \\[2pt]
Price & \shortstack[c]{$-0.7930^{***}$\\[-1pt]{\scriptsize (0.0536)}} & \shortstack[c]{$-0.8453^{***}$\\[-1pt]{\scriptsize (0.0662)}} & \shortstack[c]{$-0.4662^{***}$\\[-1pt]{\scriptsize (0.0452)}} \\
GDP/pc & \shortstack[c]{$0.0355$\\[-1pt]{\scriptsize (0.0357)}} & \shortstack[c]{$-0.1324^{***}$\\[-1pt]{\scriptsize (0.0312)}} & \shortstack[c]{$-0.0350$\\[-1pt]{\scriptsize (0.0347)}} \\
Eff & \shortstack[c]{$0.0015^{*}$\\[-1pt]{\scriptsize (0.0008)}} & \shortstack[c]{$0.0016^{**}$\\[-1pt]{\scriptsize (0.0008)}} & \shortstack[c]{$-0.0056^{***}$\\[-1pt]{\scriptsize (0.0010)}} \\
VMT/pc & \shortstack[c]{$0.4105^{***}$\\[-1pt]{\scriptsize (0.0344)}} & \shortstack[c]{$0.5001^{***}$\\[-1pt]{\scriptsize (0.0404)}} & \shortstack[c]{$0.2284^{***}$\\[-1pt]{\scriptsize (0.0339)}} \\
ComSurf/pc & \shortstack[c]{$-0.0563^{*}$\\[-1pt]{\scriptsize (0.0291)}} & \shortstack[c]{$-0.0532$\\[-1pt]{\scriptsize (0.0354)}} & \shortstack[c]{$-0.0332$\\[-1pt]{\scriptsize (0.0310)}} \\
ResSurf/pc & \shortstack[c]{$-0.2463^{***}$\\[-1pt]{\scriptsize (0.0598)}} & \shortstack[c]{$-0.5880^{***}$\\[-1pt]{\scriptsize (0.0658)}} & \shortstack[c]{$0.1604^{***}$\\[-1pt]{\scriptsize (0.0490)}} \\
E-Int Ind & \shortstack[c]{$0.0583^{***}$\\[-1pt]{\scriptsize (0.0118)}} & \shortstack[c]{$0.0830^{***}$\\[-1pt]{\scriptsize (0.0127)}} & \shortstack[c]{$0.0939^{***}$\\[-1pt]{\scriptsize (0.0091)}} \\
FF/pc & \shortstack[c]{$0.0419^{***}$\\[-1pt]{\scriptsize (0.0027)}} & \shortstack[c]{$0.0476^{***}$\\[-1pt]{\scriptsize (0.0027)}} & \shortstack[c]{$0.0322^{***}$\\[-1pt]{\scriptsize (0.0028)}} \\
FFelec/pc & \textemdash & \textemdash & \textemdash \\
HDD & \shortstack[c]{$0.0542^{***}$\\[-1pt]{\scriptsize (0.0050)}} & \shortstack[c]{$0.0734^{***}$\\[-1pt]{\scriptsize (0.0058)}} & \shortstack[c]{$-0.0013$\\[-1pt]{\scriptsize (0.0041)}} \\
CDD & \shortstack[c]{$0.0416^{***}$\\[-1pt]{\scriptsize (0.0111)}} & \shortstack[c]{$0.0166$\\[-1pt]{\scriptsize (0.0130)}} & \shortstack[c]{$0.0629^{***}$\\[-1pt]{\scriptsize (0.0097)}} \\
\midrule
\multicolumn{4}{l}{\textit{Gap log-variance equation}} \\[2pt]
Eff & \shortstack[c]{$-0.1368^{***}$\\[-1pt]{\scriptsize (0.0081)}} & \shortstack[c]{$-0.1208^{***}$\\[-1pt]{\scriptsize (0.0075)}} & \shortstack[c]{$-0.0453$\\[-1pt]{\scriptsize (0.0400)}} \\
Price & \shortstack[c]{$0.0161$\\[-1pt]{\scriptsize (0.2818)}} & \shortstack[c]{$-0.3648$\\[-1pt]{\scriptsize (0.2556)}} & \shortstack[c]{$-3.8991^{***}$\\[-1pt]{\scriptsize (1.0506)}} \\
GDP/pc & \shortstack[c]{$2.1052^{***}$\\[-1pt]{\scriptsize (0.3277)}} & \shortstack[c]{$2.4855^{***}$\\[-1pt]{\scriptsize (0.3488)}} & \shortstack[c]{$5.4439^{***}$\\[-1pt]{\scriptsize (0.6285)}} \\
\midrule
$N$ & 867 & 867 & 867 \\
Log likelihood & 411.17 & 281.16 & 689.00 \\
AIC & -758.34 & -498.32 & -1314.01 \\
BIC & -605.86 & -345.84 & -1161.53 \\
\bottomrule
\end{tabularx}
\begin{tablenotes}
\footnotesize
\item \textit{Notes:} Standard errors are in parentheses. $^{***}$ $p<0.01$, $^{**}$ $p<0.05$, $^{*}$ $p<0.10$. All models use $N=867$, include 2007--2022 year effects with 2006 omitted, and suppress intercepts, year indicators, and the symmetric-noise constant. These columns repeat the matched delivered-total, direct-primary, and retail-electricity constructs after omitting FFelec/pc.
\end{tablenotes}}
\end{threeparttable}
\end{table}

The same-vintage checks preserve the negative ACEEE gap sign but not the
frontier sign or magnitude. In the strict delivered-end-use/TETXD
specification, which omits the fossil-generation/loss proxy by design, the
ACEEE frontier coefficient is $+0.0015$ (SE $0.0008$, $p=.055$); in the
upstream-price/TETCB specification it is $-0.0041$ (SE $0.0009$, $p<.001$).
Their ACEEE gap coefficients are $-0.1368$ (SE $0.0081$) and $-0.1390$ (SE
$0.0098$), respectively.

\subsection{Sample sensitivity}\label{sec:subset_robust}

To assess whether the results are driven by specific groups of states or particular time periods, we re-estimate the baseline model on five restricted samples: (i) excluding the five largest fossil-fuel-producing states (Wyoming, North Dakota, West Virginia, Alaska, New Mexico), which are structural outliers in per-capita energy use; (ii)~excluding crisis years (2008--2009 financial crisis and 2020 COVID-19 pandemic), which may introduce transitory shocks; (iii) excluding the five most energy-intensive industrial states (Louisiana, Indiana, North Carolina, Iowa, Kentucky); (iv)~the early sub-period 2006--2013 only; and (v)~the recent sub-period 2014--2022 only. In each case, all demeaned variables are recomputed within the restricted sample. For the recent sub-period (2014--2022), which does not include the baseline reference year 2006, the first year (2014) serves as the reference for year dummies; results for this sub-period should therefore be interpreted with some caution.

Table~\ref{tab:subset_robust_summary} reports the comparisons. Four
restricted-sample fits pass the model-health criteria. The 2006--2013 fit
reaches a zero-noise boundary convergence problem and is reported as unavailable rather than
interpreted. In the four accepted samples, the price and ACEEE frontier
associations are negative; ACEEE is distinguishable at 5\% in the three
exclusion samples but not in 2014--2022. Its gap association remains negative,
from $-0.1968$ to $-0.1334$. The fossil-state exclusion changes the price gap
association from $0.3107$ to $-3.7295$; this does not identify a bias direction.

\begin{table}[H]
\centering
\caption{Subset robustness: key coefficient comparison}
\label{tab:subset_robust_summary}
\begin{threeparttable}
\small
\begin{tabular}{lcccccc}
\toprule
& Baseline & Excl.\ & Excl.\ & Excl.\ & Early & Recent \\
& & fossil & crisis & indust. & 06--13 & 14--22 \\
\midrule
\multicolumn{7}{l}{\textit{Frontier equation}} \\[2pt]
\textit{Price} & ${-}$0.7524 & ${-}$0.6612 & ${-}$0.7740 & ${-}$0.6804 & -- & ${-}$0.7399 \\
 & (0.0459) & (0.0455) & (0.0528) & (0.0409) & -- & (0.0737) \\
\textit{Eff} & ${-}$0.0017 & ${-}$0.0026 & ${-}$0.0016 & ${-}$0.0028 & -- & ${-}$0.0010 \\
 & (0.0007) & (0.0008) & (0.0007) & (0.0006) & -- & (0.0010) \\
\midrule
\multicolumn{7}{l}{\textit{Inefficiency equation} ($\ln\hat\sigma_{u}^2$)} \\[2pt]
\textit{Eff} & ${-}$0.1555 & ${-}$0.1334 & ${-}$0.1588 & ${-}$0.1670 & -- & ${-}$0.1968 \\
 & (0.0096) & (0.0179) & (0.0111) & (0.0103) & -- & (0.0201) \\
\textit{Price} & 0.3107 & ${-}$3.7295 & 0.4830 & 1.8586 & -- & ${-}$0.4841 \\
 & (0.3447) & (0.7611) & (0.4021) & (0.3986) & -- & (0.6585) \\
\midrule
$N$        & 867 & 782 & 714 & 782 & 408 & 459 \\
\bottomrule
\end{tabular}
\begin{tablenotes}
\footnotesize
\item \textit{Notes:} Standard errors are in parentheses. The early fit is unavailable because it reaches a zero-noise boundary and fails the convergence criteria. Fossil states are WY, ND, WV, AK, and NM; industrial states are LA, IN, NC, IA, and KY; crisis years are 2008--09 and 2020.
\end{tablenotes}
\end{threeparttable}
\end{table}

\clearpage
\begin{table}[H]
\centering
\caption{Subset robustness: complete substantive coefficient comparison}\label{tab:subset_robust}
\begin{threeparttable}
{\setstretch{1}\scriptsize
\setlength{\tabcolsep}{2pt}
\begin{tabularx}{\textwidth}{@{}>{\raggedright\arraybackslash}p{2.0cm}*{6}{Y}@{}}
\toprule
Variable & Baseline & \shortstack{Excl.\\fossil} & \shortstack{Excl.\\crisis} & \shortstack{Excl.\\industrial} & \shortstack{Early\\2006--13} & \shortstack{Recent\\2014--22} \\
\midrule
\multicolumn{7}{l}{\textit{Frontier equation}} \\[1pt]
Price & \shortstack[c]{$-0.7524^{***}$\\[-1pt]{\scriptsize (0.0459)}} & \shortstack[c]{$-0.6612^{***}$\\[-1pt]{\scriptsize (0.0455)}} & \shortstack[c]{$-0.7740^{***}$\\[-1pt]{\scriptsize (0.0528)}} & \shortstack[c]{$-0.6804^{***}$\\[-1pt]{\scriptsize (0.0409)}} & \textemdash & \shortstack[c]{$-0.7399^{***}$\\[-1pt]{\scriptsize (0.0737)}} \\
GDP/pc & \shortstack[c]{$0.1931^{***}$\\[-1pt]{\scriptsize (0.0322)}} & \shortstack[c]{$0.1783^{***}$\\[-1pt]{\scriptsize (0.0338)}} & \shortstack[c]{$0.1976^{***}$\\[-1pt]{\scriptsize (0.0356)}} & \shortstack[c]{$0.1644^{***}$\\[-1pt]{\scriptsize (0.0311)}} & \textemdash & \shortstack[c]{$0.0480$\\[-1pt]{\scriptsize (0.0449)}} \\
Eff & \shortstack[c]{$-0.0017^{***}$\\[-1pt]{\scriptsize (0.0007)}} & \shortstack[c]{$-0.0026^{***}$\\[-1pt]{\scriptsize (0.0008)}} & \shortstack[c]{$-0.0016^{**}$\\[-1pt]{\scriptsize (0.0007)}} & \shortstack[c]{$-0.0028^{***}$\\[-1pt]{\scriptsize (0.0006)}} & \textemdash & \shortstack[c]{$-0.0010$\\[-1pt]{\scriptsize (0.0010)}} \\
VMT/pc & \shortstack[c]{$0.3933^{***}$\\[-1pt]{\scriptsize (0.0337)}} & \shortstack[c]{$0.3084^{***}$\\[-1pt]{\scriptsize (0.0336)}} & \shortstack[c]{$0.3923^{***}$\\[-1pt]{\scriptsize (0.0372)}} & \shortstack[c]{$0.4314^{***}$\\[-1pt]{\scriptsize (0.0315)}} & \textemdash & \shortstack[c]{$0.3178^{***}$\\[-1pt]{\scriptsize (0.0415)}} \\
ComSurf/pc & \shortstack[c]{$-0.0136$\\[-1pt]{\scriptsize (0.0246)}} & \shortstack[c]{$-0.0360$\\[-1pt]{\scriptsize (0.0266)}} & \shortstack[c]{$-0.0301$\\[-1pt]{\scriptsize (0.0277)}} & \shortstack[c]{$-0.0022$\\[-1pt]{\scriptsize (0.0230)}} & \textemdash & \shortstack[c]{$-0.1053^{***}$\\[-1pt]{\scriptsize (0.0392)}} \\
ResSurf/pc & \shortstack[c]{$0.0141$\\[-1pt]{\scriptsize (0.0534)}} & \shortstack[c]{$0.0415$\\[-1pt]{\scriptsize (0.0505)}} & \shortstack[c]{$0.0005$\\[-1pt]{\scriptsize (0.0597)}} & \shortstack[c]{$-0.0245$\\[-1pt]{\scriptsize (0.0554)}} & \textemdash & \shortstack[c]{$-0.1503^{**}$\\[-1pt]{\scriptsize (0.0711)}} \\
E-Int Ind & \shortstack[c]{$0.1077^{***}$\\[-1pt]{\scriptsize (0.0099)}} & \shortstack[c]{$0.1618^{***}$\\[-1pt]{\scriptsize (0.0107)}} & \shortstack[c]{$0.1127^{***}$\\[-1pt]{\scriptsize (0.0109)}} & \shortstack[c]{$0.1099^{***}$\\[-1pt]{\scriptsize (0.0100)}} & \textemdash & \shortstack[c]{$0.1393^{***}$\\[-1pt]{\scriptsize (0.0137)}} \\
FF/pc & \shortstack[c]{$0.0364^{***}$\\[-1pt]{\scriptsize (0.0024)}} & \shortstack[c]{$0.2244^{***}$\\[-1pt]{\scriptsize (0.0169)}} & \shortstack[c]{$0.0364^{***}$\\[-1pt]{\scriptsize (0.0027)}} & \shortstack[c]{$0.0357^{***}$\\[-1pt]{\scriptsize (0.0022)}} & \textemdash & \shortstack[c]{$0.0539^{***}$\\[-1pt]{\scriptsize (0.0037)}} \\
FFelec/pc & \shortstack[c]{$0.0094^{***}$\\[-1pt]{\scriptsize (0.0031)}} & \shortstack[c]{$-0.0007$\\[-1pt]{\scriptsize (0.0037)}} & \shortstack[c]{$0.0098^{***}$\\[-1pt]{\scriptsize (0.0035)}} & \shortstack[c]{$0.0108^{***}$\\[-1pt]{\scriptsize (0.0031)}} & \textemdash & \shortstack[c]{$0.0227^{***}$\\[-1pt]{\scriptsize (0.0047)}} \\
HDD & \shortstack[c]{$0.0456^{***}$\\[-1pt]{\scriptsize (0.0044)}} & \shortstack[c]{$0.0461^{***}$\\[-1pt]{\scriptsize (0.0041)}} & \shortstack[c]{$0.0484^{***}$\\[-1pt]{\scriptsize (0.0049)}} & \shortstack[c]{$0.0390^{***}$\\[-1pt]{\scriptsize (0.0042)}} & \textemdash & \shortstack[c]{$0.0615^{***}$\\[-1pt]{\scriptsize (0.0060)}} \\
CDD & \shortstack[c]{$0.1007^{***}$\\[-1pt]{\scriptsize (0.0090)}} & \shortstack[c]{$0.1410^{***}$\\[-1pt]{\scriptsize (0.0094)}} & \shortstack[c]{$0.1076^{***}$\\[-1pt]{\scriptsize (0.0101)}} & \shortstack[c]{$0.0766^{***}$\\[-1pt]{\scriptsize (0.0083)}} & \textemdash & \shortstack[c]{$0.1193^{***}$\\[-1pt]{\scriptsize (0.0119)}} \\
\midrule
\multicolumn{7}{l}{\textit{Gap log-variance equation}} \\[1pt]
Eff & \shortstack[c]{$-0.1555^{***}$\\[-1pt]{\scriptsize (0.0096)}} & \shortstack[c]{$-0.1334^{***}$\\[-1pt]{\scriptsize (0.0179)}} & \shortstack[c]{$-0.1588^{***}$\\[-1pt]{\scriptsize (0.0111)}} & \shortstack[c]{$-0.1670^{***}$\\[-1pt]{\scriptsize (0.0103)}} & \textemdash & \shortstack[c]{$-0.1968^{***}$\\[-1pt]{\scriptsize (0.0201)}} \\
Price & \shortstack[c]{$0.3107$\\[-1pt]{\scriptsize (0.3447)}} & \shortstack[c]{$-3.7295^{***}$\\[-1pt]{\scriptsize (0.7611)}} & \shortstack[c]{$0.4830$\\[-1pt]{\scriptsize (0.4021)}} & \shortstack[c]{$1.8586^{***}$\\[-1pt]{\scriptsize (0.3986)}} & \textemdash & \shortstack[c]{$-0.4841$\\[-1pt]{\scriptsize (0.6585)}} \\
GDP/pc & \shortstack[c]{$2.6264^{***}$\\[-1pt]{\scriptsize (0.2873)}} & \shortstack[c]{$4.4394^{***}$\\[-1pt]{\scriptsize (0.4630)}} & \shortstack[c]{$2.6499^{***}$\\[-1pt]{\scriptsize (0.3254)}} & \shortstack[c]{$3.0465^{***}$\\[-1pt]{\scriptsize (0.3254)}} & \textemdash & \shortstack[c]{$4.4733^{***}$\\[-1pt]{\scriptsize (0.6139)}} \\
\midrule
$N$ & 867 & 782 & 714 & 782 & 408 & 459 \\
\bottomrule
\end{tabularx}
\begin{tablenotes}
\footnotesize
\item \textit{Notes:} Standard errors are in parentheses. $^{***}$ $p<0.01$, $^{**}$ $p<0.05$, $^{*}$ $p<0.10$. Intercepts, year indicators, and the symmetric-noise constant are estimated but suppressed. Fossil states are WY, ND, WV, AK, and NM; industrial states are LA, IN, NC, IA, and KY; crisis years are 2008--09 and 2020. Every regressor is re-centered within its restricted sample. The omitted year is 2006 except in the recent sample, where it is 2014.
\end{tablenotes}}
\end{threeparttable}
\end{table}

\subsection{ACEEE timing sensitivity}

The common 2009--2022 sample keeps the same 714 state--year observations at
each timing. All four contemporaneous-through-three-year-lag frontier
coefficients are negative and conventionally distinguishable at 5\%. The gap
association remains negative, from $-0.1643$ to $-0.1566$. These are timing
sensitivities, not an endogeneity remedy.

\begin{table}[H]
\centering
\caption{Common-sample ACEEE timing}\label{tab:aceee_timing}
\small
\begin{tabular}{lrrrr}
\toprule
Timing & Frontier & SE & Gap log variance & SE\\
\midrule
Current & $-0.0019$ & 0.0008 & $-0.1643$ & 0.0115\\
Lag 1 & $-0.0021$ & 0.0008 & $-0.1598$ & 0.0112\\
Lag 2 & $-0.0021$ & 0.0008 & $-0.1570$ & 0.0112\\
Lag 3 & $-0.0019$ & 0.0008 & $-0.1566$ & 0.0112\\
\bottomrule
\end{tabular}
\end{table}

\subsection{ACEEE composition}

The aggregate score remains primary. A two-block UPB/Other model rejects a
common score slope, but the slopes are conditional composition evidence, not
additive mechanisms or feasible score reallocations. In the frontier, UPB is
$0.0050$ (SE 0.0012) and Other is $-0.0080$ (SE 0.0012); in gap log variance,
UPB is $-0.2210$ (SE 0.0206) and Other is $-0.0866$ (SE 0.0188). The aggregate
restriction is rejected ($\chi^2(2)=39.12$, $p<.001$).

\begin{table}[H]
\centering
\caption{Exploratory five-domain ACEEE sensitivity}\label{tab:aceee_composition}
\scriptsize
\resizebox{\linewidth}{!}{%
\begin{tabular}{llrrrrr}
\toprule
Equation & Domain & Estimate & SE & Raw $p$ & Holm $p$ & BH $p$\\
\midrule
Frontier & UPB & 0.0050 & 0.0011 & $1.066\mathrm{e}{-05}$ & $9.592\mathrm{e}{-05}$ & $5.329\mathrm{e}{-05}$\\
Frontier & Transportation & $-0.0104$ & 0.0024 & $1.702\mathrm{e}{-05}$ & 0.0001 & $5.673\mathrm{e}{-05}$\\
Frontier & Building & $-0.0097$ & 0.0027 & 0.0004 & 0.0026 & 0.0009\\
Frontier & State initiatives & $-0.0083$ & 0.0035 & 0.0187 & 0.0747 & 0.0267\\
Frontier & Appliances & 0.0085 & 0.0057 & 0.1354 & 0.4061 & 0.1692\\
Gap log variance & UPB & $-0.2229$ & 0.0200 & $6.976\mathrm{e}{-29}$ & $6.976\mathrm{e}{-28}$ & $6.976\mathrm{e}{-28}$\\
Gap log variance & Transportation & $-0.1189$ & 0.0366 & 0.0012 & 0.0070 & 0.0023\\
Gap log variance & Building & $-0.0192$ & 0.0381 & 0.6138 & 1.0000 & 0.6820\\
Gap log variance & State initiatives & $-0.0178$ & 0.0525 & 0.7349 & 1.0000 & 0.7349\\
Gap log variance & Appliances & $-0.6366$ & 0.2002 & 0.0015 & 0.0073 & 0.0024\\
\bottomrule
\end{tabular}
}
\end{table}

The five-domain model \rev{(Table~\ref{tab:aceee_composition})} improves on the two-block restriction
($\chi^2(6)=20.65$, $p=.0021$), so the four non-UPB domain slopes should not be
described as adequately pooled. Under Holm correction across the ten cells,
the UPB, Transportation, and Building frontier rows; the UPB and Transportation
gap rows; and the Appliances gap row remain distinguishable. The domain
rubrics, outcome-bearing content, and reference periods change across ACEEE
editions.

\subsection{Complete ACEEE timing and composition results}

The tables below report the complete details for every robustness model.

\subsubsection{Primary-equivalent total energy}

The timing models use the common 2009--2022 support required to observe all score timings. The composition model instead uses the full 2006--2022 panel and replaces the aggregate score with its five signed, jointly entered domains.

\clearpage
\begin{table}[H]
\centering
\caption{Complete primary-equivalent SFA across ACEEE timing and five-domain specifications}\label{tab:sfa_total_timing_composition}\label{tab:sfa_total_robustness}\label{tab:sfa_total_lag1}\label{tab:sfa_total_lag2}\label{tab:sfa_total_lag3}\label{tab:sfa_total_domains}
\begin{threeparttable}
{\setstretch{1}\scriptsize
\setlength{\tabcolsep}{2pt}
\begin{tabularx}{\textwidth}{@{}>{\raggedright\arraybackslash}p{.28\textwidth}*{4}{Y}@{}}
\toprule
Variable & Lag 1 & Lag 2 & Lag 3 & Five domains \\
\midrule
\multicolumn{5}{l}{\textit{Frontier equation}} \\[2pt]
Price & \shortstack[c]{${-}0.7180^{***}$\\[-1pt]{\scriptsize (0.0564)}} & \shortstack[c]{${-}0.7094^{***}$\\[-1pt]{\scriptsize (0.0579)}} & \shortstack[c]{${-}0.7087^{***}$\\[-1pt]{\scriptsize (0.0568)}} & \shortstack[c]{${-}0.8170^{***}$\\[-1pt]{\scriptsize (0.0436)}} \\
GDP/pc & \shortstack[c]{$0.1750^{***}$\\[-1pt]{\scriptsize (0.0346)}} & \shortstack[c]{$0.1700^{***}$\\[-1pt]{\scriptsize (0.0349)}} & \shortstack[c]{$0.1669^{***}$\\[-1pt]{\scriptsize (0.0348)}} & \shortstack[c]{$0.2935^{***}$\\[-1pt]{\scriptsize (0.0346)}} \\
Eff: aggregate score & \shortstack[c]{${-}0.0021^{***}$\\[-1pt]{\scriptsize (0.0008)}} & \shortstack[c]{${-}0.0021^{***}$\\[-1pt]{\scriptsize (0.0008)}} & \shortstack[c]{${-}0.0019^{**}$\\[-1pt]{\scriptsize (0.0008)}} & \textemdash \\
Eff: UPB & \textemdash & \textemdash & \textemdash & \shortstack[c]{$0.0050^{***}$\\[-1pt]{\scriptsize (0.0011)}} \\
Eff: Transportation & \textemdash & \textemdash & \textemdash & \shortstack[c]{${-}0.0104^{***}$\\[-1pt]{\scriptsize (0.0024)}} \\
Eff: Building & \textemdash & \textemdash & \textemdash & \shortstack[c]{${-}0.0097^{***}$\\[-1pt]{\scriptsize (0.0027)}} \\
Eff: State initiatives & \textemdash & \textemdash & \textemdash & \shortstack[c]{${-}0.0083^{**}$\\[-1pt]{\scriptsize (0.0035)}} \\
Eff: Appliances & \textemdash & \textemdash & \textemdash & \shortstack[c]{$0.0085$\\[-1pt]{\scriptsize (0.0057)}} \\
VMT/pc & \shortstack[c]{$0.3825^{***}$\\[-1pt]{\scriptsize (0.0360)}} & \shortstack[c]{$0.3820^{***}$\\[-1pt]{\scriptsize (0.0361)}} & \shortstack[c]{$0.3805^{***}$\\[-1pt]{\scriptsize (0.0361)}} & \shortstack[c]{$0.3897^{***}$\\[-1pt]{\scriptsize (0.0326)}} \\
ComSurf/pc & \shortstack[c]{${-}0.0269$\\[-1pt]{\scriptsize (0.0270)}} & \shortstack[c]{${-}0.0262$\\[-1pt]{\scriptsize (0.0272)}} & \shortstack[c]{${-}0.0256$\\[-1pt]{\scriptsize (0.0272)}} & \shortstack[c]{${-}0.0158$\\[-1pt]{\scriptsize (0.0235)}} \\
ResSurf/pc & \shortstack[c]{${-}0.0233$\\[-1pt]{\scriptsize (0.0617)}} & \shortstack[c]{${-}0.0339$\\[-1pt]{\scriptsize (0.0627)}} & \shortstack[c]{${-}0.0286$\\[-1pt]{\scriptsize (0.0599)}} & \shortstack[c]{$0.0361$\\[-1pt]{\scriptsize (0.0500)}} \\
E-Int Ind & \shortstack[c]{$0.1085^{***}$\\[-1pt]{\scriptsize (0.0111)}} & \shortstack[c]{$0.1097^{***}$\\[-1pt]{\scriptsize (0.0113)}} & \shortstack[c]{$0.1121^{***}$\\[-1pt]{\scriptsize (0.0114)}} & \shortstack[c]{$0.0917^{***}$\\[-1pt]{\scriptsize (0.0096)}} \\
FF/pc & \shortstack[c]{$0.0424^{***}$\\[-1pt]{\scriptsize (0.0028)}} & \shortstack[c]{$0.0427^{***}$\\[-1pt]{\scriptsize (0.0028)}} & \shortstack[c]{$0.0432^{***}$\\[-1pt]{\scriptsize (0.0028)}} & \shortstack[c]{$0.0338^{***}$\\[-1pt]{\scriptsize (0.0022)}} \\
FFelec/pc & \shortstack[c]{$0.0116^{***}$\\[-1pt]{\scriptsize (0.0035)}} & \shortstack[c]{$0.0116^{***}$\\[-1pt]{\scriptsize (0.0035)}} & \shortstack[c]{$0.0115^{***}$\\[-1pt]{\scriptsize (0.0035)}} & \shortstack[c]{$0.0056^{*}$\\[-1pt]{\scriptsize (0.0031)}} \\
HDD & \shortstack[c]{$0.0483^{***}$\\[-1pt]{\scriptsize (0.0050)}} & \shortstack[c]{$0.0489^{***}$\\[-1pt]{\scriptsize (0.0051)}} & \shortstack[c]{$0.0490^{***}$\\[-1pt]{\scriptsize (0.0050)}} & \shortstack[c]{$0.0335^{***}$\\[-1pt]{\scriptsize (0.0051)}} \\
CDD & \shortstack[c]{$0.1010^{***}$\\[-1pt]{\scriptsize (0.0102)}} & \shortstack[c]{$0.1009^{***}$\\[-1pt]{\scriptsize (0.0104)}} & \shortstack[c]{$0.1019^{***}$\\[-1pt]{\scriptsize (0.0104)}} & \shortstack[c]{$0.0915^{***}$\\[-1pt]{\scriptsize (0.0097)}} \\
\midrule
\multicolumn{5}{l}{\textit{Gap log-variance equation}} \\[2pt]
Eff: aggregate score & \shortstack[c]{${-}0.1598^{***}$\\[-1pt]{\scriptsize (0.0112)}} & \shortstack[c]{${-}0.1570^{***}$\\[-1pt]{\scriptsize (0.0112)}} & \shortstack[c]{${-}0.1566^{***}$\\[-1pt]{\scriptsize (0.0112)}} & \textemdash \\
Eff: UPB & \textemdash & \textemdash & \textemdash & \shortstack[c]{${-}0.2229^{***}$\\[-1pt]{\scriptsize (0.0200)}} \\
Eff: Transportation & \textemdash & \textemdash & \textemdash & \shortstack[c]{${-}0.1189^{***}$\\[-1pt]{\scriptsize (0.0366)}} \\
Eff: Building & \textemdash & \textemdash & \textemdash & \shortstack[c]{${-}0.0192$\\[-1pt]{\scriptsize (0.0381)}} \\
Eff: State initiatives & \textemdash & \textemdash & \textemdash & \shortstack[c]{${-}0.0178$\\[-1pt]{\scriptsize (0.0525)}} \\
Eff: Appliances & \textemdash & \textemdash & \textemdash & \shortstack[c]{${-}0.6366^{***}$\\[-1pt]{\scriptsize (0.2002)}} \\
Price & \shortstack[c]{$0.0741$\\[-1pt]{\scriptsize (0.4183)}} & \shortstack[c]{${-}0.0966$\\[-1pt]{\scriptsize (0.4249)}} & \shortstack[c]{${-}0.2873$\\[-1pt]{\scriptsize (0.3957)}} & \shortstack[c]{$0.6879^{**}$\\[-1pt]{\scriptsize (0.3262)}} \\
GDP/pc & \shortstack[c]{$2.7492^{***}$\\[-1pt]{\scriptsize (0.3381)}} & \shortstack[c]{$2.6813^{***}$\\[-1pt]{\scriptsize (0.3440)}} & \shortstack[c]{$2.6187^{***}$\\[-1pt]{\scriptsize (0.3333)}} & \shortstack[c]{$1.9705^{***}$\\[-1pt]{\scriptsize (0.3532)}} \\
\midrule
$N$ & 714 & 714 & 714 & 867 \\
Sample years & 2009--22 & 2009--22 & 2009--22 & 2006--22 \\
Omitted year & 2009 & 2009 & 2009 & 2006 \\
Parameters & 30 & 30 & 30 & 41 \\
Log likelihood & 504.14 & 496.76 & 488.58 & 618.10 \\
AIC & -948.29 & -933.52 & -917.17 & -1154.21 \\
BIC & -811.16 & -796.40 & -780.04 & -958.84 \\
\bottomrule
\end{tabularx}
\begin{tablenotes}
\footnotesize
\item \textit{Notes:} Outcome: log primary-equivalent total energy per capita; price: centered log aggregate end-use price. Lag columns use the common 2009--2022 sample and enter the signed aggregate ACEEE score at the indicated lag. The five-domain column uses the full 2006--2022 sample and jointly enters UPB, transportation, building, state-initiative, and appliance scores on fixed 2020 point scales (20/12/9/6/3). Standard errors are in parentheses. $^{***}$ $p<0.01$, $^{**}$ $p<0.05$, $^{*}$ $p<0.10$. All manuscript controls and year effects are retained.
\end{tablenotes}}
\end{threeparttable}
\end{table}

\subsubsection{Retail electricity}

Outcome: $\log(ESTCB/TPOPP)$; price: centered $\log(ESTCD)$, the retail electricity price per million Btu.

\begin{table}[H]
\centering
\caption{Complete retail electricity SFA: current ACEEE score}\label{tab:sfa_electricity_current}
\begin{threeparttable}
{\setstretch{1}\scriptsize
\setlength{\tabcolsep}{3pt}
\begin{tabularx}{\textwidth}{@{}>{\raggedright\arraybackslash}p{.46\textwidth}*{2}{Y}@{}}
\toprule
Variable & Estimate & SE \\
\midrule
\multicolumn{3}{l}{\textit{Frontier equation}} \\[2pt]
Retail electricity price (ESTCD) & ${-}0.4625^{***}$ & (0.0405) \\
GDP/pc & ${-}0.0362$ & (0.0343) \\
VMT/pc & $0.2192^{***}$ & (0.0347) \\
ComSurf/pc & ${-}0.0296$ & (0.0305) \\
ResSurf/pc & $0.1647^{***}$ & (0.0490) \\
E-Int Ind & $0.0947^{***}$ & (0.0091) \\
CDD & $0.0622^{***}$ & (0.0097) \\
HDD & ${-}0.0023$ & (0.0042) \\
Eff & ${-}0.0060^{***}$ & (0.0010) \\
FF/pc & $0.0332^{***}$ & (0.0029) \\
FFelec/pc & ${-}0.0042$ & (0.0041) \\
\midrule
\multicolumn{3}{l}{\textit{Gap log-variance equation}} \\[2pt]
Eff & ${-}0.0393$ & (0.0338) \\
Retail electricity price (ESTCD) & ${-}3.9378^{***}$ & (0.9131) \\
GDP/pc & $5.2862^{***}$ & (0.6130) \\
\bottomrule
\end{tabularx}
\begin{tablenotes}
\footnotesize
\item \textit{Notes:} Outcome: $\log(ESTCB/TPOPP)$; price: centered $\log(ESTCD)$, the retail electricity price per million Btu. Full sample: $N=867$, 51 jurisdictions, 2006--2022. Standard errors are shown. $^{***}$ $p<0.01$, $^{**}$ $p<0.05$, $^{*}$ $p<0.10$.
\end{tablenotes}}
\end{threeparttable}
\end{table}

\clearpage
\begin{table}[H]
\centering
\caption{Complete retail electricity SFA across ACEEE timing and five-domain specifications}\label{tab:sfa_electricity_timing_composition}\label{tab:sfa_electricity_lag1}\label{tab:sfa_electricity_lag2}\label{tab:sfa_electricity_lag3}\label{tab:sfa_electricity_domains}
\begin{threeparttable}
{\setstretch{1}\scriptsize
\setlength{\tabcolsep}{2pt}
\begin{tabularx}{\textwidth}{@{}>{\raggedright\arraybackslash}p{.28\textwidth}*{4}{Y}@{}}
\toprule
Variable & Lag 1 & Lag 2 & Lag 3 & Five domains \\
\midrule
\multicolumn{5}{l}{\textit{Frontier equation}} \\[2pt]
Price & \shortstack[c]{${-}0.4996^{***}$\\[-1pt]{\scriptsize (0.0344)}} & \shortstack[c]{${-}0.4898^{***}$\\[-1pt]{\scriptsize (0.0332)}} & \shortstack[c]{${-}0.4816^{***}$\\[-1pt]{\scriptsize (0.0358)}} & \shortstack[c]{${-}0.4264^{***}$\\[-1pt]{\scriptsize (0.0358)}} \\
GDP/pc & \shortstack[c]{${-}0.0140$\\[-1pt]{\scriptsize (0.0346)}} & \shortstack[c]{${-}0.0215$\\[-1pt]{\scriptsize (0.0346)}} & \shortstack[c]{${-}0.0280$\\[-1pt]{\scriptsize (0.0349)}} & \shortstack[c]{${-}0.0814^{**}$\\[-1pt]{\scriptsize (0.0375)}} \\
Eff: aggregate score & \shortstack[c]{${-}0.0063^{***}$\\[-1pt]{\scriptsize (0.0008)}} & \shortstack[c]{${-}0.0061^{***}$\\[-1pt]{\scriptsize (0.0008)}} & \shortstack[c]{${-}0.0060^{***}$\\[-1pt]{\scriptsize (0.0009)}} & \textemdash \\
Eff: UPB & \textemdash & \textemdash & \textemdash & \shortstack[c]{${-}0.0101^{***}$\\[-1pt]{\scriptsize (0.0017)}} \\
Eff: Transportation & \textemdash & \textemdash & \textemdash & \shortstack[c]{${-}0.0090^{***}$\\[-1pt]{\scriptsize (0.0030)}} \\
Eff: Building & \textemdash & \textemdash & \textemdash & \shortstack[c]{$0.0015$\\[-1pt]{\scriptsize (0.0034)}} \\
Eff: State initiatives & \textemdash & \textemdash & \textemdash & \shortstack[c]{$0.0081$\\[-1pt]{\scriptsize (0.0052)}} \\
Eff: Appliances & \textemdash & \textemdash & \textemdash & \shortstack[c]{${-}0.0180^{*}$\\[-1pt]{\scriptsize (0.0106)}} \\
VMT/pc & \shortstack[c]{$0.2203^{***}$\\[-1pt]{\scriptsize (0.0358)}} & \shortstack[c]{$0.2301^{***}$\\[-1pt]{\scriptsize (0.0360)}} & \shortstack[c]{$0.2359^{***}$\\[-1pt]{\scriptsize (0.0364)}} & \shortstack[c]{$0.2176^{***}$\\[-1pt]{\scriptsize (0.0342)}} \\
ComSurf/pc & \shortstack[c]{${-}0.0431$\\[-1pt]{\scriptsize (0.0307)}} & \shortstack[c]{${-}0.0380$\\[-1pt]{\scriptsize (0.0307)}} & \shortstack[c]{${-}0.0341$\\[-1pt]{\scriptsize (0.0311)}} & \shortstack[c]{${-}0.0184$\\[-1pt]{\scriptsize (0.0308)}} \\
ResSurf/pc & \shortstack[c]{$0.1626^{***}$\\[-1pt]{\scriptsize (0.0519)}} & \shortstack[c]{$0.1604^{***}$\\[-1pt]{\scriptsize (0.0522)}} & \shortstack[c]{$0.1624^{***}$\\[-1pt]{\scriptsize (0.0525)}} & \shortstack[c]{$0.1605^{***}$\\[-1pt]{\scriptsize (0.0515)}} \\
E-Int Ind & \shortstack[c]{$0.0882^{***}$\\[-1pt]{\scriptsize (0.0096)}} & \shortstack[c]{$0.0900^{***}$\\[-1pt]{\scriptsize (0.0098)}} & \shortstack[c]{$0.0925^{***}$\\[-1pt]{\scriptsize (0.0099)}} & \shortstack[c]{$0.1035^{***}$\\[-1pt]{\scriptsize (0.0091)}} \\
FF/pc & \shortstack[c]{$0.0357^{***}$\\[-1pt]{\scriptsize (0.0032)}} & \shortstack[c]{$0.0364^{***}$\\[-1pt]{\scriptsize (0.0032)}} & \shortstack[c]{$0.0372^{***}$\\[-1pt]{\scriptsize (0.0032)}} & \shortstack[c]{$0.0364^{***}$\\[-1pt]{\scriptsize (0.0030)}} \\
FFelec/pc & \shortstack[c]{${-}0.0044$\\[-1pt]{\scriptsize (0.0043)}} & \shortstack[c]{${-}0.0037$\\[-1pt]{\scriptsize (0.0044)}} & \shortstack[c]{${-}0.0032$\\[-1pt]{\scriptsize (0.0045)}} & \shortstack[c]{${-}0.0067^{*}$\\[-1pt]{\scriptsize (0.0041)}} \\
HDD & \shortstack[c]{${-}0.0001$\\[-1pt]{\scriptsize (0.0045)}} & \shortstack[c]{$0.0013$\\[-1pt]{\scriptsize (0.0046)}} & \shortstack[c]{$0.0022$\\[-1pt]{\scriptsize (0.0046)}} & \shortstack[c]{${-}0.0002$\\[-1pt]{\scriptsize (0.0048)}} \\
CDD & \shortstack[c]{$0.0606^{***}$\\[-1pt]{\scriptsize (0.0104)}} & \shortstack[c]{$0.0641^{***}$\\[-1pt]{\scriptsize (0.0105)}} & \shortstack[c]{$0.0670^{***}$\\[-1pt]{\scriptsize (0.0106)}} & \shortstack[c]{$0.0673^{***}$\\[-1pt]{\scriptsize (0.0106)}} \\
\midrule
\multicolumn{5}{l}{\textit{Gap log-variance equation}} \\[2pt]
Eff: aggregate score & \shortstack[c]{${-}0.0667^{*}$\\[-1pt]{\scriptsize (0.0364)}} & \shortstack[c]{${-}0.0470$\\[-1pt]{\scriptsize (0.0307)}} & \shortstack[c]{${-}0.0329$\\[-1pt]{\scriptsize (0.0314)}} & \textemdash \\
Eff: UPB & \textemdash & \textemdash & \textemdash & \shortstack[c]{$0.0469$\\[-1pt]{\scriptsize (0.0524)}} \\
Eff: Transportation & \textemdash & \textemdash & \textemdash & \shortstack[c]{$0.0405$\\[-1pt]{\scriptsize (0.0795)}} \\
Eff: Building & \textemdash & \textemdash & \textemdash & \shortstack[c]{${-}0.1425$\\[-1pt]{\scriptsize (0.1004)}} \\
Eff: State initiatives & \textemdash & \textemdash & \textemdash & \shortstack[c]{${-}0.4093^{***}$\\[-1pt]{\scriptsize (0.1401)}} \\
Eff: Appliances & \textemdash & \textemdash & \textemdash & \shortstack[c]{$0.6059^{**}$\\[-1pt]{\scriptsize (0.2509)}} \\
Price & \shortstack[c]{${-}6.7737^{***}$\\[-1pt]{\scriptsize (1.6517)}} & \shortstack[c]{${-}6.6455^{***}$\\[-1pt]{\scriptsize (1.4619)}} & \shortstack[c]{${-}6.4812^{***}$\\[-1pt]{\scriptsize (1.3615)}} & \shortstack[c]{${-}3.7240^{***}$\\[-1pt]{\scriptsize (0.7693)}} \\
GDP/pc & \shortstack[c]{$6.6886^{***}$\\[-1pt]{\scriptsize (0.9304)}} & \shortstack[c]{$6.2642^{***}$\\[-1pt]{\scriptsize (0.8484)}} & \shortstack[c]{$5.9962^{***}$\\[-1pt]{\scriptsize (0.8235)}} & \shortstack[c]{$5.2426^{***}$\\[-1pt]{\scriptsize (0.5665)}} \\
\midrule
$N$ & 714 & 714 & 714 & 867 \\
Sample years & 2009--22 & 2009--22 & 2009--22 & 2006--22 \\
Omitted year & 2009 & 2009 & 2009 & 2006 \\
Parameters & 30 & 30 & 30 & 41 \\
Log likelihood & 596.97 & 589.55 & 583.73 & 704.34 \\
AIC & -1133.95 & -1119.11 & -1107.46 & -1326.68 \\
BIC & -996.82 & -981.98 & -970.34 & -1131.31 \\
\bottomrule
\end{tabularx}
\begin{tablenotes}
\footnotesize
\item \textit{Notes:} Outcome: $\log(ESTCB/TPOPP)$; price: centered $\log(ESTCD)$, the retail electricity price per million Btu. Lag columns use the common 2009--2022 sample and enter the signed aggregate ACEEE score at the indicated lag. The five-domain column uses the full 2006--2022 sample and jointly enters UPB, transportation, building, state-initiative, and appliance scores on fixed 2020 point scales (20/12/9/6/3). Standard errors are in parentheses. $^{***}$ $p<0.01$, $^{**}$ $p<0.05$, $^{*}$ $p<0.10$. All manuscript controls and year effects are retained.
\end{tablenotes}}
\end{threeparttable}
\end{table}

\subsubsection{Direct primary energy excluding retail electricity}

The outcome is log direct primary energy per resident:
\[\log((TNTCB-ESTCB)/TPOPP).\]
The price regressor is centered $\log(PETXD)$. The physical outcome and price are conceptually aligned but not denominator-identical.

\begin{table}[H]
\centering
\caption{Complete direct primary energy excluding retail electricity SFA: current ACEEE score}\label{tab:sfa_primary_current}
\begin{threeparttable}
{\setstretch{1}\scriptsize
\setlength{\tabcolsep}{3pt}
\begin{tabularx}{\textwidth}{@{}>{\raggedright\arraybackslash}p{.46\textwidth}*{2}{Y}@{}}
\toprule
Variable & Estimate & SE \\
\midrule
\multicolumn{3}{l}{\textit{Frontier equation}} \\[2pt]
Direct-primary price (PETXD) & ${-}0.8186^{***}$ & (0.0671) \\
GDP/pc & ${-}0.1255^{***}$ & (0.0308) \\
VMT/pc & $0.5453^{***}$ & (0.0479) \\
ComSurf/pc & ${-}0.0529$ & (0.0349) \\
ResSurf/pc & ${-}0.5937^{***}$ & (0.0650) \\
E-Int Ind & $0.0704^{***}$ & (0.0150) \\
CDD & $0.0130$ & (0.0130) \\
HDD & $0.0749^{***}$ & (0.0057) \\
Eff & $0.0019^{**}$ & (0.0008) \\
FF/pc & $0.0455^{***}$ & (0.0030) \\
FFelec/pc & $0.0065^{*}$ & (0.0038) \\
\midrule
\multicolumn{3}{l}{\textit{Gap log-variance equation}} \\[2pt]
Eff & ${-}0.1206^{***}$ & (0.0074) \\
Direct-primary price (PETXD) & ${-}0.4179^{*}$ & (0.2532) \\
GDP/pc & $2.4888^{***}$ & (0.3482) \\
\bottomrule
\end{tabularx}
\begin{tablenotes}
\footnotesize
\item \textit{Notes:} Outcome: $\log((TNTCB-ESTCB)/TPOPP)$; price: centered $\log(PETXD)$. The constructs are conceptually aligned but not denominator-identical. Full sample: $N=867$, 51 jurisdictions, 2006--2022. Year effects are included with 2006 omitted. Standard errors are shown. $^{***}$ $p<0.01$, $^{**}$ $p<0.05$, $^{*}$ $p<0.10$. 
\end{tablenotes}}
\end{threeparttable}
\end{table}

\clearpage
\begin{table}[H]
\centering
\caption{Complete direct-primary-energy SFA across ACEEE timing and five-domain specifications}\label{tab:sfa_primary_timing_composition}\label{tab:sfa_primary_lag1}\label{tab:sfa_primary_lag2}\label{tab:sfa_primary_lag3}\label{tab:sfa_primary_domains}
\begin{threeparttable}
{\setstretch{1}\scriptsize
\setlength{\tabcolsep}{2pt}
\begin{tabularx}{\textwidth}{@{}>{\raggedright\arraybackslash}p{.28\textwidth}*{4}{Y}@{}}
\toprule
Variable & Lag 1 & Lag 2 & Lag 3 & Five domains \\
\midrule
\multicolumn{5}{l}{\textit{Frontier equation}} \\[2pt]
Price & \shortstack[c]{${-}0.7462^{***}$\\[-1pt]{\scriptsize (0.0736)}} & \shortstack[c]{${-}0.7539^{***}$\\[-1pt]{\scriptsize (0.0751)}} & \shortstack[c]{${-}0.7634^{***}$\\[-1pt]{\scriptsize (0.0765)}} & \shortstack[c]{${-}0.8103^{***}$\\[-1pt]{\scriptsize (0.0642)}} \\
GDP/pc & \shortstack[c]{${-}0.1571^{***}$\\[-1pt]{\scriptsize (0.0313)}} & \shortstack[c]{${-}0.1631^{***}$\\[-1pt]{\scriptsize (0.0314)}} & \shortstack[c]{${-}0.1633^{***}$\\[-1pt]{\scriptsize (0.0316)}} & \shortstack[c]{${-}0.0816^{***}$\\[-1pt]{\scriptsize (0.0294)}} \\
Eff: aggregate score & \shortstack[c]{$0.0021^{***}$\\[-1pt]{\scriptsize (0.0008)}} & \shortstack[c]{$0.0019^{**}$\\[-1pt]{\scriptsize (0.0008)}} & \shortstack[c]{$0.0019^{**}$\\[-1pt]{\scriptsize (0.0008)}} & \textemdash \\
Eff: UPB & \textemdash & \textemdash & \textemdash & \shortstack[c]{$0.0033^{**}$\\[-1pt]{\scriptsize (0.0015)}} \\
Eff: Transportation & \textemdash & \textemdash & \textemdash & \shortstack[c]{$0.0015$\\[-1pt]{\scriptsize (0.0035)}} \\
Eff: Building & \textemdash & \textemdash & \textemdash & \shortstack[c]{${-}0.0017$\\[-1pt]{\scriptsize (0.0039)}} \\
Eff: State initiatives & \textemdash & \textemdash & \textemdash & \shortstack[c]{${-}0.0186^{***}$\\[-1pt]{\scriptsize (0.0060)}} \\
Eff: Appliances & \textemdash & \textemdash & \textemdash & \shortstack[c]{$0.0433^{***}$\\[-1pt]{\scriptsize (0.0083)}} \\
VMT/pc & \shortstack[c]{$0.5339^{***}$\\[-1pt]{\scriptsize (0.0468)}} & \shortstack[c]{$0.5267^{***}$\\[-1pt]{\scriptsize (0.0468)}} & \shortstack[c]{$0.5258^{***}$\\[-1pt]{\scriptsize (0.0475)}} & \shortstack[c]{$0.5055^{***}$\\[-1pt]{\scriptsize (0.0497)}} \\
ComSurf/pc & \shortstack[c]{$0.0003$\\[-1pt]{\scriptsize (0.0422)}} & \shortstack[c]{${-}0.0008$\\[-1pt]{\scriptsize (0.0407)}} & \shortstack[c]{${-}0.0034$\\[-1pt]{\scriptsize (0.0403)}} & \shortstack[c]{${-}0.1149^{***}$\\[-1pt]{\scriptsize (0.0371)}} \\
ResSurf/pc & \shortstack[c]{${-}0.5907^{***}$\\[-1pt]{\scriptsize (0.0672)}} & \shortstack[c]{${-}0.5960^{***}$\\[-1pt]{\scriptsize (0.0676)}} & \shortstack[c]{${-}0.5940^{***}$\\[-1pt]{\scriptsize (0.0675)}} & \shortstack[c]{${-}0.5704^{***}$\\[-1pt]{\scriptsize (0.0651)}} \\
E-Int Ind & \shortstack[c]{$0.0549^{***}$\\[-1pt]{\scriptsize (0.0166)}} & \shortstack[c]{$0.0551^{***}$\\[-1pt]{\scriptsize (0.0165)}} & \shortstack[c]{$0.0555^{***}$\\[-1pt]{\scriptsize (0.0167)}} & \shortstack[c]{$0.0915^{***}$\\[-1pt]{\scriptsize (0.0138)}} \\
FF/pc & \shortstack[c]{$0.0522^{***}$\\[-1pt]{\scriptsize (0.0032)}} & \shortstack[c]{$0.0525^{***}$\\[-1pt]{\scriptsize (0.0032)}} & \shortstack[c]{$0.0524^{***}$\\[-1pt]{\scriptsize (0.0032)}} & \shortstack[c]{$0.0424^{***}$\\[-1pt]{\scriptsize (0.0033)}} \\
FFelec/pc & \shortstack[c]{$0.0044$\\[-1pt]{\scriptsize (0.0038)}} & \shortstack[c]{$0.0040$\\[-1pt]{\scriptsize (0.0038)}} & \shortstack[c]{$0.0040$\\[-1pt]{\scriptsize (0.0039)}} & \shortstack[c]{$0.0096^{**}$\\[-1pt]{\scriptsize (0.0038)}} \\
HDD & \shortstack[c]{$0.0741^{***}$\\[-1pt]{\scriptsize (0.0056)}} & \shortstack[c]{$0.0735^{***}$\\[-1pt]{\scriptsize (0.0057)}} & \shortstack[c]{$0.0732^{***}$\\[-1pt]{\scriptsize (0.0058)}} & \shortstack[c]{$0.0842^{***}$\\[-1pt]{\scriptsize (0.0070)}} \\
CDD & \shortstack[c]{$0.0024$\\[-1pt]{\scriptsize (0.0130)}} & \shortstack[c]{$0.0009$\\[-1pt]{\scriptsize (0.0131)}} & \shortstack[c]{$0.0006$\\[-1pt]{\scriptsize (0.0134)}} & \shortstack[c]{$0.0366^{**}$\\[-1pt]{\scriptsize (0.0146)}} \\
\midrule
\multicolumn{5}{l}{\textit{Gap log-variance equation}} \\[2pt]
Eff: aggregate score & \shortstack[c]{${-}0.1215^{***}$\\[-1pt]{\scriptsize (0.0077)}} & \shortstack[c]{${-}0.1204^{***}$\\[-1pt]{\scriptsize (0.0077)}} & \shortstack[c]{${-}0.1179^{***}$\\[-1pt]{\scriptsize (0.0078)}} & \textemdash \\
Eff: UPB & \textemdash & \textemdash & \textemdash & \shortstack[c]{${-}0.1255^{***}$\\[-1pt]{\scriptsize (0.0160)}} \\
Eff: Transportation & \textemdash & \textemdash & \textemdash & \shortstack[c]{${-}0.1348^{***}$\\[-1pt]{\scriptsize (0.0346)}} \\
Eff: Building & \textemdash & \textemdash & \textemdash & \shortstack[c]{${-}0.0555$\\[-1pt]{\scriptsize (0.0348)}} \\
Eff: State initiatives & \textemdash & \textemdash & \textemdash & \shortstack[c]{${-}0.0587$\\[-1pt]{\scriptsize (0.0564)}} \\
Eff: Appliances & \textemdash & \textemdash & \textemdash & \shortstack[c]{${-}0.8641^{***}$\\[-1pt]{\scriptsize (0.1856)}} \\
Price & \shortstack[c]{${-}0.4857^{*}$\\[-1pt]{\scriptsize (0.2508)}} & \shortstack[c]{${-}0.5637^{**}$\\[-1pt]{\scriptsize (0.2485)}} & \shortstack[c]{${-}0.7076^{***}$\\[-1pt]{\scriptsize (0.2473)}} & \shortstack[c]{${-}0.1934$\\[-1pt]{\scriptsize (0.2590)}} \\
GDP/pc & \shortstack[c]{$2.5550^{***}$\\[-1pt]{\scriptsize (0.3703)}} & \shortstack[c]{$2.6311^{***}$\\[-1pt]{\scriptsize (0.3717)}} & \shortstack[c]{$2.6344^{***}$\\[-1pt]{\scriptsize (0.3705)}} & \shortstack[c]{$2.3092^{***}$\\[-1pt]{\scriptsize (0.3936)}} \\
\midrule
$N$ & 714 & 714 & 714 & 867 \\
Sample years & 2009--22 & 2009--22 & 2009--22 & 2006--22 \\
Omitted year & 2009 & 2009 & 2009 & 2006 \\
Parameters & 30 & 30 & 30 & 41 \\
Log likelihood & 248.50 & 243.28 & 236.58 & 304.70 \\
AIC & -437.00 & -426.57 & -413.16 & -527.40 \\
BIC & -299.87 & -289.44 & -276.03 & -332.03 \\
\bottomrule
\end{tabularx}
\begin{tablenotes}
\footnotesize
\item \textit{Notes:} Outcome: $\log((TNTCB-ESTCB)/TPOPP)$; price: centered $\log(PETXD)$. The constructs are conceptually aligned but not denominator-identical. Lag columns use the common 2009--2022 sample and enter the signed aggregate ACEEE score at the indicated lag. The five-domain column uses the full 2006--2022 sample and jointly enters UPB, transportation, building, state-initiative, and appliance scores on fixed 2020 point scales (20/12/9/6/3). Standard errors are in parentheses. $^{***}$ $p<0.01$, $^{**}$ $p<0.05$, $^{*}$ $p<0.10$.
\end{tablenotes}}
\end{threeparttable}
\end{table}

% Generated by review/feasibility/eia-components/expanded-tables/

% build_lmg_appendix_tables.py. Do not hand-edit this file.

\subsection{Complete pooled-model LMG attribution across energy formulations}

\label{sec:lmg_variants}

These tables extend the variable attribution in ~\ref{sec:imp} to every pooled timing, policy-composition, and energy-component formulation reported above. For the frontier, exact LMG averages sequential $R^2$ increments over all subsets of the model's substantive regressors after removing calendar-year means from both the fitted frontier and regressors. The fitted frontier is linear in those regressors, so its full auxiliary-model $R^2$ equals one.

The second part uses the model's generated composite gap,
\[G^{\mathrm{JLMS}}_{st}=-\log(\mathrm{JLMS}_{st}).\] Its observed-predictor allocations sum to the auxiliary regression's $R^2$; the unexplained remainder is $1-R^2$. The five-domain columns aggregate the five policy components only for compact cross-model comparison, and Table~\ref{tab:lmg_domain_detail_variants} retains every individual component.

The frontier rankings vary across outcomes and formulations. Price ranks first in 8 of the 15 formulations and is in the top three in 12; its share ranges from 13.2\% to 28.1\%. VMT/pc is among the top three contributors in 13 models and ranges from 13.1\% to 18.1\%. The grouped ACEEE-policy share is positive in every formulation but non-invariant: it ranges from 4.6\% to 30.1\%, has a median of 11.5\%, and is in the top three in 7 models. Its share under the five-domain parameterization is 24.3\% for total energy, 30.1\% for retail electricity, and 14.2\% for direct primary. These differences are not causal gains: splitting the score changes the predictor set and the way shared explanatory power is divided among correlated domains.

Within each outcome family the frontier ordering is comparatively robust to timing and policy parameterization: the median pairwise Spearman rank correlation is 0.982 for total energy, 0.982 for retail electricity, and 0.964 for direct primary energy. The generated-gap allocation is less transportable across outcomes: ACEEE policy is the largest observed driver in 10 models, GDP/pc in 5, and price in 0. The gap auxiliary $R^2$ ranges from 22.6\% to 68.1\%, leaving 31.9\%--77.4\% unexplained. Therefore, frontier rankings are fairly stable within outcome families, ACEEE retains nontrivial but non-invariant frontier weight, and gap-channel rankings remain outcome- and specification-dependent.

\subsubsection{Primary-equivalent total energy}

\begin{table}[H]
\centering
\caption{Complete pooled-model LMG fitted-frontier shares (percent): primary-equivalent total energy}\label{tab:lmg_total_frontier_variants}
\begin{threeparttable}
\scriptsize
\setlength{\tabcolsep}{4pt}
\begin{tabularx}{\textwidth}{@{}>{\raggedright\arraybackslash}p{.30\textwidth}*{5}{Y}@{}}
\toprule
Component & \shortstack{Current\\score} & \shortstack{Lag\\1} & \shortstack{Lag\\2} & \shortstack{Lag\\3} & \shortstack{Five\\domains} \\
\midrule
Price & 25.50 & 24.38 & 24.21 & 24.21 & 24.15 \\
GDP/pc & 2.95 & 2.83 & 2.85 & 2.88 & 2.98 \\
ACEEE policy & 11.30 & 11.47 & 11.24 & 10.75 & 24.32 \\
VMT/pc & 15.95 & 15.88 & 15.95 & 16.01 & 13.06 \\
ComSurf/pc & 2.53 & 2.75 & 2.72 & 2.72 & 2.21 \\
ResSurf/pc & 3.78 & 3.55 & 3.53 & 3.58 & 2.76 \\
E-Int Ind & 9.56 & 9.13 & 9.20 & 9.48 & 7.35 \\
FF/pc & 15.07 & 16.20 & 16.34 & 16.41 & 13.69 \\
FFelec/pc & 7.89 & 8.25 & 8.30 & 8.32 & 5.35 \\
HDD & 3.66 & 3.77 & 3.84 & 3.83 & 2.68 \\
CDD & 1.81 & 1.79 & 1.80 & 1.81 & 1.44 \\
\bottomrule
\end{tabularx}
\begin{tablenotes}
\footnotesize
\item \textit{Notes:} The outcome is log primary-equivalent total energy per capita; Price is the centered log aggregate end-use energy price. Entries are exact LMG shares of the full-model $R^2$ in the auxiliary regression. Entries are percent shares and every column sums to 100. In the five-domain column, ACEEE policy is the sum of the five individual domain allocations reported in Table~\ref{tab:lmg_domain_detail_variants}.
\end{tablenotes}
\end{threeparttable}
\end{table}

\begin{table}[H]
\centering
\caption{Complete pooled-model LMG generated-gap shares (percent): primary-equivalent total energy}\label{tab:lmg_total_gap_variants}
\begin{threeparttable}
\scriptsize
\setlength{\tabcolsep}{4pt}
\begin{tabularx}{\textwidth}{@{}>{\raggedright\arraybackslash}p{.30\textwidth}*{5}{Y}@{}}
\toprule
Component & \shortstack{Current\\score} & \shortstack{Lag\\1} & \shortstack{Lag\\2} & \shortstack{Lag\\3} & \shortstack{Five\\domains} \\
\midrule
ACEEE policy & 22.42 & 23.59 & 23.02 & 22.69 & 25.35 \\
GDP/pc & 11.04 & 11.75 & 11.69 & 11.65 & 6.34 \\
Price & 1.71 & 2.10 & 2.19 & 2.18 & 0.97 \\
Unexplained remainder & 64.83 & 62.55 & 63.10 & 63.48 & 67.34 \\
\bottomrule
\end{tabularx}
\begin{tablenotes}
\footnotesize
\item \textit{Notes:} The outcome is log primary-equivalent total energy per capita; Price is the centered log aggregate end-use energy price. The target is the residual-conditional generated composite gap, $-\log(\mathrm{JLMS})$. The observed rows are exact LMG $R^2$ allocations expressed as shares of total generated-gap variance.
\end{tablenotes}
\end{threeparttable}
\end{table}

\subsubsection{Retail electricity}

\begin{table}[H]
\centering
\caption{Complete pooled-model LMG fitted-frontier shares (percent): retail electricity}\label{tab:lmg_electricity_frontier_variants}
\begin{threeparttable}
\scriptsize
\setlength{\tabcolsep}{4pt}
\begin{tabularx}{\textwidth}{@{}>{\raggedright\arraybackslash}p{.30\textwidth}*{5}{Y}@{}}
\toprule
Component & \shortstack{Current\\score} & \shortstack{Lag\\1} & \shortstack{Lag\\2} & \shortstack{Lag\\3} & \shortstack{Five\\domains} \\
\midrule
Price & 26.93 & 28.11 & 27.32 & 26.63 & 21.17 \\
GDP/pc & 6.41 & 5.67 & 5.87 & 6.06 & 6.05 \\
ACEEE policy & 16.83 & 17.22 & 16.75 & 16.32 & 30.15 \\
VMT/pc & 17.78 & 17.46 & 17.83 & 18.12 & 14.72 \\
ComSurf/pc & 1.43 & 1.82 & 1.84 & 1.85 & 0.92 \\
ResSurf/pc & 2.80 & 2.80 & 2.83 & 2.86 & 2.29 \\
E-Int Ind & 8.82 & 7.92 & 8.13 & 8.36 & 9.05 \\
FF/pc & 8.25 & 8.66 & 8.91 & 9.10 & 6.91 \\
FFelec/pc & 7.90 & 7.82 & 7.98 & 8.12 & 6.35 \\
HDD & 1.10 & 0.98 & 0.96 & 0.95 & 1.12 \\
CDD & 1.75 & 1.54 & 1.58 & 1.63 & 1.28 \\
\bottomrule
\end{tabularx}
\begin{tablenotes}
\footnotesize
\item \textit{Notes:} The outcome is $\log(ESTCB/TPOPP)$; Price is centered $\log(ESTCD)$, the retail electricity price per million Btu. Entries are exact LMG shares of the full-model $R^2$ in the auxiliary regression.
\end{tablenotes}
\end{threeparttable}
\end{table}

\begin{table}[H]
\centering
\caption{Complete pooled-model LMG generated-gap shares (percent): retail electricity}\label{tab:lmg_electricity_gap_variants}
\begin{threeparttable}
\scriptsize
\setlength{\tabcolsep}{4pt}
\begin{tabularx}{\textwidth}{@{}>{\raggedright\arraybackslash}p{.30\textwidth}*{5}{Y}@{}}
\toprule
Component & \shortstack{Current\\score} & \shortstack{Lag\\1} & \shortstack{Lag\\2} & \shortstack{Lag\\3} & \shortstack{Five\\domains} \\
\midrule
ACEEE policy & 2.80 & 3.57 & 3.32 & 3.01 & 8.27 \\
GDP/pc & 58.53 & 57.43 & 57.59 & 57.71 & 56.58 \\
Price & 4.23 & 4.17 & 4.62 & 4.93 & 3.22 \\
Unexplained remainder & 34.44 & 34.84 & 34.47 & 34.35 & 31.94 \\
\bottomrule
\end{tabularx}
\begin{tablenotes}
\footnotesize
\item \textit{Notes:} The outcome is $\log(ESTCB/TPOPP)$; Price is centered $\log(ESTCD)$, the retail electricity price per million Btu. The target is the residual-conditional generated composite gap, $-\log(\mathrm{JLMS})$. The observed rows are exact LMG $R^2$ allocations expressed as shares of total generated-gap variance.
\end{tablenotes}
\end{threeparttable}
\end{table}

\subsubsection{Direct primary energy excluding retail electricity}

\begin{table}[H]
\centering
\caption{Complete pooled-model LMG fitted-frontier shares (percent): direct primary energy excluding retail electricity}\label{tab:lmg_primary_frontier_variants}
\begin{threeparttable}
\scriptsize
\setlength{\tabcolsep}{4pt}
\begin{tabularx}{\textwidth}{@{}>{\raggedright\arraybackslash}p{.30\textwidth}*{5}{Y}@{}}
\toprule
Component & \shortstack{Current\\score} & \shortstack{Lag\\1} & \shortstack{Lag\\2} & \shortstack{Lag\\3} & \shortstack{Five\\domains} \\
\midrule
Price & 15.24 & 13.16 & 13.34 & 13.56 & 14.09 \\
GDP/pc & 3.72 & 3.55 & 3.65 & 3.67 & 2.80 \\
ACEEE policy & 4.88 & 4.72 & 4.68 & 4.58 & 14.23 \\
VMT/pc & 17.36 & 17.47 & 17.44 & 17.45 & 14.07 \\
ComSurf/pc & 1.01 & 1.38 & 1.37 & 1.38 & 0.88 \\
ResSurf/pc & 4.49 & 4.77 & 4.74 & 4.73 & 3.88 \\
E-Int Ind & 6.38 & 5.47 & 5.55 & 5.66 & 6.46 \\
FF/pc & 20.74 & 22.25 & 22.21 & 22.12 & 18.64 \\
FFelec/pc & 4.80 & 4.42 & 4.47 & 4.49 & 4.16 \\
HDD & 14.18 & 14.93 & 14.75 & 14.63 & 13.65 \\
CDD & 7.20 & 7.89 & 7.80 & 7.75 & 7.13 \\
\bottomrule
\end{tabularx}
\begin{tablenotes}
\footnotesize
\item \textit{Notes:} The outcome is $\log((TNTCB-ESTCB)/TPOPP)$; Price is centered $\log(PETXD)$. Entries are exact LMG shares of the full-model $R^2$ in the auxiliary regression.
\end{tablenotes}
\end{threeparttable}
\end{table}

\begin{table}[H]
\centering
\caption{Complete pooled-model LMG generated-gap shares (percent): direct primary energy excluding retail electricity}\label{tab:lmg_primary_gap_variants}
\begin{threeparttable}
\scriptsize
\setlength{\tabcolsep}{4pt}
\begin{tabularx}{\textwidth}{@{}>{\raggedright\arraybackslash}p{.30\textwidth}*{5}{Y}@{}}
\toprule
Component & \shortstack{Current\\score} & \shortstack{Lag\\1} & \shortstack{Lag\\2} & \shortstack{Lag\\3} & \shortstack{Five\\domains} \\
\midrule
ACEEE policy & 18.85 & 20.05 & 18.95 & 17.92 & 20.80 \\
GDP/pc & 2.33 & 2.10 & 2.04 & 1.94 & 2.39 \\
Price & 2.16 & 2.66 & 2.66 & 2.74 & 1.25 \\
Unexplained remainder & 76.66 & 75.20 & 76.35 & 77.39 & 75.55 \\
\bottomrule
\end{tabularx}
\begin{tablenotes}
\footnotesize
\item \textit{Notes:} The outcome is $\log((TNTCB-ESTCB)/TPOPP)$; Price is centered $\log(PETXD)$. The target is the residual-conditional generated composite gap, $-\log(\mathrm{JLMS})$. The observed rows are exact LMG $R^2$ allocations expressed as shares of total generated-gap variance.
\end{tablenotes}
\end{threeparttable}
\end{table}

\subsubsection{Individual ACEEE-domain allocations}

\begin{table}[H]
\centering
\caption{Individual five-domain ACEEE LMG shares (percent) across energy formulations}
\label{tab:lmg_domain_detail_variants}
\begin{threeparttable}
\scriptsize
\setlength{\tabcolsep}{3.5pt}
\begin{tabularx}{\textwidth}{@{}>{\raggedright\arraybackslash}p{.28\textwidth}*{6}{Y}@{}}
\toprule
& \multicolumn{2}{c}{Total energy} & \multicolumn{2}{c}{Retail electricity} & \multicolumn{2}{c}{Direct primary} \\
\cmidrule(lr){2-3}\cmidrule(lr){4-5}\cmidrule(lr){6-7}
Domain & Frontier & \shortstack{Generated\\gap} & Frontier & \shortstack{Generated\\gap} & Frontier & \shortstack{Generated\\gap} \\
\midrule
Utilities/public benefits & 3.91 & 15.76 & 10.32 & 0.76 & 1.52 & 9.90 \\
Transportation & 7.86 & 3.32 & 9.55 & 2.03 & 4.26 & 3.59 \\
Building energy codes & 5.59 & 3.01 & 3.38 & 0.26 & 3.80 & 3.68 \\
State government initiatives & 4.60 & 1.39 & 3.25 & 4.48 & 3.58 & 1.37 \\
Appliance standards & 2.36 & 1.87 & 3.64 & 0.74 & 1.07 & 2.25 \\
\bottomrule
\end{tabularx}
\begin{tablenotes}
\footnotesize
\item \textit{Notes:} Frontier entries are shares of full-model $R^2$. Generated-gap entries are shares of total generated-gap variance.
\end{tablenotes}
\end{threeparttable}
\end{table}

% Generated by build_sunburst_comparison_tables.py. Do not hand-edit.
\subsection{Sunburst-style contribution accounting across specifications}
\label{sec:sunburst_specification_comparisons}

Tables~\ref{tab:sunburst_distribution_contributions}--\ref{tab:sunburst_matched_outcomes_contributions} put the sunburst figures on a common scale. Every displayed regression has exactly one outcome-specific price. Each column first reports the inner frontier--gap--noise split, then expresses every outer slice as a percentage point of the same column's total $\Var(\log q)$. The one-sided distribution changes the parent balance more than it changes the focal ACEEE allocation. Relative to the half-normal, the fitted-frontier share changes by +27.10 percentage points under the truncated normal and -0.42 points under the exponential; the exponential gap share changes by -1.56 points. Yet the combined ACEEE frontier-plus-gap contribution is 15.83\%, 16.19\%, and 14.63\%. In the truncated-normal column, the final generated gap combines the pre-truncation location and log-variance equations; their coefficients are not treated as additive contributions. On the fixed timing samples, the total-energy frontier spans 57.33--58.95\%; the direct-primary frontier spans 41.66--46.04\%. The retail electricity frontier spans 43.76--72.69\%, with gap shares of 18.71--47.31\%. Five-domain formulations assign more frontier variation to the grouped ACEEE block, especially for retail electricity. That reflects a more flexible correlated predictor set, not a larger policy effect. Among matched pairs, the preferred aggregate, delivered-total, and direct-primary specifications have frontier shares of 45.00--56.35\%. The retail electricity specification assigns 20.43\% to the one-sided gap and 70.96\% to the frontier.

\begin{table}[H]
\centering
\caption{Sunburst accounting across one-sided distributions (\% of total outcome variance)}\label{tab:sunburst_distribution_contributions}
\begin{threeparttable}
{\setstretch{1}\footnotesize
\setlength{\tabcolsep}{3pt}
\renewcommand{\arraystretch}{0.94}
\begin{tabularx}{\textwidth}{@{}>{\raggedright\arraybackslash}p{.34\textwidth}*{3}{Y}@{}}
\toprule
Component & Half-normal & \shortstack{Truncated\\normal} & Exponential \\
\midrule
\multicolumn{4}{l}{\textit{Panel A: inner-ring split of $\Var(\log q)$}} \\[1pt]
Fitted frontier & 56.35 & 83.45 & 55.93 \\
One-sided gap & 42.20 & 15.84 & 40.65 \\
Symmetric noise & 1.45 & 0.71 & 3.42 \\
\midrule
\multicolumn{4}{l}{\textit{Panel B: fitted-frontier contributions to $\Var(\log q)$}} \\[1pt]
Price & 14.37 & 25.58 & 14.04 \\
GDP/pc & 1.66 & 9.43 & 1.73 \\
ACEEE policy & 6.37 & 15.96 & 7.73 \\
VMT/pc & 8.98 & 6.38 & 9.04 \\
ComSurf/pc & 1.43 & 3.77 & 1.27 \\
ResSurf/pc & 2.13 & 2.12 & 1.67 \\
E-Int Ind & 5.39 & 3.77 & 5.00 \\
FF/pc & 8.49 & 7.37 & 7.74 \\
FFelec/pc & 4.45 & 5.03 & 4.94 \\
HDD & 2.06 & 2.41 & 1.80 \\
CDD & 1.02 & 1.63 & 0.96 \\
\midrule
\multicolumn{4}{l}{\textit{Panel C: generated-gap contributions to $\Var(\log q)$}} \\[1pt]
ACEEE policy & 9.46 & 0.23 & 6.90 \\
Price & 0.72 & 1.96 & 0.58 \\
GDP/pc & 4.66 & 1.52 & 7.10 \\
Unexplained remainder & 27.36 & 12.13 & 26.07 \\
\midrule
\multicolumn{4}{l}{\textit{Panel D: symmetric-noise contribution to $\Var(\log q)$}} \\[1pt]
Symmetric noise & 1.45 & 0.71 & 3.42 \\
\bottomrule
\end{tabularx}
\begin{tablenotes}
\footnotesize
\item \textit{Notes:} Entries are percentage points of that column's own $\Var(\log q)$. Panel A sums to 100. Panels B--D provide the corresponding outer-ring accounting and jointly sum to 100. The covariance residual required for exact closure is divided equally between the fitted frontier and one-sided gap.
\end{tablenotes}
}
\end{threeparttable}
\end{table}

\begin{table}[H]
\centering
\caption{Sunburst accounting across ACEEE timing: primary-equivalent total energy (percent of total outcome variance)}\label{tab:sunburst_total_timing_contributions}
\begin{threeparttable}
{\setstretch{1}\footnotesize
\setlength{\tabcolsep}{3pt}
\renewcommand{\arraystretch}{0.94}
\begin{tabularx}{\textwidth}{@{}>{\raggedright\arraybackslash}p{.29\textwidth}*{4}{Y}@{}}
\toprule
Component & \shortstack{Current\\(same sample)} & \shortstack{Lag\\1} & \shortstack{Lag\\2} & \shortstack{Lag\\3} \\
\midrule
\multicolumn{5}{l}{\textit{Panel A: inner-ring split of $\Var(\log q)$}} \\[1pt]
Fitted frontier & 58.95 & 58.24 & 57.79 & 57.33 \\
One-sided gap & 39.31 & 40.14 & 40.61 & 41.07 \\
Symmetric noise & 1.73 & 1.62 & 1.60 & 1.59 \\
\midrule
\multicolumn{5}{l}{\textit{Panel B: fitted-frontier contributions to $\Var(\log q)$}} \\[1pt]
Price & 14.44 & 14.20 & 13.99 & 13.88 \\
GDP/pc & 1.68 & 1.65 & 1.64 & 1.65 \\
ACEEE policy & 6.82 & 6.68 & 6.50 & 6.16 \\
VMT/pc & 9.32 & 9.25 & 9.22 & 9.18 \\
ComSurf/pc & 1.62 & 1.60 & 1.57 & 1.56 \\
ResSurf/pc & 2.10 & 2.07 & 2.04 & 2.05 \\
E-Int Ind & 5.43 & 5.31 & 5.32 & 5.44 \\
FF/pc & 9.42 & 9.43 & 9.44 & 9.41 \\
FFelec/pc & 4.87 & 4.81 & 4.80 & 4.77 \\
HDD & 2.19 & 2.19 & 2.22 & 2.20 \\
CDD & 1.07 & 1.04 & 1.04 & 1.04 \\
\midrule
\multicolumn{5}{l}{\textit{Panel C: generated-gap contributions to $\Var(\log q)$}} \\[1pt]
ACEEE policy & 9.45 & 9.47 & 9.35 & 9.32 \\
Price & 0.80 & 0.84 & 0.89 & 0.89 \\
GDP/pc & 4.83 & 4.72 & 4.75 & 4.79 \\
Unexplained remainder & 24.23 & 25.11 & 25.62 & 26.07 \\
\midrule
\multicolumn{5}{l}{\textit{Panel D: symmetric-noise contribution to $\Var(\log q)$}} \\[1pt]
Symmetric noise & 1.73 & 1.62 & 1.60 & 1.59 \\
\bottomrule
\end{tabularx}
\begin{tablenotes}
\footnotesize
\item \textit{Notes:} Entries are percentage points of that column's own $\Var(\log q)$. Panel A sums to 100. Panels B--D provide the corresponding outer-ring accounting and jointly sum to 100. The covariance residual required for exact closure is divided equally between the fitted frontier and one-sided gap.
\end{tablenotes}
}
\end{threeparttable}
\end{table}

\begin{table}[H]
\centering
\caption{Sunburst accounting across ACEEE score composition: primary-equivalent total energy (percent of total outcome variance)}\label{tab:sunburst_total_composition_contributions}
\begin{threeparttable}
{\setstretch{1}\footnotesize
\setlength{\tabcolsep}{3pt}
\renewcommand{\arraystretch}{0.94}
\begin{tabularx}{\textwidth}{@{}>{\raggedright\arraybackslash}p{.42\textwidth}*{2}{Y}@{}}
\toprule
Component & \shortstack{Current\\total} & \shortstack{Five\\domains} \\
\midrule
\multicolumn{3}{l}{\textit{Panel A: inner-ring split of $\Var(\log q)$}} \\[1pt]
Fitted frontier & 56.35 & 58.99 \\
One-sided gap & 42.20 & 39.99 \\
Symmetric noise & 1.45 & 1.02 \\
\midrule
\multicolumn{3}{l}{\textit{Panel B: fitted-frontier contributions to $\Var(\log q)$}} \\[1pt]
Price & 14.37 & 14.25 \\
GDP/pc & 1.66 & 1.76 \\
ACEEE policy & 6.37 & 14.35 \\
\hspace{1em}of which: utility/public benefits & \textemdash & 2.30 \\
\hspace{1em}of which: transportation & \textemdash & 4.64 \\
\hspace{1em}of which: building codes & \textemdash & 3.30 \\
\hspace{1em}of which: state initiatives & \textemdash & 2.71 \\
\hspace{1em}of which: appliance standards & \textemdash & 1.39 \\
VMT/pc & 8.98 & 7.71 \\
ComSurf/pc & 1.43 & 1.31 \\
ResSurf/pc & 2.13 & 1.63 \\
E-Int Ind & 5.39 & 4.34 \\
FF/pc & 8.49 & 8.08 \\
FFelec/pc & 4.45 & 3.15 \\
HDD & 2.06 & 1.58 \\
CDD & 1.02 & 0.85 \\
\midrule
\multicolumn{3}{l}{\textit{Panel C: generated-gap contributions to $\Var(\log q)$}} \\[1pt]
ACEEE policy & 9.46 & 10.14 \\
\hspace{1em}of which: utility/public benefits & \textemdash & 6.30 \\
\hspace{1em}of which: transportation & \textemdash & 1.33 \\
\hspace{1em}of which: building codes & \textemdash & 1.21 \\
\hspace{1em}of which: state initiatives & \textemdash & 0.55 \\
\hspace{1em}of which: appliance standards & \textemdash & 0.75 \\
Price & 0.72 & 0.39 \\
GDP/pc & 4.66 & 2.53 \\
Unexplained remainder & 27.36 & 26.93 \\
\midrule
\multicolumn{3}{l}{\textit{Panel D: symmetric-noise contribution to $\Var(\log q)$}} \\[1pt]
Symmetric noise & 1.45 & 1.02 \\
\bottomrule
\end{tabularx}
\begin{tablenotes}
\footnotesize
\item \textit{Notes:} Entries are percentage points of that column's own $\Var(\log q)$. Panel A sums to 100. Panels B--D provide the corresponding outer-ring accounting and jointly sum to 100\rev{; five-domain ``of which'' rows are informational and excluded from that sum. The grouped ACEEE-policy row in the five-domain column is the sum of the five individual allocations shown below it.}
\end{tablenotes}
}
\end{threeparttable}
\end{table}

\begin{table}[H]
\centering
\caption{Sunburst accounting across ACEEE timing: retail electricity (percent of total outcome variance)}\label{tab:sunburst_electricity_timing_contributions}
\begin{threeparttable}
{\setstretch{1}\footnotesize
\setlength{\tabcolsep}{3pt}
\renewcommand{\arraystretch}{0.94}
\begin{tabularx}{\textwidth}{@{}>{\raggedright\arraybackslash}p{.29\textwidth}*{4}{Y}@{}}
\toprule
Component & \shortstack{Current\\(same sample)} & \shortstack{Lag\\1} & \shortstack{Lag\\2} & \shortstack{Lag\\3} \\
\midrule
\multicolumn{5}{l}{\textit{Panel A: inner-ring split of $\Var(\log q)$}} \\[1pt]
Fitted frontier & 43.76 & 72.00 & 72.60 & 72.69 \\
One-sided gap & 47.31 & 19.35 & 18.76 & 18.71 \\
Symmetric noise & 8.93 & 8.65 & 8.63 & 8.60 \\
\midrule
\multicolumn{5}{l}{\textit{Panel B: fitted-frontier contributions to $\Var(\log q)$}} \\[1pt]
Price & 13.96 & 20.24 & 19.83 & 19.35 \\
GDP/pc & 2.23 & 4.08 & 4.26 & 4.41 \\
ACEEE policy & 7.34 & 12.40 & 12.16 & 11.86 \\
VMT/pc & 7.15 & 12.57 & 12.95 & 13.17 \\
ComSurf/pc & 0.78 & 1.31 & 1.34 & 1.34 \\
ResSurf/pc & 1.22 & 2.02 & 2.05 & 2.08 \\
E-Int Ind & 3.34 & 5.70 & 5.90 & 6.08 \\
FF/pc & 3.38 & 6.24 & 6.47 & 6.61 \\
FFelec/pc & 3.22 & 5.63 & 5.79 & 5.90 \\
HDD & 0.49 & 0.70 & 0.70 & 0.69 \\
CDD & 0.67 & 1.11 & 1.15 & 1.18 \\
\midrule
\multicolumn{5}{l}{\textit{Panel C: generated-gap contributions to $\Var(\log q)$}} \\[1pt]
ACEEE policy & 2.23 & 0.69 & 0.62 & 0.56 \\
Price & 0.88 & 0.81 & 0.87 & 0.92 \\
GDP/pc & 26.66 & 11.11 & 10.81 & 10.80 \\
Unexplained remainder & 17.53 & 6.74 & 6.47 & 6.43 \\
\midrule
\multicolumn{5}{l}{\textit{Panel D: symmetric-noise contribution to $\Var(\log q)$}} \\[1pt]
Symmetric noise & 8.93 & 8.65 & 8.63 & 8.60 \\
\bottomrule
\end{tabularx}
\begin{tablenotes}
\footnotesize
\item \textit{Notes:} Entries are percentage points of that column's own $\Var(\log q)$. Panel A sums to 100. Panels B--D provide the corresponding outer-ring accounting and jointly sum to 100.
\end{tablenotes}
}
\end{threeparttable}
\end{table}

\clearpage
\begin{table}[H]
\centering
\caption{Sunburst accounting across ACEEE score composition: retail electricity (percent of total outcome variance)}\label{tab:sunburst_electricity_composition_contributions}
\begin{threeparttable}
{\setstretch{1}\footnotesize
\setlength{\tabcolsep}{3pt}
\renewcommand{\arraystretch}{0.94}
\begin{tabularx}{\textwidth}{@{}>{\raggedright\arraybackslash}p{.42\textwidth}*{2}{Y}@{}}
\toprule
Component & \shortstack{Current\\total} & \shortstack{Five\\domains} \\
\midrule
\multicolumn{3}{l}{\textit{Panel A: inner-ring split of $\Var(\log q)$}} \\[1pt]
Fitted frontier & 70.96 & 65.99 \\
One-sided gap & 20.43 & 26.07 \\
Symmetric noise & 8.61 & 7.95 \\
\midrule
\multicolumn{3}{l}{\textit{Panel B: fitted-frontier contributions to $\Var(\log q)$}} \\[1pt]
Price & 19.11 & 13.97 \\
GDP/pc & 4.55 & 3.99 \\
ACEEE policy & 11.94 & 19.89 \\
\hspace{1em}of which: utility/public benefits & \textemdash & 6.81 \\
\hspace{1em}of which: transportation & \textemdash & 6.30 \\
\hspace{1em}of which: building codes & \textemdash & 2.23 \\
\hspace{1em}of which: state initiatives & \textemdash & 2.15 \\
\hspace{1em}of which: appliance standards & \textemdash & 2.40 \\
VMT/pc & 12.61 & 9.72 \\
ComSurf/pc & 1.01 & 0.60 \\
ResSurf/pc & 1.99 & 1.51 \\
E-Int Ind & 6.26 & 5.97 \\
FF/pc & 5.85 & 4.56 \\
FFelec/pc & 5.61 & 4.19 \\
HDD & 0.78 & 0.74 \\
CDD & 1.24 & 0.84 \\
\midrule
\multicolumn{3}{l}{\textit{Panel C: generated-gap contributions to $\Var(\log q)$}} \\[1pt]
ACEEE policy & 0.57 & 2.16 \\
\hspace{1em}of which: utility/public benefits & \textemdash & 0.20 \\
\hspace{1em}of which: transportation & \textemdash & 0.53 \\
\hspace{1em}of which: building codes & \textemdash & 0.07 \\
\hspace{1em}of which: state initiatives & \textemdash & 1.17 \\
\hspace{1em}of which: appliance standards & \textemdash & 0.19 \\
Price & 0.86 & 0.84 \\
GDP/pc & 11.96 & 14.75 \\
Unexplained remainder & 7.04 & 8.32 \\
\midrule
\multicolumn{3}{l}{\textit{Panel D: symmetric-noise contribution to $\Var(\log q)$}} \\[1pt]
Symmetric noise & 8.61 & 7.95 \\
\bottomrule
\end{tabularx}
\begin{tablenotes}
\footnotesize
\item \textit{Notes:} Entries are percentage points of that column's own $\Var(\log q)$. Panel A sums to 100. Panels B--D provide the corresponding outer-ring accounting and jointly sum to 100; five-domain ``of which'' rows are informational and excluded from that sum. The covariance residual required for exact closure is divided equally between the fitted frontier and one-sided gap. Frontier rows use exact point LMG of the fitted frontier; gap rows use exact point LMG of the final residual-conditional $-\log(\mathrm{JLMS})$ target and retain its unexplained remainder. These are descriptive, model-dependent allocations, not coefficients, causal effects, mechanism shares, or uncertainty estimates. No bootstrap is used. The outcome is $\log(ESTCB/TPOPP)$ and price is centered $\log(ESTCD)$. Both columns use $N=867$ (2006--2022). The grouped ACEEE-policy row in the five-domain column is the sum of the five individual allocations shown below it.
\end{tablenotes}
}
\end{threeparttable}
\end{table}

\clearpage
\begin{table}[H]
\centering
\caption{Sunburst accounting across ACEEE timing: direct primary energy (percent of total outcome variance)}\label{tab:sunburst_primary_timing_contributions}
\begin{threeparttable}
{\setstretch{1}\footnotesize
\setlength{\tabcolsep}{3pt}
\renewcommand{\arraystretch}{0.94}
\begin{tabularx}{\textwidth}{@{}>{\raggedright\arraybackslash}p{.29\textwidth}*{4}{Y}@{}}
\toprule
Component & \shortstack{Current\\(same sample)} & \shortstack{Lag\\1} & \shortstack{Lag\\2} & \shortstack{Lag\\3} \\
\midrule
\multicolumn{5}{l}{\textit{Panel A: inner-ring split of $\Var(\log q)$}} \\[1pt]
Fitted frontier & 46.04 & 43.77 & 42.12 & 41.66 \\
One-sided gap & 53.31 & 55.65 & 57.29 & 57.73 \\
Symmetric noise & 0.64 & 0.58 & 0.59 & 0.61 \\
\midrule
\multicolumn{5}{l}{\textit{Panel B: fitted-frontier contributions to $\Var(\log q)$}} \\[1pt]
Price & 6.05 & 5.76 & 5.62 & 5.65 \\
GDP/pc & 1.59 & 1.55 & 1.54 & 1.53 \\
ACEEE policy & 2.18 & 2.07 & 1.97 & 1.91 \\
VMT/pc & 8.04 & 7.64 & 7.34 & 7.27 \\
ComSurf/pc & 0.62 & 0.60 & 0.58 & 0.57 \\
ResSurf/pc & 2.20 & 2.09 & 2.00 & 1.97 \\
E-Int Ind & 2.56 & 2.39 & 2.34 & 2.36 \\
FF/pc & 10.22 & 9.74 & 9.35 & 9.21 \\
FFelec/pc & 2.03 & 1.94 & 1.88 & 1.87 \\
HDD & 6.94 & 6.53 & 6.21 & 6.09 \\
CDD & 3.62 & 3.45 & 3.29 & 3.23 \\
\midrule
\multicolumn{5}{l}{\textit{Panel C: generated-gap contributions to $\Var(\log q)$}} \\[1pt]
ACEEE policy & 11.29 & 11.16 & 10.85 & 10.35 \\
Price & 1.44 & 1.48 & 1.52 & 1.58 \\
GDP/pc & 1.17 & 1.17 & 1.17 & 1.12 \\
Unexplained remainder & 39.41 & 41.85 & 43.74 & 44.68 \\
\midrule
\multicolumn{5}{l}{\textit{Panel D: symmetric-noise contribution to $\Var(\log q)$}} \\[1pt]
Symmetric noise & 0.64 & 0.58 & 0.59 & 0.61 \\
\bottomrule
\end{tabularx}
\begin{tablenotes}
\footnotesize
\item \textit{Notes:} Entries are percentage points of that column's own $\Var(\log q)$. Panel A sums to 100. Panels B--D provide the corresponding outer-ring accounting and jointly sum to 100. The outcome is $\log((TNTCB-ESTCB)/TPOPP)$ and price is centered $\log(PETXD)$.
\end{tablenotes}
}
\end{threeparttable}
\end{table}

\begin{table}[H]
\centering
\caption{Sunburst accounting across ACEEE score composition: direct primary energy (percent of total outcome variance)}\label{tab:sunburst_primary_composition_contributions}
\begin{threeparttable}
{\setstretch{1}\footnotesize
\setlength{\tabcolsep}{3pt}
\renewcommand{\arraystretch}{0.94}
\begin{tabularx}{\textwidth}{@{}>{\raggedright\arraybackslash}p{.42\textwidth}*{2}{Y}@{}}
\toprule
Component & \shortstack{Current\\total} & \shortstack{Five\\domains} \\
\midrule
\multicolumn{3}{l}{\textit{Panel A: inner-ring split of $\Var(\log q)$}} \\[1pt]
Fitted frontier & 45.00 & 49.76 \\
One-sided gap & 53.98 & 49.02 \\
Symmetric noise & 1.01 & 1.21 \\
\midrule
\multicolumn{3}{l}{\textit{Panel B: fitted-frontier contributions to $\Var(\log q)$}} \\[1pt]
Price & 6.86 & 7.01 \\
GDP/pc & 1.67 & 1.39 \\
ACEEE policy & 2.20 & 7.08 \\
\hspace{1em}of which: utility/public benefits & \textemdash & 0.75 \\
\hspace{1em}of which: transportation & \textemdash & 2.12 \\
\hspace{1em}of which: building codes & \textemdash & 1.89 \\
\hspace{1em}of which: state initiatives & \textemdash & 1.78 \\
\hspace{1em}of which: appliance standards & \textemdash & 0.53 \\
VMT/pc & 7.81 & 7.00 \\
ComSurf/pc & 0.45 & 0.44 \\
ResSurf/pc & 2.02 & 1.93 \\
E-Int Ind & 2.87 & 3.22 \\
FF/pc & 9.33 & 9.28 \\
FFelec/pc & 2.16 & 2.07 \\
HDD & 6.38 & 6.79 \\
CDD & 3.24 & 3.55 \\
\midrule
\multicolumn{3}{l}{\textit{Panel C: generated-gap contributions to $\Var(\log q)$}} \\[1pt]
ACEEE policy & 10.17 & 10.20 \\
\hspace{1em}of which: utility/public benefits & \textemdash & 4.86 \\
\hspace{1em}of which: transportation & \textemdash & 1.76 \\
\hspace{1em}of which: building codes & \textemdash & 1.81 \\
\hspace{1em}of which: state initiatives & \textemdash & 0.67 \\
\hspace{1em}of which: appliance standards & \textemdash & 1.10 \\
Price & 1.16 & 0.61 \\
GDP/pc & 1.26 & 1.17 \\
Unexplained remainder & 41.39 & 37.04 \\
\midrule
\multicolumn{3}{l}{\textit{Panel D: symmetric-noise contribution to $\Var(\log q)$}} \\[1pt]
Symmetric noise & 1.01 & 1.21 \\
\bottomrule
\end{tabularx}
\begin{tablenotes}
\footnotesize
\item \textit{Notes:} Entries are percentage points of that column's own $\Var(\log q)$. Panel A sums to 100. Panels B--D provide the corresponding outer-ring accounting and jointly sum to 100\rev{; five-domain ``of which'' rows are informational and excluded from that sum}. The grouped ACEEE-policy row in the five-domain column is the sum of the five individual allocations shown below it.
\end{tablenotes}
}
\end{threeparttable}
\end{table}

\begin{table}[H]
\centering
\caption{Sunburst accounting across outcome-specific single-price pairs (percent of total outcome variance)}\label{tab:sunburst_matched_outcomes_contributions}
\begin{threeparttable}
{\setstretch{1}\footnotesize
\setlength{\tabcolsep}{3pt}
\renewcommand{\arraystretch}{0.94}
\begin{tabularx}{\textwidth}{@{}>{\raggedright\arraybackslash}p{.29\textwidth}*{4}{Y}@{}}
\toprule
Component & \shortstack{Preferred\\aggregate} & \shortstack{Delivered\\total} & \shortstack{Direct\\primary} & \shortstack{Retail\\electricity} \\
\midrule
\multicolumn{5}{l}{\textit{Panel A: inner-ring split of $\Var(\log q)$}} \\[1pt]
Fitted frontier & 56.35 & 48.45 & 45.00 & 70.96 \\
One-sided gap & 42.20 & 50.47 & 53.98 & 20.43 \\
Symmetric noise & 1.45 & 1.08 & 1.01 & 8.61 \\
\midrule
\multicolumn{5}{l}{\textit{Panel B: fitted-frontier contributions to $\Var(\log q)$}} \\[1pt]
Price & 14.37 & 12.95 & 6.86 & 19.11 \\
GDP/pc & 1.66 & 1.07 & 1.67 & 4.55 \\
ACEEE policy & 6.37 & 3.08 & 2.20 & 11.94 \\
VMT/pc & 8.98 & 7.20 & 7.81 & 12.61 \\
ComSurf/pc & 1.43 & 0.63 & 0.45 & 1.01 \\
ResSurf/pc & 2.13 & 1.87 & 2.02 & 1.99 \\
E-Int Ind & 5.39 & 3.03 & 2.87 & 6.26 \\
FF/pc & 8.49 & 9.79 & 9.33 & 5.85 \\
FFelec/pc & 4.45 & 2.18 & 2.16 & 5.61 \\
HDD & 2.06 & 4.46 & 6.38 & 0.78 \\
CDD & 1.02 & 2.19 & 3.24 & 1.24 \\
\midrule
\multicolumn{5}{l}{\textit{Panel C: generated-gap contributions to $\Var(\log q)$}} \\[1pt]
ACEEE policy & 9.46 & 11.11 & 10.17 & 0.57 \\
Price & 0.72 & 1.51 & 1.16 & 0.86 \\
GDP/pc & 4.66 & 1.63 & 1.26 & 11.96 \\
Unexplained remainder & 27.36 & 36.22 & 41.39 & 7.04 \\
\midrule
\multicolumn{5}{l}{\textit{Panel D: symmetric-noise contribution to $\Var(\log q)$}} \\[1pt]
Symmetric noise & 1.45 & 1.08 & 1.01 & 8.61 \\
\bottomrule
\end{tabularx}
\begin{tablenotes}
\footnotesize
\item \textit{Notes:} Entries are percentage points of that column's own $\Var(\log q)$. Panel A sums to 100. Panels B--D provide the corresponding outer-ring accounting and jointly sum to 100. Columns use the primary-equivalent pair, delivered total energy with TETCD, direct primary energy with PETXD, and retail electricity with ESTCD.
\end{tablenotes}
}
\end{threeparttable}
\end{table}

\subsection{Cross-specification stability of ranks and total contributions}
\label{sec:robustness_stability}

The following figures synthesize the outcome-specific robustness checks into
two descriptive summaries. Within each accepted pooled specification, we add
each main-model driver's exact LMG allocation from the fitted frontier to its
exact LMG allocation from the generated composite gap, after scaling both by
their parent shares of that model's own $\Var(\log q)$. Drivers are then ranked
by this combined contribution. The attribution rule is held fixed: these
figures compare SFA specifications and do not pool LMG ranks with Random Forest
importance. Rows follow the main-specification ordering. All quartiles use the
type-7 empirical quantile convention; the displayed ranges summarize the
finite, designed specification universe rather than a sampling distribution.

\begin{table}[H]
\centering
\caption{Specification universe underlying the robustness-stability figures}
\label{tab:robustness_stability_scope}
\begin{threeparttable}
\footnotesize
\setlength{\tabcolsep}{3pt}
\renewcommand{\arraystretch}{1.05}
\begin{tabularx}{\textwidth}{@{}p{.16\textwidth}c>{\raggedright\arraybackslash}X>{\raggedright\arraybackslash}p{.29\textwidth}@{}}
\toprule
Outcome & $m$ & Included accepted pooled specifications & Full-detail locations \\
\midrule
Primary-equivalent total
& 21
& Main; common-sample current and lags 1--3; five domains and two policy blocks; truncated normal and exponential; matched delivered-total outcome--price; four accepted sample restrictions; strict delivered-end-use and upstream-price measurement; fixed-2006, strict-sample, and annual IOU context; omit both floor controls; no floor-series endpoint extrapolation.
& Regression and design details: Tables~\ref{tab:measurement_distribution}, \ref{tab:measurement_matched}, \ref{tab:measurement_no_ff_matched}, \ref{tab:subset_robust}, \ref{tab:sfa_total_timing_composition}, \ref{tab:iou_context}, and \ref{tab:floor_sensitivity}. Component accounts: Tables~\ref{tab:sunburst_distribution_contributions}, \ref{tab:sunburst_total_timing_contributions}, \ref{tab:sunburst_total_composition_contributions}, and \ref{tab:sunburst_matched_outcomes_contributions}. \\
\addlinespace
Direct primary
& 7
& Main; common-sample current and lags 1--3; five domains; omit fossil-electricity generation.
& Tables~\ref{tab:sfa_primary_current}, \ref{tab:sfa_primary_timing_composition}, \ref{tab:measurement_no_ff_matched}, \ref{tab:lmg_primary_frontier_variants}, \ref{tab:lmg_primary_gap_variants}, \ref{tab:sunburst_primary_timing_contributions}, and \ref{tab:sunburst_primary_composition_contributions}. \\
\addlinespace
Retail electricity
& 7
& Main; common-sample current and lags 1--3; five domains; omit fossil-electricity generation.
& Tables~\ref{tab:sfa_electricity_current}, \ref{tab:sfa_electricity_timing_composition}, \ref{tab:measurement_no_ff_matched}, \ref{tab:lmg_electricity_frontier_variants}, \ref{tab:lmg_electricity_gap_variants}, \ref{tab:sunburst_electricity_timing_contributions}, and \ref{tab:sunburst_electricity_composition_contributions}. \\
\bottomrule
\end{tabularx}
\begin{tablenotes}
\footnotesize
\item \textit{Notes:} $m$ is the number of unique accepted specifications.
The unavailable 2006--2013 boundary fit,
canonical Mundlak models, and Random Forest method variants are outside this
set of specifications. For the summary figures, every model is re-expressed under the same
frontier-plus-generated-composite-gap accounting. A main-model driver enters
as zero only when that specification deliberately omits it and the remaining
decomposition has been recomputed. Added context variables that do not belong
to the common 11-driver set are excluded from the rankings.
\end{tablenotes}
\end{threeparttable}
\end{table}

\begin{figure}[H]
  \centering
  \caption{Rank stability of total contributions: direct primary energy}
  \includegraphics[width=0.98\textwidth]{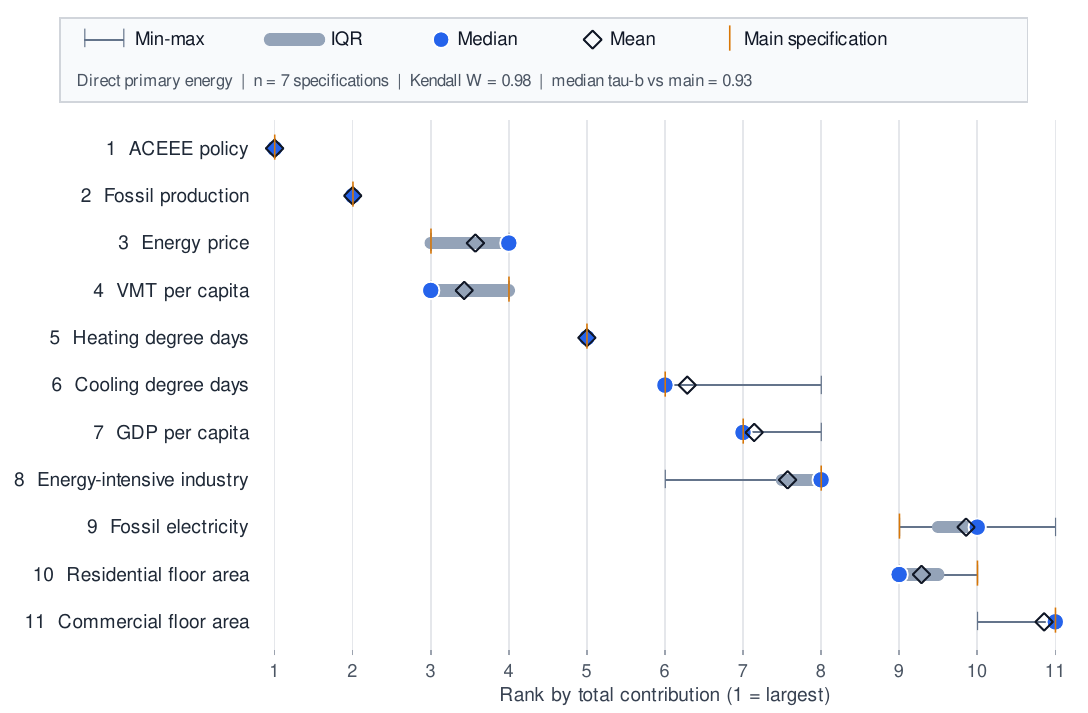}
  \label{fig:robustness_rank_primary}
  \vspace{-0.4em}
  \begin{minipage}{0.98\textwidth}
  \footnotesize
  \textit{Notes:} Rank 1 is the largest combined frontier-plus-generated-
  composite-gap contribution. The thin capped line shows the minimum--maximum,
  the thick line the IQR, the filled circle the median, the open diamond the
  mean, and the gold tick the main specification; exact ties use midranks.  Table~\ref{tab:robustness_stability_scope}
  inventories the set of specifications and points to Tables~\ref{tab:sfa_primary_current},
  \ref{tab:sfa_primary_timing_composition},
  \ref{tab:lmg_primary_frontier_variants}--\ref{tab:lmg_primary_gap_variants},
  and \ref{tab:sunburst_primary_timing_contributions}--
  \ref{tab:sunburst_primary_composition_contributions}.
  \end{minipage}
\end{figure}

\begin{figure}[H]
  \centering
  \caption{Stability of total contribution shares: direct primary energy}
  \includegraphics[width=0.98\textwidth]{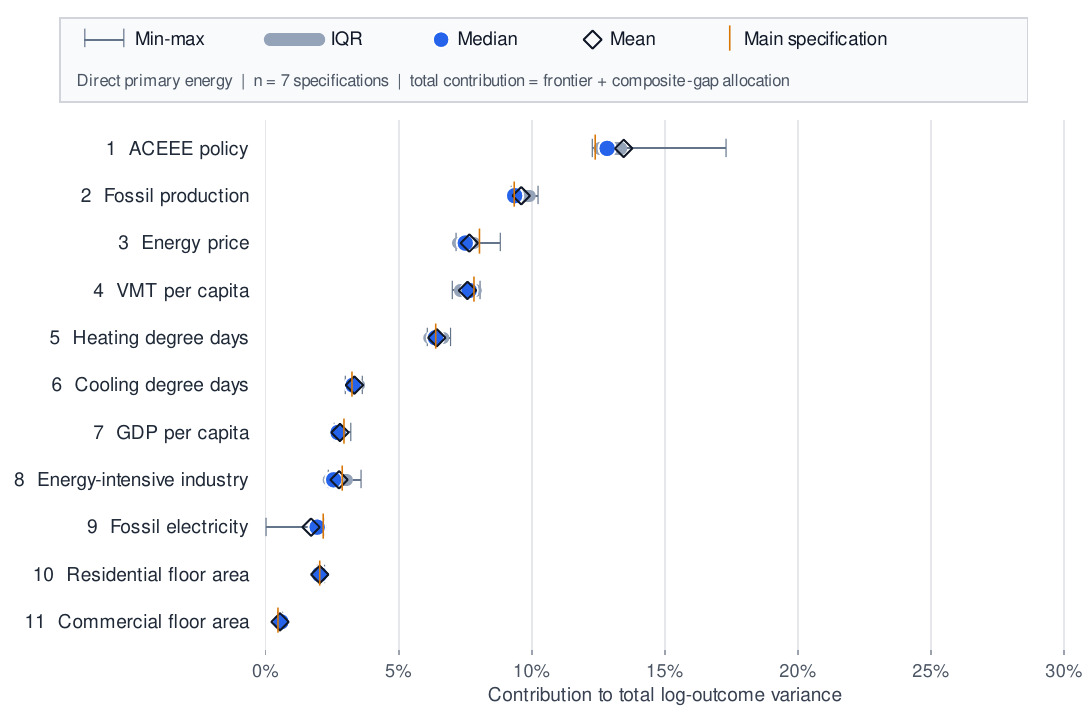}
  \label{fig:robustness_contribution_primary}
  \vspace{-0.4em}
  \begin{minipage}{0.98\textwidth}
  \footnotesize
  \textit{Notes:} Marks and the seven-specification battery are defined in
  Figure~\ref{fig:robustness_rank_primary}. A contribution is the sum of a
  driver's fitted-frontier and generated-composite-gap allocations, in
  percentage points of each model's own $\Var(\log q)$. The common scale does
  not imply a common absolute variance across outcomes. The IQR and full range
  describe designed specification dispersion, not sampling uncertainty or a
  confidence interval. Tables~\ref{tab:lmg_primary_frontier_variants}--
  \ref{tab:lmg_primary_gap_variants} report within-parent allocations, while
  Tables~\ref{tab:sunburst_primary_timing_contributions}--
  \ref{tab:sunburst_primary_composition_contributions} report the corresponding
  total-variance component accounts; Table~\ref{tab:robustness_stability_scope}
  links the remaining model details.
  \end{minipage}
\end{figure}

For direct primary energy, the rank ordering is particularly stable. ACEEE
policy is rank 1 and fossil production rank 2 in all seven specifications.
Their median total contributions are 12.83 and 9.35 percentage points,
respectively. Price and VMT exchange ranks 3 and 4, but their contribution IQRs
remain narrow at 7.23--7.82 and 7.31--7.89 points. The figures therefore show
high agreement about the leading direct-primary contributors without implying
that their coefficients or causal effects are stable.

\begin{figure}[H]
  \centering
  \caption{Rank stability of total contributions: retail electricity}
  \includegraphics[width=0.98\textwidth]{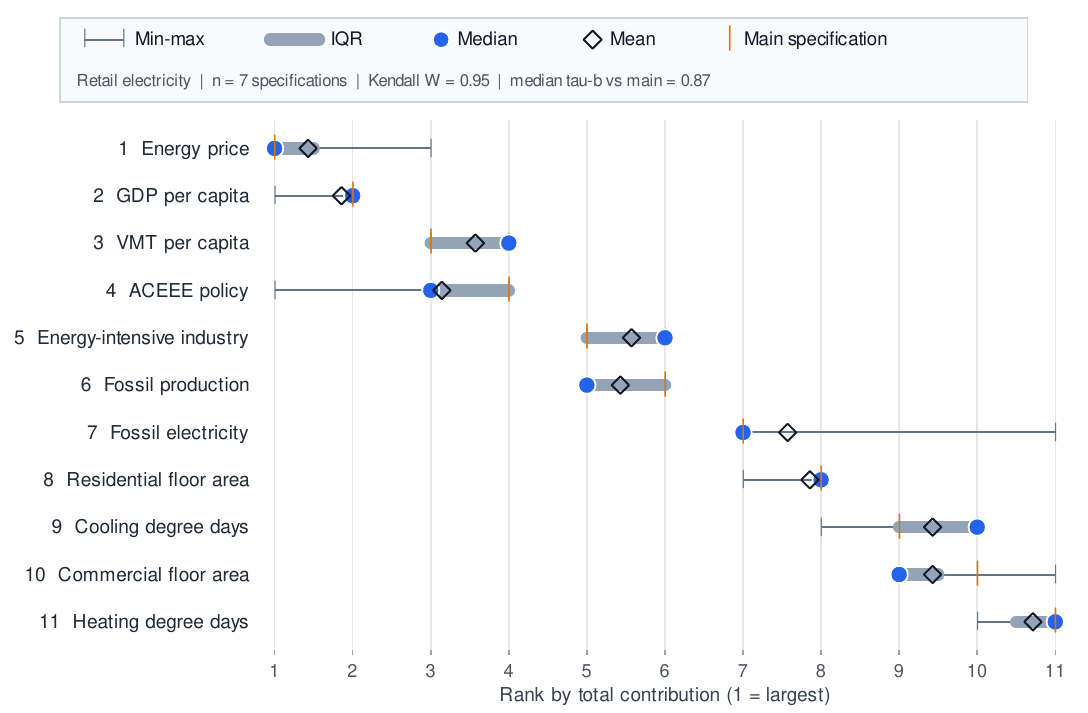}
  \label{fig:robustness_rank_retail}
  \vspace{-0.4em}
  \begin{minipage}{0.98\textwidth}
  \footnotesize
  \textit{Notes:} Rank 1 is the largest combined frontier-plus-generated-
  composite-gap contribution. The thin capped line shows the minimum--maximum,
  the thick line the IQR, the filled circle the median, the open diamond the
  mean, and the gold tick the main specification; exact ties use midranks. Table~\ref{tab:robustness_stability_scope}
  inventories the set of specs and points to Tables~
  \ref{tab:sfa_electricity_current},
  \ref{tab:sfa_electricity_timing_composition},
  \ref{tab:lmg_electricity_frontier_variants}--
  \ref{tab:lmg_electricity_gap_variants}, and
  \ref{tab:sunburst_electricity_timing_contributions}--
  \ref{tab:sunburst_electricity_composition_contributions}.
  \end{minipage}
\end{figure}

\begin{figure}[H]
  \centering
  \caption{Stability of total contribution shares: retail electricity}
  \includegraphics[width=0.98\textwidth]{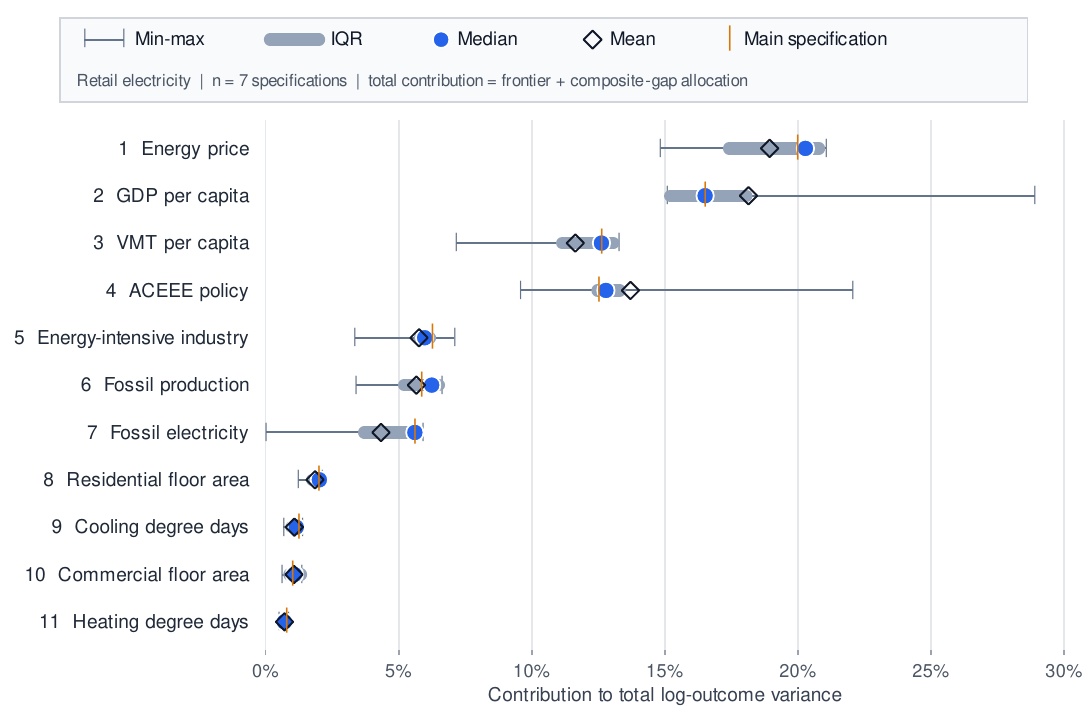}
  \label{fig:robustness_contribution_retail}
  \vspace{-0.4em}
  \begin{minipage}{0.98\textwidth}
  \footnotesize
  \textit{Notes:} Marks and the seven-specification battery are defined in
  Figure~\ref{fig:robustness_rank_retail}. A contribution is the sum of a
  driver's fitted-frontier and generated-composite-gap allocations, in
  percentage points of each model's own $\Var(\log q)$. The common scale does
  not imply a common absolute variance across outcomes. The IQR and full range
  describe designed specification dispersion, not sampling uncertainty or a
  confidence interval. Tables~\ref{tab:lmg_electricity_frontier_variants}--
  \ref{tab:lmg_electricity_gap_variants} report within-parent allocations,
  while Tables~\ref{tab:sunburst_electricity_timing_contributions}--
  \ref{tab:sunburst_electricity_composition_contributions} report the
  corresponding total-variance component accounts; Table~
  \ref{tab:robustness_stability_scope} links the remaining model details.
  \end{minipage}
\end{figure}

Retail electricity also has strong overall rank agreement, but the leading
pair is less invariant in magnitude. Price is rank 1 in the main model and has
median rank 1; GDP is rank 2 in the main model and has median rank 2. Their
median contributions are 20.28 and 16.51 percentage points, but GDP reaches
28.89 points in one specification. VMT and ACEEE typically occupy ranks 3 and
4 (median ranks 4 and 3, respectively), with median contributions of 12.61 and
12.78 points; ACEEE rises to rank 1 in one specification. The high concordance
therefore coexists with meaningful specification sensitivity in the size of
some outer contribution shares.

\clearpage
% Generated by review/feasibility/eia-components/expanded-tables/

% rf-presentation/build_rf_appendix_tables.py. Do not hand-edit.

\subsection{Random Forest relative-importance sensitivities}

\label{sec:rf_variants}

We extend the frontier-only Random Forest diagnostic to the same fifteen pooled SFA formulations used in the coefficient and LMG robustness checks. In every case the predictive target is the frontier and the inputs are predictors; the intercept and year indicators are partialled out. We report signed marginal permutation importance only.

The calculation uses the direct legacy \texttt{party}~1.3-21 implementation retained by the preferred-model diagnostic. For every formulation, three fixed-seed forests (101, 202, and 303) use 500 trees and $mtry=3$. The raw score is the all-tree mean increase in per-tree out-of-bag squared error, with zero assigned when a predictor is absent from a tree. Conditional permutation importance is not rerun in this expanded set of checks.

To make rankings legible despite outcome-specific raw-loss scales, each table reports the fixed-seed mean raw score after scaling the largest positive value within that model to 100. Negative scores remain negative. This normalization is purely an index transformation: it does not turn Random Forest importance into percentages, additive shares, or an $R^2$ decomposition. It also does not make raw predictive-loss magnitudes comparable across models. The five ACEEE domains are retained separately throughout and are never summed.

\subsubsection{Primary-equivalent total energy}

\begin{table}[H]
\centering
\caption{Random Forest signed marginal importance index: primary-equivalent total energy}
\label{tab:rf_total_variants}
\begin{threeparttable}
\scriptsize
\setlength{\tabcolsep}{4pt}
\begin{tabularx}{\textwidth}{@{}>{\raggedright\arraybackslash}p{.30\textwidth}*{5}{Y}@{}}
\toprule
Predictor & \shortstack{Current\\score} & \shortstack{Lag\\1} & \shortstack{Lag\\2} & \shortstack{Lag\\3} & \shortstack{Five\\domains} \\
\midrule
Price & 100.0 & 100.0 & 100.0 & 100.0 & 100.0 \\
GDP/pc & 8.0 & 7.7 & 7.6 & 8.5 & \rev{9.0} \\
VMT/pc & 43.4 & 41.6 & 44.0 & 43.7 & \rev{46.5} \\
ComSurf/pc & 4.8 & 3.8 & 3.9 & 3.8 & \rev{6.0} \\
ResSurf/pc & 9.6 & 10.0 & 10.2 & 9.7 & \rev{10.0} \\
E-Int Ind & 23.3 & 22.1 & 23.8 & 23.0 & \rev{24.3} \\
CDD & 3.1 & 3.3 & 3.7 & 3.4 & \rev{4.5} \\
HDD & 7.2 & 6.3 & 7.2 & 6.5 & \rev{7.0} \\
ACEEE total score & 32.1 & 33.2 & 32.3 & 30.9 & -- \\
ACEEE: utilities/public benefits & -- & -- & -- & -- & \rev{8.3} \\
ACEEE: transportation & -- & -- & -- & -- & \rev{29.1} \\
ACEEE: building energy codes & -- & -- & -- & -- & \rev{13.9} \\
ACEEE: state initiatives & -- & -- & -- & -- & \rev{11.3} \\
ACEEE: appliance standards & -- & -- & -- & -- & \rev{3.0} \\
FF/pc & 36.5 & 36.6 & 38.9 & 36.2 & \rev{38.1} \\
FFelec/pc & 33.2 & 36.9 & 37.1 & 39.7 & \rev{34.2} \\
\bottomrule
\end{tabularx}
\begin{tablenotes}
\footnotesize
\item \textit{Notes:} The SFA outcome is log primary-equivalent total energy per capita; Price is the centered log aggregate end-use energy price. Each entry is the fixed-seed mean signed marginal permutation importance, divided by that model's largest positive seed-mean importance and multiplied by 100. The leading positive predictor equals 100 within each column. The entries are indices, not percentages. 
\end{tablenotes}
\end{threeparttable}
\end{table}

\subsubsection{Retail electricity}

\begin{table}[H]
\centering
\caption{Random Forest signed marginal importance index: retail electricity}
\label{tab:rf_electricity_variants}
\begin{threeparttable}
\scriptsize
\setlength{\tabcolsep}{4pt}
\begin{tabularx}{\textwidth}{@{}>{\raggedright\arraybackslash}p{.30\textwidth}*{5}{Y}@{}}
\toprule
Predictor & \shortstack{Current\\score} & \shortstack{Lag\\1} & \shortstack{Lag\\2} & \shortstack{Lag\\3} & \shortstack{Five\\domains} \\
\midrule
Price & 100.0 & 100.0 & 100.0 & 100.0 & 100.0 \\
GDP/pc & 10.7 & 7.0 & 8.0 & 8.9 & \rev{18.2} \\
VMT/pc & 58.9 & 55.8 & 54.5 & 58.6 & \rev{76.1} \\
ComSurf/pc & 1.7 & 1.4 & 1.5 & 1.6 & \rev{2.1} \\
ResSurf/pc & 9.5 & 9.3 & 9.2 & 9.4 & \rev{12.1} \\
E-Int Ind & 19.1 & 16.6 & 16.7 & 18.6 & \rev{29.1} \\
CDD & 6.3 & 5.6 & 6.0 & 6.1 & \rev{8.4} \\
HDD & 4.2 & 3.7 & 3.6 & 3.8 & \rev{8.2} \\
ACEEE total score & 45.4 & 43.0 & 43.4 & 42.6 & -- \\
ACEEE: utilities/public benefits & -- & -- & -- & -- & \rev{36.1} \\
ACEEE: transportation & -- & -- & -- & -- & \rev{31.5} \\
ACEEE: building energy codes & -- & -- & -- & -- & \rev{5.7} \\
ACEEE: state initiatives & -- & -- & -- & -- & \rev{5.0} \\
ACEEE: appliance standards & -- & -- & -- & -- & \rev{5.4} \\
FF/pc & 19.0 & 18.1 & 17.3 & 20.0 & \rev{21.3} \\
FFelec/pc & 37.2 & 34.2 & 38.4 & 41.0 & \rev{46.4} \\
\bottomrule
\end{tabularx}
\begin{tablenotes}
\footnotesize
\item \textit{Notes:} The SFA outcome is $\log(ESTCB/TPOPP)$; Price is centered $\log(ESTCD)$. Each entry is the fixed-seed mean signed marginal permutation importance, divided by that model's largest positive seed-mean importance and multiplied by 100.
\end{tablenotes}
\end{threeparttable}
\end{table}

\subsubsection{Direct primary energy excluding retail electricity}

\begin{table}[H]
\centering
\caption{Random Forest signed marginal importance index: direct primary energy excluding retail electricity}
\label{tab:rf_primary_variants}
\begin{threeparttable}
\scriptsize
\setlength{\tabcolsep}{4pt}
\begin{tabularx}{\textwidth}{@{}>{\raggedright\arraybackslash}p{.30\textwidth}*{5}{Y}@{}}
\toprule
Predictor & \shortstack{Current\\score} & \shortstack{Lag\\1} & \shortstack{Lag\\2} & \shortstack{Lag\\3} & \shortstack{Five\\domains} \\
\midrule
Price & 94.9 & 75.7 & \rev{73.4} & 79.0 & \rev{95.3} \\
GDP/pc & 25.7 & 20.6 & 20.3 & 20.7 & \rev{29.2} \\
VMT/pc & 100.0 & 100.0 & 100.0 & 100.0 & 100.0 \\
ComSurf/pc & 2.4 & 2.7 & 2.9 & 2.8 & \rev{4.2} \\
ResSurf/pc & 14.7 & 14.0 & 13.8 & 13.9 & \rev{17.7} \\
E-Int Ind & 44.2 & 37.4 & \rev{37.2} & 39.9 & \rev{53.3} \\
CDD & 30.2 & 31.9 & 31.0 & 31.3 & \rev{37.3} \\
HDD & 79.1 & 76.9 & \rev{73.2} & 74.0 & \rev{89.2} \\
ACEEE total score & 18.5 & 16.3 & 16.7 & 16.1 & -- \\
ACEEE: utilities/public benefits & -- & -- & -- & -- & \rev{7.1} \\
ACEEE: transportation & -- & -- & -- & -- & \rev{32.9} \\
ACEEE: building energy codes & -- & -- & -- & -- & \rev{15.5} \\
ACEEE: state initiatives & -- & -- & -- & -- & \rev{17.7} \\
ACEEE: appliance standards & -- & -- & -- & -- & \rev{3.1} \\
FF/pc & 91.9 & 91.9 & \rev{89.0} & 90.8 & \rev{95.1} \\
FFelec/pc & 44.8 & 34.3 & 34.9 & 37.8 & \rev{51.4} \\
\bottomrule
\end{tabularx}
\begin{tablenotes}
\footnotesize
\item \textit{Notes:} The SFA outcome is $\log((TNTCB-ESTCB)/TPOPP)$; Price is centered $\log(PETXD)$. Each entry is the fixed-seed mean signed marginal permutation importance, divided by that model's largest positive seed-mean importance and multiplied by 100.
\end{tablenotes}
\end{threeparttable}
\end{table}

\subsubsection{Individual ACEEE domains}

\begin{table}[H]
\centering
\caption{Five-domain ACEEE Random Forest importance indices across energy formulations}
\label{tab:rf_domain_detail_variants}
\begin{threeparttable}
\scriptsize
\setlength{\tabcolsep}{5pt}
\begin{tabularx}{\textwidth}{@{}>{\raggedright\arraybackslash}p{.38\textwidth}*{3}{Y}@{}}
\toprule
ACEEE domain & Total energy & Retail electricity & Direct primary \\
\midrule
Utilities/public benefits & \rev{8.3} & \rev{36.1} & \rev{7.1} \\
Transportation & \rev{29.1} & \rev{31.5} & \rev{32.9} \\
Building energy codes & \rev{13.9} & \rev{5.7} & \rev{15.5} \\
State initiatives & \rev{11.3} & \rev{5.0} & \rev{17.7} \\
Appliance standards & \rev{3.0} & \rev{5.4} & \rev{3.1} \\
\bottomrule
\end{tabularx}
\begin{tablenotes}
\footnotesize
\item \textit{Notes:} These are the same signed leader-scaled indices shown in the five-domain columns of Tables~\ref{tab:rf_total_variants}--\ref{tab:rf_primary_variants}. Scaling is performed separately within each full model, against that model's leading positive predictor. Values are not percentages or additive shares.
\end{tablenotes}
\end{threeparttable}
\end{table}

\subsubsection{Ranking stability across formulations}

For primary-equivalent total energy, pairwise correlations of the complete 11-predictor ranking across the current and three lag specifications have median 0.991 and range from 0.991 to 1.000. Price leads all four timing models, while the ACEEE-total index ranges from 30.9 to 33.2. When the five-domain model is added, comparisons are deliberately restricted to the ten unchanged non-policy predictors rather than summing its domains; those pairwise rank correlations have median 0.988 (range 0.988--1.000).

For retail electricity, pairwise correlations of the complete 11-predictor ranking across the current and three lag specifications have median 0.991 and range from 0.982 to 1.000. Price leads all four timing models, while the ACEEE-total index ranges from 42.6 to 45.4. When the five-domain model is added, comparisons are deliberately restricted to the ten unchanged non-policy predictors rather than summing its domains; those pairwise rank correlations have median 0.976 (range 0.976--1.000).

For direct primary energy excluding retail electricity, pairwise correlations of the complete 11-predictor ranking across the current and three lag specifications have median 0.986 and range from 0.964 to 1.000. VMT/pc leads all four timing models, while the ACEEE-total index ranges from 16.1 to 18.5. When the five-domain model is added, comparisons are deliberately restricted to the ten unchanged non-policy predictors rather than summing its domains; those pairwise rank correlations have median 0.988 (range 0.952--1.000).

Domain rankings also vary with the energy outcome: Transportation is the highest-ranked domain for primary-equivalent total energy (\rev{29.1}); Utilities and public benefits is the highest-ranked domain for retail electricity (\rev{36.1}); Transportation is the highest-ranked domain for direct primary energy excluding retail electricity (\rev{32.9}). Because each index is scaled against the leading predictor in its own full model, these values do not compare raw predictive-loss magnitudes across outcomes.

\subsection{Investor-owned utilities}

Table~\ref{tab:iou_context} pairs the full fixed-2006 and strict-annual
same-row samples. IOU is the share of identified investor-owned retail-
provider customer accounts.

\begin{table}[H]
\centering
\caption{Paired IOU provider-account context}\label{tab:iou_context}
\small
\begin{tabular}{llrrrrr}
\toprule
Panel & ACEEE term & Baseline & SE & + IOU & SE & $N$\\
\midrule
A: fixed 2006 & Frontier & $-0.0017$ & 0.0007 & $-0.0020$ & 0.0007 & 867\\
 & Gap log variance & $-0.1555$ & 0.0096 & $-0.1468$ & 0.0102 & 867\\
B: strict annual & Frontier & $-0.0024$ & 0.0008 & $-0.0029$ & 0.0008 & 755\\
 & Gap log variance & $-0.1433$ & 0.0105 & $-0.1312$ & 0.0112 & 755\\
\bottomrule
\end{tabular}
\end{table}

\subsection{Floor construction sensitivity}

Table~\ref{tab:floor_sensitivity} provides the compact comparison, while
Table~\ref{tab:mundlak_no_surfaces} reports the complete pooled and Mundlak
results obtained after omitting both floor-area controls.

\begin{table}[H]
\centering
\caption{Floor construction sensitivity}\label{tab:floor_sensitivity}
\small
\resizebox{\linewidth}{!}{%
\begin{tabular}{lrrrr}
\toprule
Specification & Price frontier & ACEEE frontier & $p$ & ACEEE gap\\
\midrule
Preferred annual-population floors & $-0.7524$ & $-0.0017$ & 0.0096 & $-0.1555$\\
Drop both floor controls & $-0.7457$ & $-0.0017$ & 0.0041 & $-0.1548$\\
No endpoint extrapolation & $-0.7489$ & $-0.0017$ & 0.0100 & $-0.1547$\\
\bottomrule
\end{tabular}
}
\end{table}

The ACEEE frontier and gap associations are stable across the drop-both and
no-endpoint stresses: all three frontier coefficients are negative and
conventionally distinguishable, and the gap coefficients differ only slightly.
The preferred annual-population construction remains a design choice, not a
likelihood-selected specification. In the preferred model, ComSurf and ResSurf
are individually imprecise; in the no-endpoint model they remain imprecise,
with ComSurf $+0.0129$ (SE $0.0311$, $p=.678$) and ResSurf $+0.0088$ (SE
$0.0538$, $p=.870$). The canonical Mundlak floor rows are likewise imprecise:
ComSurf within $-0.0013$ ($p=.963$), total between $-0.0060$ ($p=.926$);
ResSurf within $+0.0039$ ($p=.987$), total between $-0.0133$ ($p=.825$).
These checks do not identify causal floor-area effects.

\begin{table}[H]
\centering
\caption{Pooled and Mundlak results after omitting both floor-area controls}
\label{tab:mundlak_no_surfaces}
\begin{threeparttable}
{\setstretch{1}\small
\begin{tabular}{lccc}
\toprule
& \multicolumn{1}{c}{Pooled }
& \multicolumn{2}{c}{Mundlak } \\
\cmidrule(lr){2-2}\cmidrule(lr){3-4}
& Estimate
& Within $\beta$
& Total between $\beta+\delta$ \\
\midrule

\multicolumn{4}{l}{\textit{Frontier equation}} \\[4pt]

\textit{Price}
& ${-}0.7457^{***}$
& ${-}0.1733^{**}$
& ${-}0.7844^{***}$ \\
& (0.0412) & (0.0836) & (0.0379) \\[2pt]

\textit{GDP/pc}
& $0.1894^{***}$
& $0.2269^{***}$
& $0.2432^{***}$ \\
& (0.0314) & (0.0809) & (0.0303) \\[2pt]

\textit{Eff}
& ${-}0.0017^{***}$
& $0.0043^{***}$
& ${-}0.0025^{***}$ \\
& (0.0006) & (0.0010) & (0.0007) \\[2pt]

\textit{VMT/pc}
& $0.3951^{***}$
& $0.1337$
& $0.3842^{***}$ \\
& (0.0321) & (0.0842) & (0.0319) \\[2pt]

\textit{E-Int Ind}
& $0.1074^{***}$
& $0.0038$
& $0.1064^{***}$ \\
& (0.0095) & (0.0271) & (0.0093) \\[2pt]

\textit{FF/pc}
& $0.0365^{***}$
& ${-}0.0103$
& $0.0454^{***}$ \\
& (0.0022) & (0.0074) & (0.0022) \\[2pt]

\textit{FFelec/pc}
& $0.0095^{***}$
& $0.0456^{***}$
& $0.0029$ \\
& (0.0030) & (0.0159) & (0.0028) \\[2pt]

\textit{HDD}
& $0.0456^{***}$
& $0.0345^{***}$
& $0.0437^{***}$ \\
& (0.0032) & (0.0105) & (0.0031) \\[2pt]

\textit{CDD}
& $0.0991^{***}$
& $0.0286$
& $0.1004^{***}$ \\
& (0.0083) & (0.0277) & (0.0082) \\[3pt]

\midrule
\multicolumn{2}{l}{\textit{Composite-gap log-variance equation}}
& \multicolumn{2}{c}{\textit{Mundlak-model estimate}} \\[4pt]

\textit{Eff}
& ${-}0.1548^{***}$
& \multicolumn{2}{c}{${-}0.1595^{***}$} \\
& (0.0095)
& \multicolumn{2}{c}{(0.0090)} \\[2pt]

\textit{Price}
& $0.2920$
& \multicolumn{2}{c}{$0.0649$} \\
& (0.3127)
& \multicolumn{2}{c}{(0.2941)} \\[2pt]

\textit{GDP/pc}
& $2.6346^{***}$
& \multicolumn{2}{c}{$2.4520^{***}$} \\
& (0.2745)
& \multicolumn{2}{c}{(0.2745)} \\[3pt]

\midrule
$N$
& 867
& \multicolumn{2}{c}{867} \\

\bottomrule
\end{tabular}

\begin{tablenotes}
\footnotesize

\item \textit{Notes:}
This sensitivity omits ComSurf/pc and ResSurf/pc from both frontiers and
omits their two state means from the Mundlak frontier. All other terms and
the fixed $N=867$ sample are unchanged. The specification retaining both
floor controls remains the main specification.
\end{tablenotes}}
\end{threeparttable}
\end{table}

\subsection{Climate variables}

The table below keeps two related but non-nested estimands separate. Panel A is LMDI change accounting based on a preferred state-and-year-effect climate normalization; Panel B is a pooled SFA conditional log-frontier association that retains within- and between-state information. No generated LMDI component enters the SFA, and neither panel structurally decomposes the other.

\begin{table}[H]
\centering
\caption{Climate accounting inputs, conditional LMDI ranges, and SFA climate associations}
\label{tab:climate_bridge}
\begin{threeparttable}
\scriptsize
\resizebox{\linewidth}{!}{%
\begin{tabular}{lccc}
\toprule
\multicolumn{4}{l}{\textit{Panel A. LMDI climate normalization and conditional allocation}} \\[2pt]
\multicolumn{4}{l}{\textit{First-stage slopes}} \\
Sector and weather & Estimate & State-clustered HC1 SE & \\
\midrule
Residential HDD & 0.1197 & 0.0614 & \\
Residential CDD & 0.0135 & 0.0222 & \\
Commercial HDD & 0.0721 & 0.0318 & \\
Commercial CDD & 0.0205 & 0.0295 & \\
\midrule
\multicolumn{4}{l}{\textit{National 2006--2022 allocation (percent)}} \\
Factor & Point & \multicolumn{2}{c}{Conditional 95\% range} \\
Climate & $+0.271$ & \multicolumn{2}{c}{$[+0.024,+0.515]$} \\
Intensity & $-15.304$ & \multicolumn{2}{c}{$[-15.509,-15.095]$} \\
\midrule
\multicolumn{4}{l}{\textit{Building-sector allocation (percent)}} \\
Sector and factor & Point & \multicolumn{2}{c}{Conditional 95\% range} \\
Residential Climate & $+0.160$ & \multicolumn{2}{c}{$[-0.005,+0.329]$} \\
Residential Intensity & $-3.360$ & \multicolumn{2}{c}{$[-3.523,-3.200]$} \\
Commercial Climate & $+0.111$ & \multicolumn{2}{c}{$[-0.044,+0.269]$} \\
Commercial Intensity & $-5.882$ & \multicolumn{2}{c}{$[-6.031,-5.736]$} \\
\midrule
\multicolumn{4}{l}{\textit{Panel B. SFA conditional log-frontier associations}} \\[2pt]
Weather & Pooled SFA & Mundlak within & Mundlak total between \\
\midrule
HDD & 0.0456 (0.0044) & 0.0345 (0.0110) & 0.0447 (0.0062) \\
CDD & 0.1007 (0.0090) & 0.0285 (0.0286) & 0.1013 (0.0132) \\
\bottomrule
\end{tabular}
}
\begin{tablenotes}
\footnotesize
\item \textit{Notes:} Panel A estimates the residential and commercial first stages on 51 jurisdictions over 2006--2022. The national Climate allocations under pooled, year-effect, and preferred two-way-effect are $+0.338\%$, $+0.379\%$, and $+0.271\%$, respectively. Panel B reports standard errors in parentheses. State means account for 97.09\% of HDD variation and 97.87\% of CDD variation in the panel.
\end{tablenotes}
\end{threeparttable}
\end{table}

\section{Policy association, scale, and state-context diagnostics}
\label{sec:policy_diagnostics}

The policy discussion uses the pooled half-normal model together
with alternative specifications.

\begin{table}[H]
\centering
\caption{Model-implied cross-state association and scale illustrations}
\label{tab:policy_illustrations_appendix}
\small
\resizebox{\textwidth}{!}{%
\begin{tabular}{lrrrr}
\toprule
Illustration & Jurisd. & Reduction (\%) & MMBtu/person & Quads \\
\midrule
10\% higher aggregate price & 51 & 6.920 [4.068, 10.780] & 19.656 & 6.551 \\
12 low-price jurisdictions to 2022 median & 12 & 13.567 [8.179, 20.435] & 58.008 & 4.628 \\
10-point higher \rev{harmonized} ACEEE score & 51 & 1.685 [$-2.463$, 7.897] & 4.787 & 1.595 \\
Other 46 to 2022 high-five ACEEE mean & 46 & 4.427 [$-6.740$, 19.503] & 13.783 & 3.664 \\
Half of fixed-2022 expected composite gap & 51 & 10.769 [1.195, 13.829] & 30.588 & 10.194 \\
\bottomrule
\end{tabular}%
}
\begin{minipage}{\textwidth}
\footnotesize
\textit{Notes:} Brackets are 95\% state-block percentile intervals from 500 reps, with the 2022 groups and evaluation grid fixed. Every percentage uses the exact exponential mapping and observed population $\times$ energy-use weights. MMBtu/person uses the affected-set population; quads are physical MMBtu divided by $10^9$. The first four rows are frontier conditional associations; the last is a mechanical scale for half of $E[u\mid Z]=\sigma_u(Z)\sqrt{2/\pi}$ on the fixed 2022 grid.
\end{minipage}
\end{table}

\begin{table}[H]
\centering
\caption{U.S. state contexts}
\label{tab:state_contexts_appendix}
\begin{threeparttable}
\scriptsize
\begin{tabularx}{\textwidth}{p{3.0cm}p{4.0cm}Xp{3.5cm}}
\toprule
Context & Released evidence & Diagnostic use & Boundary \\
\midrule
Resource extraction and energy-intensive industry & Fossil/extraction sensitivity set: WY, ND, WV, AK, NM; industry sensitivity set: LA, IN, NC, IA, KY. & Separate extraction and process activity from demand-side interpretations before reading price or gap comparisons. & Sensitivity sets are not heterogeneous effects or state types. \\
Low-price or weaker policy environments & The derived 2022 low-price set is LA, ND, IA, TX, WY, IN, NE, OH, AL, SD, MI, OK; ACEEE remains continuous. & Review price composition, score domains, implementation constraints, affordability, and distributional safeguards. & Neither price nor ACEEE is an identified policy dose; contexts overlap. \\
Higher income or larger modeled gaps & Income and the released composite-gap ranks remain continuous diagnostic fields. & Prompt a mechanism audit and collection of service, technology, allocation, and preference data. & A rank is not a quota, welfare loss, or mechanism estimate. \\
Transport-, building-, or industry-shaped demand & VMT, constructed floor-area, industry controls, and state LMDI entries identify dimensions for local diagnosis. & Diagnose sectors before considering instruments the aggregate model cannot select. & The model does not identify a sectoral intervention or treatment effect. \\
\bottomrule
\end{tabularx}
\begin{tablenotes}
\footnotesize
\item \textit{Notes:} Contexts can overlap and are not a 51-jurisdiction classification.
\end{tablenotes}
\end{threeparttable}
\end{table}

\begin{figure}[H]
  \centering
  \caption{Overlapping descriptive state contexts. Price and ACEEE remain continuous; fossil production and energy-intensive-industry share provide two additional diagnostic dimensions.}
  \includegraphics[width=0.90\textwidth]{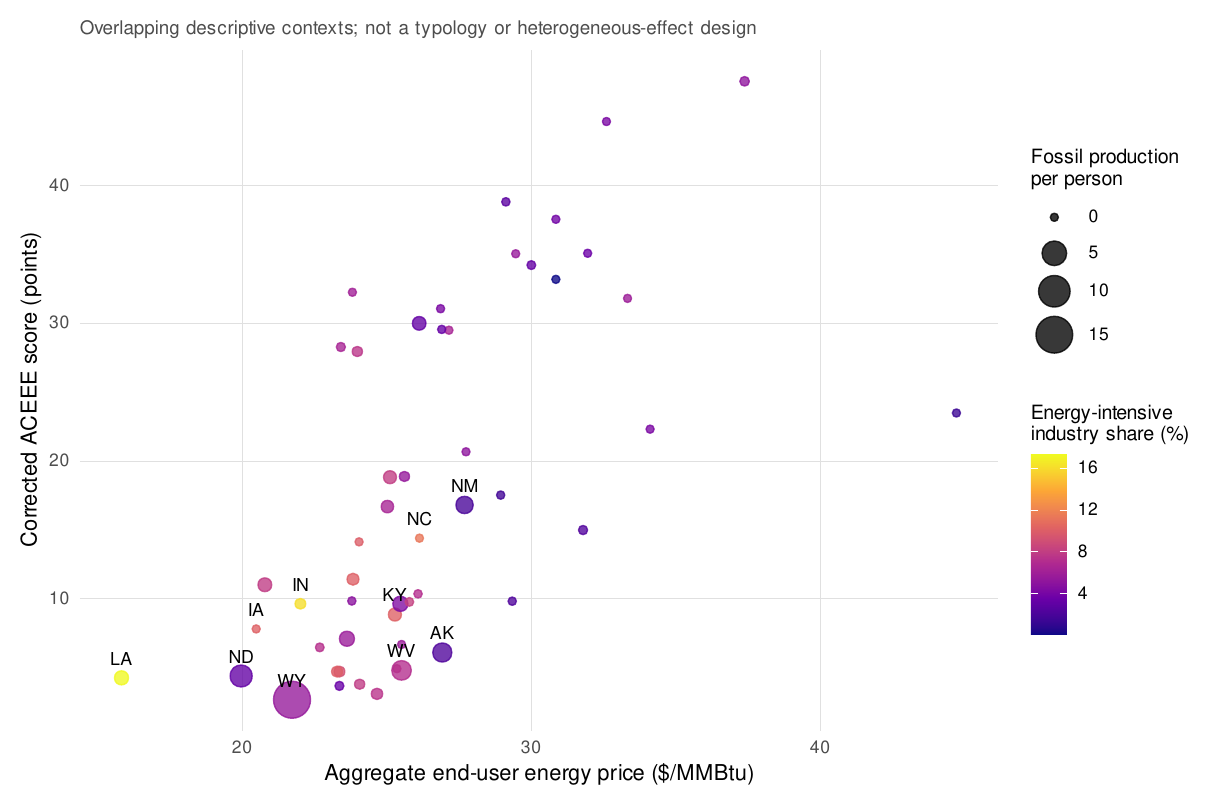}
  \label{fig:state_context_diagnostic}
\end{figure}

\subsection{Result-specific policy-coefficient sensitivities}

Tables~\ref{tab:price_policy_sensitivity}--\ref{tab:aceee_gap_policy_sensitivity}
summarize released coefficients without pooling different estimands or
specifications into one uncertainty interval. Parentheses contain standard errors. The Mundlak within slope is
$\beta$ and total between is $\beta+\delta$.

\begin{table}[H]
\centering
\caption{Price--frontier coefficients across specifications}
\label{tab:price_policy_sensitivity}
\begin{threeparttable}
\scriptsize
\begin{tabularx}{\textwidth}{lrX}
\toprule
Specification & Estimate (SE) & Estimand boundary \\
\midrule
Preferred pooled half-normal & $-0.7524$ (0.0459) & Pooled conditional association; not causal. \\
Mundlak total between & $-0.7793$ (0.0466) & Total between is $\beta+\delta$. \\
Mundlak within & $-0.1708$ (0.0853) & Within slope; not pooled or total between. \\
Strict TNTCB/TETXD, FFelec omitted & $-0.7930$ (0.0536) & Different outcome/price and controls. \\
Upstream TETCB/PETCD & $-0.1673$ (0.0373) & Different outcome/price pairing. \\
Exponential & $-0.7578$ (0.0639) & Distribution-specific parameter. \\
Truncated normal & $-1.2238$ (0.1198) & Distribution-specific parameter. \\
Drop both floor controls & $-0.7457$ (0.0412) & Construction sensitivity. \\
No endpoint extrapolation & $-0.7489$ (0.0456) & Construction sensitivity. \\
\bottomrule
\end{tabularx}
\begin{tablenotes}
\footnotesize
\item \textit{Notes:} Measurement, distribution, and floor rows are different specifications or estimands.
\end{tablenotes}
\end{threeparttable}
\end{table}

\begin{table}[H]
\centering
\caption{ACEEE--frontier association across specifications}
\label{tab:aceee_frontier_policy_sensitivity}
\begin{threeparttable}
\scriptsize
\begin{tabularx}{\textwidth}{lrX}
\toprule
Specification & Estimate (SE) & Estimand boundary \\
\midrule
Preferred pooled half-normal & $-0.0017$ (0.0007) & Pooled conditional association; not causal. \\
Mundlak total between & $-0.0027$ (0.0010) & Total between is $\beta+\delta$. \\
Mundlak within & $0.0043$ (0.0011) & Within slope; sign differs from pooled and total between. \\
Strict TNTCB/TETXD, FFelec omitted & $+0.0015$ (0.0008) & Different outcome/price pairing and control set. \\
Upstream TETCB/PETCD & $-0.0041$ (0.0009) & Different outcome/price pairing. \\
Exponential & $-0.0037$ (0.0008) & Distribution-specific parameter. \\
Truncated normal & $-0.0108$ (0.0016) & Distribution-specific parameter. \\
Common-sample timing $t$ & $-0.0019$ (0.0008) & Timing sensitivity, not an endogeneity remedy. \\
Common-sample timing $t-1$ & $-0.0021$ (0.0008) & Timing sensitivity, not an endogeneity remedy. \\
Common-sample timing $t-2$ & $-0.0021$ (0.0008) & Timing sensitivity, not an endogeneity remedy. \\
Common-sample timing $t-3$ & $-0.0019$ (0.0008) & Timing sensitivity, not an endogeneity remedy. \\
Drop both floor controls & $-0.0017$ (0.0006) & Construction sensitivity; main specification retained. \\
No endpoint extrapolation & $-0.0017$ (0.0006) & Construction sensitivity; main specification retained. \\
\bottomrule
\end{tabularx}
\begin{tablenotes}
\footnotesize
\item \textit{Notes:} The within-state sign differs from the pooled and total-between signs.
\end{tablenotes}
\end{threeparttable}
\end{table}

\begin{table}[H]
\centering
\caption{ACEEE association with the modeled composite-gap scale}
\label{tab:aceee_gap_policy_sensitivity}
\begin{threeparttable}
\scriptsize
\begin{tabularx}{\textwidth}{lrX}
\toprule
Specification & Estimate (SE) & Estimand boundary \\
\midrule
Preferred pooled half-normal & $-0.1555$ (0.0096) & Pooled gap log-variance association; not causal. \\
Strict TNTCB/TETXD, FFelec omitted & $-0.1368$ (0.0081) & Different outcome/price pairing and control set. \\
Upstream TETCB/PETCD & $-0.1390$ (0.0098) & Different outcome/price pairing. \\
Exponential & $-0.1751$ (0.0158) & Distribution-specific gap parameter. \\
Truncated normal & $-0.1841$ (0.0102) & Distribution-specific gap parameter. \\
Common-sample timing $t$ & $-0.1643$ (0.0115) & Timing sensitivity, not an endogeneity remedy. \\
Common-sample timing $t-1$ & $-0.1598$ (0.0112) & Timing sensitivity, not an endogeneity remedy. \\
Common-sample timing $t-2$ & $-0.1570$ (0.0112) & Timing sensitivity, not an endogeneity remedy. \\
Common-sample timing $t-3$ & $-0.1566$ (0.0112) & Timing sensitivity, not an endogeneity remedy. \\
Drop both floor controls & $-0.1548$ (0.0095) & Construction sensitivity; main specification retained. \\
No endpoint extrapolation & $-0.1547$ (0.0096) & Construction sensitivity; main specification retained. \\
\bottomrule
\end{tabularx}
\begin{tablenotes}
\footnotesize
\item \textit{Notes:} Negative signs recur with meaningful magnitude variation. Each coefficient belongs to a model-specific gap log-scale or log-variance equation.
\end{tablenotes}
\end{threeparttable}
\end{table}

\clearpage

\section{Theoretical microfoundation}
\label{app:microfoundation}

We develop a parsimonious microfoundation for the economic framework. In this section, we show that the conditional equilibrium quantities of this model aggregate exactly to
$q_{st}=q_{st}^{*}I_{st}$ with
$q_{st}^{*}=S(Y_{st},X_{st})/a(P_{st},\textit{Eff}_{st},X_{st})$ and
$I_{st}\geq1$, and confirm the key comparative statics: the frontier rises with $Y$, falls with $P$ and
$\textit{Eff}$, the gap weakly falls with $P$ and conditionally with
$\textit{Eff}$, and its income response is ambiguous.  Taking logs and adding
the stated symmetric-noise and half-normal reduced-form assumptions then gives
the stochastic-frontier specification used in our analysis.

\subsection{Setup \& preferences}

A state contains a unit mass of atomistic energy-using decision units
$i\in[0,1]$, representing household, business, and public-service activities.
The unit-mass normalization expresses state aggregates.
Let $Y>0$ denote the period resource endowment of this normalized economy,
proxied empirically by real GDP per capita. Decision unit $i$ has the endowment $y_i=\omega_iY$,
with 
$\omega_i>0,$
$\int_0^1\omega_i\,di=1.$ Thus, $\int_0^1y_i\,di=Y$.

Following the characteristics approach of consumer theory, decision units
derive value from energy services rather than from energy inputs themselves.
Let $z_i\geq0$ denote the energy services \rev{realized} by unit $i$ and let $c_i$
denote a numeraire good. We adopt a Stone--Geary preference specification:
\begin{equation}
 U_i(c_i,z_i\mid y_i,X)
 =
 u_i(c_i)
 +\vartheta_i b_i
   \ln\!\left(1+\frac{z_i}{b_i}\right)
 -\lambda_i(b_i-z_i)_{+},
 \label{app:eq:service_preferences}
\end{equation}
with $b_i\equiv B_i(y_i,X)>0$ and 
 $(x)_{+}\equiv\max\{x,0\}$, $\vartheta_i,\lambda_i>0$, and $u_i$
is increasing and concave. The function $B_i$ is increasing in the unit's
resource endowment $y_i=\omega_iY$.

The term $b_i$ is the unit's \emph{baseline service need}: it is the point at
which the marginal value of an additional unit of service falls discretely by
$\lambda_i$. Unlike a hard requirement, it softens the subsistence quantity
in the conventional Stone--Geary formulation \citep{Stone1954}; related
piecewise preference representations explicitly allow consumption both below
and above a subsistence threshold
\citep{BaumgaertnerDruppQuaas2017}. Baseline service needs aggregate into 
\begin{equation}
 S(Y,X)
 \equiv
 \int_0^1 B_i(\omega_iY,X)\,di,
 \label{app:eq:aggregate_service_needs}
\end{equation}

The dependence on $Y$ and $X$ allows baseline needs to vary with the
persistent scale and structure of state activity. Higher resource endowments
may be associated with greater expectations for heating, cooling, mobility,
productive activity, and public services, while climate, buildings,
transportation patterns, and industrial composition shift the services needed
to support a given scale of activity. Through these variables, $B_i$ captures the service-need channel. Some elements of
$X$ may also condition the production possibilities introduced below; when
this occurs, the corresponding empirical coefficient combines the two roles.
Current energy prices and efficiency policy do not enter $B_i$.

It is important to distinguish baseline \emph{service needs} from the energy
required to satisfy them. The object $b_i=B_i(y_i,X)$ is expressed in units of
energy services, such as thermal comfort, mobility, lighting, or productive
and public-service activity. Technology does not enter this reference need
directly. Instead, technology determines the energy input required to deliver
a given amount of services.

To illustrate, suppose that all units have the same baseline service need
$b>0$ and that two service-equivalent technologies are available, with
service-to-energy efficiencies $a_H>a_L>0$. Under the unit-mass normalization,
aggregate baseline services are $S=b$. If the baseline is fully attained, the
corresponding energy requirements are
\[
 q_H=\frac{S}{a_H}
 \qquad\text{and}\qquad
 q_L=\frac{S}{a_L},
\]
so the more efficient technology requires less energy to deliver the same
services.

More generally, let $a^{F}$ denote the efficiency of the best technology
feasibly available in the state. 

\subsection{Technology and timing}

We assume a two-period model distinguishing an investment date, indexed by $0$, from an operating
date, indexed by $1$. At date $0$, decision unit $i$ undertakes efficiency
investment $x_i\geq0$. This investment augments the installed efficiency-capital
stock available at date $1$ according to
\begin{equation}
 k_{i1}=(1-\delta_K)k_{i0}+x_i,
 \qquad
 0\leq\delta_K\leq1,
 \label{app:eq:capital_accumulation}
\end{equation}
where $k_{i0}$ is the inherited stock and $\delta_K$ is its depreciation rate.
The distinction between $x_i$ and $k_{i1}$ is important: $x_i$ is an investment
flow, whereas $k_{i1}$ is the stock of insulation, efficient equipment,
vehicles, buildings, and production processes available during operation.

At date $1$, installed efficiency capital and the composite energy input
$e_i\geq0$ jointly produce energy services according to the production function 
\begin{equation}
 z_i
 \leq
 A(E,X)k_{i1}^{\rho}e_i^{1-\rho},
 \qquad
 0<\rho<1,
 \label{app:eq:service_technology}
\end{equation}
where $E\equiv\textit{Eff}$ is the policy index and $A(E,X)>0$ is the
productivity of the available technology. The technology is
increasing, concave, constant-returns-to-scale, and strictly quasi-concave for
positive inputs. We allow $A_E(E,X)\geq0$ when policy supports the innovation, diffusion,
quality, or availability of more efficient technologies. Standards and investment incentives will enter separately when
policy and adoption choices are introduced.

For $e_i>0$, define the capital--energy ratio
\begin{equation}
 \kappa_i\equiv\frac{k_{i1}}{e_i}.
 \label{app:eq:capital_energy_ratio}
\end{equation}
On the technological boundary, the corresponding service-to-energy efficiency
is
\begin{equation}
 a(\kappa_i;E,X)
 \equiv
 \frac{z_i}{e_i}
 =
 A(E,X)\kappa_i^{\rho}.
 \label{app:eq:technological_efficiency}
\end{equation}
Thus, greater efficiency capital per unit of energy raises the services
obtained from that energy. At this stage, $a(\kappa_i;E,X)$ is a purely
technological object: it contains no prices. The subsequent cost-minimization problem will
determine the chosen ratio $\kappa_i^{*}$ and hence the economically relevant
benchmark efficiency. Finally, define the service-need attainment ratio as
\begin{equation}
 m_i\equiv\frac{z_i}{b_i(Y,X)}\geq0.
 \label{app:eq:need_attainment}
\end{equation}
Because the baseline is a need rather than a hard requirement, $m_i$ may lie
below, equal to, or above one.

\subsection{Frictionless benchmark}
\label{app:subsec:lifecycle_benchmark}

Let $P>0$ denote the price of the composite energy input and let
$R(E,X)>0$ denote the private user cost of one unit of efficiency-capital
services, net of any investment subsidy. Date-0 investment costs and date-1
operating costs are expressed in common present-value units, with discounting,
depreciation, and residual value incorporated into $R(E,X)$.

To define the state energy frontier, consider a frictionless environment
 in which date-1 efficiency capital can be
freely adjusted at user cost $R(E,X)$. Conditional on delivering any service quantity $z>0$, an agent 
chooses its capital--energy mix to minimize lifetime cost:
\begin{equation}
 \min_{k,e>0}
 \left\{
   R(E,X)k+Pe
 \right\}
 \quad\text{subject to}\quad
 z\leq A(E,X)k^\rho e^{1-\rho}.
 \label{app:eq:conditional_cost_problem}
\end{equation}
Positive input prices imply that the technological constraint binds. The
first-order conditions give the unique cost-\rev{minimizing} capital--energy ratio
\begin{equation}
 \kappa^{F}(P,E,X)
 \equiv
 \frac{k^{F}}{e^{F}}
 =
 \frac{\rho P}
      {(1-\rho)R(E,X)}.
 \label{app:eq:frontier_input_ratio}
\end{equation}
Consequently, a higher energy price induces substitution toward efficiency
capital, whereas a higher user cost of efficiency capital induces substitution
toward energy.

Evaluating the technological efficiency function
\eqref{app:eq:technological_efficiency} at this optimal ratio defines the
economically relevant benchmark efficiency:
\begin{align}
 a^{F}(P,E,X)
 &\equiv
 a\!\left(\kappa^{F}(P,E,X);E,X\right)
 \notag\\
 &=
 A(E,X)
 \left[
   \frac{\rho P}
        {(1-\rho)R(E,X)}
 \right]^\rho .
 \label{app:eq:frontier_efficiency}
\end{align}
Thus, $a(\kappa;E,X)$ is a price-free technological relationship, whereas
$a^{F}(P,E,X)$ is the service-to-energy ratio associated with the
lifetime-cost-\rev{minimizing} input mix.

The corresponding conditional factor demands are
\begin{equation}
 e^{F}(z)
 =
 \frac{z}{a^{F}(P,E,X)},
 \qquad
 k^{F}(z)
 =
 \kappa^{F}(P,E,X)
 \frac{z}{a^{F}(P,E,X)}.
 \label{app:eq:frontier_factor_demands}
\end{equation}
The minimum lifetime cost of delivering $z$ units of services is
$C^{F}(P,E,X)z$, where
\begin{align}
 C^{F}(P,E,X)
 &=
 \frac{1}{A(E,X)}
 \left[
   \frac{R(E,X)}{\rho}
 \right]^\rho
 \left[
   \frac{P}{1-\rho}
 \right]^{1-\rho}
 \notag\\
 &=
 \frac{P}
      {(1-\rho)a^{F}(P,E,X)}.
 \label{app:eq:frontier_unit_cost}
\end{align}
At the optimum,
\begin{equation}
 R(E,X)k^{F}(z)=\rho C^{F}(P,E,X)z,
 \qquad
 Pe^{F}(z)=(1-\rho)C^{F}(P,E,X)z.
 \label{app:eq:frontier_cost_shares}
\end{equation}

The reference-need frontier evaluates these conditional demands at
$z_i=b_i$. Because all units face the same state-level benchmark technology
and factor prices, the resulting aggregate energy requirement is
\begin{align}
 q^{*}(Y,P,E,X)
 &\equiv
 \int_0^1 e^{F}\!\left(b_i\right)\,di
 \notag\\
 &=
 \frac{\int_0^1 b_i\,di}
      {a^{F}(P,E,X)}
 =
 \frac{S(Y,X)}
      {a^{F}(P,E,X)}.
 \label{app:eq:aggregate_energy_frontier}
\end{align}
Aggregate benchmark efficiency-capital demand is correspondingly
\begin{equation}
 K^{F}(Y,P,E,X)
 =
 \int_0^1 k^{F}\!\left(b_i\right)\,di
 =
 \kappa^{F}(P,E,X)q^{*}(Y,P,E,X).
 \label{app:eq:aggregate_frontier_capital}
\end{equation}
This is a conditional economic energy frontier: it measures the energy that
would be required to meet aggregate baseline service needs using the
full-adjustment, lifetime-cost-\rev{minimizing} input mix.

Since $B_i$ is increasing in $y_i=\omega_iY$,
\begin{equation}
 \varepsilon_{S,Y}
 \equiv
 \frac{\partial\ln S(Y,X)}{\partial\ln Y}
 >0.
 \label{app:eq:service_need_elasticity}
\end{equation}
Holding $X$ fixed, the frontier therefore satisfies
\begin{equation}
 \frac{\partial\ln q^{*}}{\partial\ln Y}
 =\varepsilon_{S,Y}>0,
 \qquad
 \frac{\partial\ln q^{*}}{\partial\ln P}
 =-\rho<0.
 \label{app:eq:frontier_income_price_effects}
\end{equation}
The negative price effect is the conditional substitution effect: a higher
energy price raises the cost-\rev{minimizing} capital--energy ratio and reduces the
energy required to supply a fixed reference-service quantity.

Policy can also affect the frontier through technology availability and the
private user cost of efficiency investment. Suppose
\begin{equation}
 A_E(E,X)\geq0,
 \qquad
 R_E(E,X)\leq0,
 \label{app:eq:frontier_policy_assumptions}
\end{equation}
with at least one inequality strict. Then
\begin{equation}
 \frac{\partial\ln q^{*}}{\partial E}
 =
 -\frac{A_E(E,X)}{A(E,X)}
 +\rho\frac{R_E(E,X)}{R(E,X)}
 <0.
 \label{app:eq:frontier_policy_effect}
\end{equation}
Here the first term captures policy-supported technological quality,
diffusion, or availability, while the second captures subsidies or financing
that reduce the private user cost of efficiency capital.

\subsection{Actual service use}
\label{app:subsec:service_attainment}

Let $C_i(P,E,X)>0$ denote the lifetime cost per unit of service under unit
$i$'s installed technology. Thus, $C_i z_i$ is shorthand for the efficiency-capital and energy expenditures required to produce $z_i$; it is not an
additional good. A unit operating at the full-adjustment benchmark faces
$C_i=C^{F}(P,E,X)$, whereas a unit using a legacy technology may face a
different cost.

Conditional on its installed technology, unit $i$ allocates its resource
endowment between the numeraire and energy services:
\begin{equation}
 V_i(C_i,y_i;b_i)
 \equiv
 \max_{c_i,z_i\geq0}
 \left\{
  u_i(c_i)
  +\vartheta_i b_i
    \ln\!\left(1+\frac{z_i}{b_i}\right)
  -\lambda_i(b_i-z_i)_+
 \right\},
 \label{app:eq:actual_service_problem}
\end{equation}
subject to 
 $c_i+C_i z_i\leq y_i$.
Because preferences are increasing in the numeraire, the budget constraint is binding. Using
$m_i=z_i/b_i$, define the maximum affordable service-attainment ratio
\begin{equation}
 \overline m_i(C_i,y_i,b_i)
 \equiv
 \frac{y_i}{C_i b_i}.
 \label{app:eq:affordable_attainment}
\end{equation}
The problem can therefore be written as
\begin{equation}
 \max_{0\leq m_i\leq\overline m_i}
 \left\{
  u_i\!\left(y_i-C_i b_i m_i\right)
  +\vartheta_i b_i\ln(1+m_i)
  -\lambda_i b_i(1-m_i)_+
 \right\}.
 \label{app:eq:reduced_service_problem}
\end{equation}
The objective is concave. At an interior solution below the baseline need,
\begin{equation}
 C_i
 u_i'\!\left(y_i-C_i b_i m_i\right)
 =
 \frac{\vartheta_i}{1+m_i}+\lambda_i,
 \qquad 0<m_i<1,
 \label{app:eq:service_foc_below}
\end{equation}
whereas above the baseline need,
\begin{equation}
 C_i
 u_i'\!\left(y_i-C_i b_i m_i\right)
 =
 \frac{\vartheta_i}{1+m_i},
 \qquad m_i>1.
 \label{app:eq:service_foc_above}
\end{equation}
Provided that $m_i=1$ is affordable, exact attainment of the baseline need is
optimal when
\begin{equation}
 \frac{\vartheta_i}{2}
 \leq
 C_i u_i'\!\left(y_i-C_i b_i\right)
 \leq
 \frac{\vartheta_i}{2}+\lambda_i.
 \label{app:eq:service_kink_condition}
\end{equation}
The interval arises because the marginal value of services falls discretely
at $z_i=b_i$.

For closed-form results, consider the quasi-linear specialization
$u_i(c_i)=c_i$. Let $\widehat m_i(C_i)$ denote the desired attainment ratio
in the absence of the  constraint. It is
\begin{equation}
 \widehat m_i(C_i)
 =
 \begin{cases}
 \displaystyle
 \frac{\vartheta_i}{C_i}-1,
 & 0<C_i<\vartheta_i/2,
 \\[3mm]
 1,
 & \vartheta_i/2
   \leq C_i
   \leq\vartheta_i/2+\lambda_i,
 \\[2mm]
 \displaystyle
 \frac{\vartheta_i}{C_i-\lambda_i}-1,
 & \vartheta_i/2+\lambda_i
   <C_i
   <\vartheta_i+\lambda_i,
 \\[3mm]
 0,
 & C_i\geq\vartheta_i+\lambda_i.
 \end{cases}
 \label{app:eq:desired_attainment}
\end{equation}
The feasible solution is therefore
\begin{equation}
 m_i^*
 =
 \min\left\{
   \widehat m_i(C_i),
   \overline m_i(C_i,y_i,b_i)
 \right\},
 \qquad
 z_i^*=b_i m_i^*,
 \qquad
 c_i^*=y_i-C_i b_i m_i^*.
 \label{app:eq:optimal_service_attainment}
\end{equation}

The solution has four economic regimes. When service provision is sufficiently
inexpensive, the unit consumes services above its baseline need. For an
intermediate range of costs, the kink makes exact attainment optimal. At
higher costs, the need is only partially fulfilled, and at sufficiently high
cost the unit may choose no service. In any regime, the affordability
constraint may further reduce attainment.

Thus, holding $y_i$ and $b_i$ fixed, $m_i^*$ is weakly decreasing in $C_i$.
A reduction in service cost can therefore have two conceptually
different effects: it can help achieve baseline needs when
$m_i^*<1$, or increase discretionary service use when $m_i^*>1$. The latter
is the direct utilization or rebound channel.

\subsection{Legacy technology, standards, and adoption}
\label{app:subsec:technology_adoption}

The full-adjustment technology need not be installed by every agent.
Before upgrading, agent $i$ operates a legacy technology characterized by the
Hicks-neutral productivity wedge $d_i\geq1$:
\begin{equation}
 z_i
 \leq
 \frac{A(E,X)}{d_i}
 k_{i1}^{\rho}e_i^{1-\rho}.
 \label{app:eq:legacy_technology}
\end{equation}
The wedge summarizes differences in equipment vintages, buildings, vehicles,
production processes, and other inherited technology characteristics. Because
it is Hicks neutral, it does not change the cost-\rev{minimizing} capital--energy
ratio. Instead, relative to the benchmark, it proportionally raises both
capital and energy requirements per unit of service.

A minimum-performance standard restricts the least efficient technology that
may remain in operation. Let $\overline d(E)\geq1$ denote the largest
permissible technology wedge, with
\begin{equation}
 \overline d'(E)\leq0.
 \label{app:eq:standard_stringency}
\end{equation}
The effective post-compliance wedge of a nonadopter is therefore
\begin{equation}
 \widetilde d_i(E)
 \equiv
 \min\{d_i,\overline d(E)\}
 \geq1.
 \label{app:eq:compliant_legacy_wedge}
\end{equation}
Thus, a standard may require a partial improvement without requiring adoption
of the full benchmark technology. The object $\widetilde d_i(E)$ should be
interpreted as the effective installed-stock constraint after accounting for
compliance, enforcement, and capital turnover, rather than as instantaneous
replacement of every legacy asset.

Let $D_i\in\{0,1\}$ indicate full adoption of the benchmark technology. Define
the installed-technology wedge as
\begin{equation}
 \delta_i(D_i,E)
 \equiv
 D_i+(1-D_i)\widetilde d_i(E).
 \label{app:eq:installed_technology_wedge}
\end{equation}
Hence $\delta_i(1,E)=1$, whereas
$\delta_i(0,E)=\widetilde d_i(E)\geq1$. The corresponding technology can be
written as
\begin{equation}
 z_i
 \leq
 \frac{A(E,X)}{\delta_i(D_i,E)}
 k_{i1}^{\rho}e_i^{1-\rho}.
 \label{app:eq:regime_technology}
\end{equation}

Because the technology wedge is Hicks neutral, conditional cost minimization
within either regime preserves the benchmark ratio $\kappa^F$. Consequently,
the regime-specific service-to-energy efficiency and lifetime unit cost are
\begin{equation}
 a_i(D_i;P,E,X)
 =
 \frac{a^F(P,E,X)}{\delta_i(D_i,E)},
 \qquad
 C_i(D_i;P,E,X)
 =
 \delta_i(D_i,E)C^F(P,E,X).
 \label{app:eq:regime_efficiency_cost}
\end{equation}

Full benchmark adoption entails a fixed date-0 expenditure $f_i\geq0$.
Let $T_i(E)$ be a fixed adoption subsidy satisfying
$0\leq T_i(E)\leq f_i$ and $T_i'(E)\geq0$. The net fixed adoption expenditure is
\begin{equation}
 \Phi_i(E)
 \equiv
 f_i-T_i(E)
 \geq0.
 \label{app:eq:net_adoption_expenditure}
\end{equation}
This expenditure captures fixed installation, certification, search,
contracting, or disruption costs. It does not include the scalable
efficiency-capital expenditure already contained in $C_i(D_i)z_i$. Similarly,
let $g_i(E)\geq0$ denote any fixed expenditure required to bring a legacy
technology into compliance, with $g_i(E)=0$ whenever $d_i\leq\overline d(E)$.

Spendable resources under regime $D_i$ are
\begin{equation}
 \widetilde y_i(D_i,E)
 \equiv
 y_i-D_i\Phi_i(E)-(1-D_i)g_i(E),
 \label{app:eq:regime_resources}
\end{equation}
and are assumed to be nonnegative for every feasible regime. The baseline need
$b_i=B_i(y_i,X)$ remains fixed across the two regimes: adoption expenditures
change available consumption resources, not the predetermined reference need.

Using the indirect value function in
equation~\eqref{app:eq:actual_service_problem}, the optimized value of regime
$D_i$ is
\begin{equation}
 W_i(D_i)
 \equiv
 V_i\!\left(
   C_i(D_i;P,E,X),
   \widetilde y_i(D_i,E);
   b_i
 \right).
 \label{app:eq:technology_regime_value}
\end{equation}
The optimal adoption decision is
\begin{equation}
 D_i^*
 \in
 \arg\max_{D_i\in\{0,1\}}W_i(D_i).
 \label{app:eq:optimal_adoption}
\end{equation}
Equivalently, with ties resolved in favor of adoption,
\begin{equation}
 D_i^*=1
 \quad\Longleftrightarrow\quad
 V_i\!\left(
   C^F,
   y_i-\Phi_i;
   b_i
 \right)
 \geq
 V_i\!\left(
   \widetilde d_i C^F,
   y_i-g_i;
   b_i
 \right),
 \label{app:eq:adoption_threshold}
\end{equation}
where the arguments $(P,E,X)$ have been suppressed for readability. This is a
forward-looking adoption rule: the unit compares optimized utility under the
two technologies, allowing technology choice to affect both service use and
numeraire consumption.

Define the equilibrium installed wedge, resources, and service-attainment
ratio by
\begin{equation}
 \delta_i^*
 \equiv
 \delta_i(D_i^*,E),
 \qquad
 \widetilde y_i^*
 \equiv
 \widetilde y_i(D_i^*,E),
 \label{app:eq:equilibrium_regime_objects}
\end{equation}
and
\begin{equation}
 m_i^*
 =
 \min\left\{
   \widehat m_i\!\left(\delta_i^*C^F\right),
   \frac{\widetilde y_i^*}
        {\delta_i^*C^F b_i}
 \right\}.
 \label{app:eq:equilibrium_attainment}
\end{equation}
Actual services and energy use are then
\begin{equation}
 z_i^*=b_i m_i^*
 \label{app:eq:actual_services}
\end{equation}
and
\begin{align}
 e_i^*
 &=
 \frac{z_i^*}{a_i(D_i^*)}
 =
 \frac{b_i m_i^*}{a^F/\delta_i^*}
 \notag\\
 &=
 \frac{b_i}{a^F(P,E,X)}
 \delta_i^*m_i^*.
 \label{app:eq:individual_energy_identity}
\end{align}
Equation~\eqref{app:eq:individual_energy_identity} is the individual
micro-to-macro building block. The factor $\delta_i^*$ captures the installed
technology, compliance, and adoption margin, while $m_i^*$ captures incomplete
need attainment and additional service use.

The policy channels are distinct. Changes in $A(E,X)$ and $R(E,X)$ shift the
 frontier; standards reduce the permissible legacy wedge
$\widetilde d_i(E)$; and fixed subsidies reduce $\Phi_i(E)$ and thereby
encourage benchmark adoption. A tighter standard can nevertheless reduce the
incremental private value of full adoption by improving the nonadoption
alternative. Consequently, stronger policy reduces the composite energy gap only when
its technology, compliance, adoption, and conservation effects dominate any
crowding out of full adoption and induced service use.

\subsection{Exact aggregation and deterministic comparative statics}
\label{app:subsec:exact_aggregation}

For notational correspondence with the main text, define
\begin{equation}
 a(P,E,X)\equiv a^F(P,E,X).
 \label{app:eq:efficiency_notation_bridge}
\end{equation}
Assume that $b_i$, $\delta_i^*$, and $m_i^*$ are measurable across decision
units and that
\begin{equation}
 0<S(Y,X)=\int_0^1b_i\,di<\infty,
 \qquad
 \int_0^1b_i\delta_i^*m_i^*\,di<\infty.
 \label{app:eq:aggregation_regularity}
\end{equation}
Define the baseline-service weights
\begin{equation}
 w_i(Y,X)
 \equiv
 \frac{b_i}{S(Y,X)},
 \qquad
 w_i>0,
 \qquad
 \int_0^1w_i\,di=1.
 \label{app:eq:baseline_service_weights}
\end{equation}

Using equation~\eqref{app:eq:individual_energy_identity}, aggregate \rev{realized}
energy use is
\begin{align}
 q^e(Y,P,E,X)
 &\equiv
 \int_0^1e_i^*\,di
 \notag\\
 &=
 \frac{1}{a(P,E,X)}
 \int_0^1b_i\delta_i^*m_i^*\,di
 \notag\\
 &=
 \frac{S(Y,X)}{a(P,E,X)}
 \int_0^1w_i\delta_i^*m_i^*\,di.
 \label{app:eq:aggregate_realised_energy}
\end{align}
Because
\[
 q^*(Y,P,E,X)
 =
 \frac{S(Y,X)}{a(P,E,X)},
\]
equation~\eqref{app:eq:aggregate_realised_energy} gives the exact
decomposition
\begin{equation}
 q^e(Y,P,E,X)
 =
 q^*(Y,P,E,X)I(Y,P,E,X),
 \label{app:eq:exact_frontier_gap_decomposition}
\end{equation}
where
\begin{equation}
 I(Y,P,E,X)
 \equiv
 \int_0^1w_i(Y,X)\delta_i^*m_i^*\,di.
 \label{app:eq:composite_energy_gap}
\end{equation}

The object $I$ is a baseline-service-weighted composite of two margins.
The installed-technology wedge $\delta_i^*\geq1$ captures legacy technology,
compliance, and adoption, while $m_i^*$ captures actual service attainment and
additional service use.

\paragraph{Aggregate lower-envelope condition.}

Because the soft-needs problem permits $m_i^*<1$, the inequality $I\geq1$
does not follow from individual optimization. We therefore maintain the
aggregate lower-envelope condition
\begin{equation}
 \int_0^1w_i\delta_i^*m_i^*\,di\geq1.
 \label{app:eq:aggregate_lower_envelope}
\end{equation}
Equivalently,
\begin{equation}
 I-1
 =
 \int_0^1w_i
 \left[
   (\delta_i^*-1)m_i^*
   +(m_i^*-1)
 \right]di
 \geq0.
 \label{app:eq:lower_envelope_decomposition}
\end{equation}
Thus, excess energy associated with legacy technologies, incomplete adoption,
and above-baseline service use must, in the aggregate, be at least as large as
the reduction in energy associated with unmet service needs. A stronger
sufficient condition is
\begin{equation}
 \delta_i^*m_i^*\geq1
 \quad\text{for almost every }i,
 \label{app:eq:pointwise_lower_envelope}
\end{equation}
but equation~\eqref{app:eq:aggregate_lower_envelope} permits some individual
units to fall below the reference benchmark.

Under this maintained condition,
\begin{equation}
 I\geq1,
 \qquad
 q^e\geq q^*.
 \label{app:eq:one_sided_deterministic_gap}
\end{equation}
The condition is a restriction on the aggregate support of the model.

\paragraph{Price and the composite gap}

Let $m_i^1(P)$ and $m_i^0(P)$ denote optimal service attainment under,
respectively, benchmark adoption and compliant nonadoption. The corresponding
relative energy requirements are
\begin{equation}
 r_i^1(P)\equiv m_i^1(P),
 \qquad
 r_i^0(P)\equiv\widetilde d_i(E)m_i^0(P).
 \label{app:eq:regime_relative_energy}
\end{equation}
Within either fixed technology regime, a higher energy price raises lifetime
service cost because
\begin{equation}
 \frac{\partial\ln C^F}{\partial\ln P}
 =1-\rho>0.
 \label{app:eq:service_cost_price_effect}
\end{equation}
The service-choice solution therefore implies that $m_i^D(P)$ is weakly
decreasing in $P$, holding the technology regime fixed.

For the adoption margin, impose the following sufficient conditions:
\begin{equation}
 P'\geq P
 \quad\Longrightarrow\quad
 D_i^*(P')\geq D_i^*(P)
 \quad\text{for almost every }i,
 \label{app:eq:price_monotone_adoption}
\end{equation}
and
\begin{equation}
 m_i^1(P)
 \leq
 \widetilde d_i(E)m_i^0(P).
 \label{app:eq:energy_saving_adoption}
\end{equation}
The first is a single-crossing condition under which higher energy prices do
not discourage benchmark adoption. The second is a no-backfire condition:
the service response following adoption cannot be sufficiently large to make
the adopter use more energy than under the compliant legacy technology.

Under these conditions, $\delta_i^*m_i^*$ is weakly decreasing in $P$ for
almost every unit. Since the weights $w_i(Y,X)$ do not depend on current
energy prices,
\begin{equation}
 \frac{\partial I}{\partial P}\leq0.
 \label{app:eq:gap_price_direction}
\end{equation}
This is a sufficient, rather than unconditional, price-gap result.

\paragraph{Policy and the composite gap.}

Stronger policy can reduce $\delta_i^*$ through standards, compliance, and
subsidized adoption. At the same time, improvements in technology and lower
private investment costs reduce the price of services and may increase
$m_i^*$ through greater need attainment or rebound.

A transparent sufficient condition for the technology and adoption effects to
dominate the service-use response is that, for every $E'\geq E$,
\begin{equation}
 \delta_i^*(E')m_i^*(E')
 \leq
 \delta_i^*(E)m_i^*(E)
 \quad\text{for almost every }i.
 \label{app:eq:policy_dominance_condition}
\end{equation}
Where the relevant functions are differentiable and strictly positive, the
same condition can be written locally as
\begin{equation}
 -\frac{\partial\ln\delta_i^*}{\partial E}
 \geq
 \frac{\partial\ln m_i^*}{\partial E}.
 \label{app:eq:local_policy_dominance}
\end{equation}
This condition states that the proportional improvement in installed
technology must be at least as large as the induced proportional increase in
service use. Because $w_i$ does not depend on current policy, it implies
\begin{equation}
 \frac{\partial I}{\partial E}\leq0.
 \label{app:eq:gap_policy_direction}
\end{equation}
Without condition~\eqref{app:eq:policy_dominance_condition}, the policy effect
on the composite gap is ambiguous.

\paragraph{Resource endowments and the composite gap}

The effect of $Y$ on the composite gap has no determinate sign. Wherever the
relevant derivatives exist,
\begin{equation}
 \frac{\partial I}{\partial Y}
 =
 \int_0^1
 \frac{\partial w_i}{\partial Y}
 \delta_i^*m_i^*\,di
 +
 \int_0^1
 w_i
 \frac{\partial(\delta_i^*m_i^*)}{\partial Y}\,di.
 \label{app:eq:gap_resource_derivative}
\end{equation}
Greater resources may make fixed adoption expenditures easier to bear and
thereby reduce $\delta_i^*$. Conversely, they may relax affordability
constraints, increase service attainment, and alter the composition of
baseline-service weights, thereby increasing $m_i^*$ or placing more weight on
service-intensive units. Consequently,
\begin{equation}
 \frac{\partial I}{\partial Y}
 \gtreqless0.
 \label{app:eq:gap_resource_ambiguity}
\end{equation}

\paragraph{Deterministic model implications.}

Combining the frontier and gap results gives
\begin{equation}
 \frac{\partial\ln q^*}{\partial\ln Y}>0,
 \qquad
 \frac{\partial\ln q^*}{\partial\ln P}=-\rho<0,
 \qquad
 \frac{\partial\ln q^*}{\partial E}<0,
 \label{app:eq:final_frontier_directions}
\end{equation}
and, under the sufficient conditions stated above,
\begin{equation}
 \frac{\partial I}{\partial P}\leq0,
 \qquad
 \frac{\partial I}{\partial E}\leq0,
 \qquad
 \frac{\partial I}{\partial Y}\gtreqless0.
 \label{app:eq:final_gap_directions}
\end{equation}
Because $q^e=q^*I$, the corresponding total deterministic effects satisfy
\begin{align}
 \frac{\partial\ln q^e}{\partial\ln P}
 &=
 -\rho+
 \frac{\partial\ln I}{\partial\ln P}
 <0,
 \label{app:eq:total_price_effect}\\
 \frac{\partial\ln q^e}{\partial E}
 &=
 -\frac{A_E(E,X)}{A(E,X)}
 +\rho\frac{R_E(E,X)}{R(E,X)}
 +\frac{\partial\ln I}{\partial E}
 <0
 \label{app:eq:total_policy_effect}
\end{align}
whenever the respective gap conditions hold. The total resource-endowment
effect remains potentially ambiguous because the positive frontier effect may
be accompanied by either sign of the composite-gap response.

Restoring state and year subscripts, the deterministic model therefore delivers
\begin{equation}
 q_{st}
 =
 q_{st}^{*}I_{st},
 \qquad
 q_{st}^{*}
 =
 \frac{S(Y_{st},X_{st})}
      {a(P_{st},\textit{Eff}_{st},X_{st})},
 \qquad
 I_{st}\geq1.
 \label{app:eq:final_state_relationship}
\end{equation}
At adoption thresholds, binding standards, and service-demand kinks, the
comparative statics should be interpreted as weak finite-difference
statements rather than ordinary derivatives.

The model thus separates three deterministic objects: baseline service needs,
the benchmark energy requirement for satisfying those needs, and the composite
difference between benchmark and \rev{realized} energy use. The latter combines
installed technology, compliance, adoption, incomplete need attainment, and
additional service use. It is therefore a reference-baseline energy-use gap,
not a pure measure of technical inefficiency or a welfare loss. This completes
the deterministic microfoundation; the statistical assumptions used to
estimate the frontier are introduced separately in the empirical section.

%\pagebreak
%\listoffigures

\end{document}